%% file: Battery_v8_arxiv.tex
\documentclass[11pt]{article}

\usepackage[top=1.0in,right=1.0in,left=1.0in,bottom=1.75in]{geometry}
\usepackage{latexsym}
\usepackage{amssymb,amsbsy}
\usepackage{bm}
\usepackage{amsmath}
\usepackage{amsthm}
\usepackage{graphics,graphicx}
\usepackage[T1]{fontenc}
\usepackage{color}
\usepackage[affil-it,auth-sc]{authblk}
\usepackage{stmaryrd}
\usepackage{verbatimbox}
\usepackage{subcaption}
\usepackage{dsfont}
\usepackage{booktabs}
\usepackage{tabularx}
\usepackage{boldline,multirow}
\usepackage{longtable}
\usepackage{psfrag}
\usepackage{array}

\usepackage[]{jabbrv}

\usepackage[numbers,sort&compress]{natbib}

\input{definitions}

\newcommand{\bluecom}[1]{{\color{blue} {#1}}}

\newcommand{\trace}[1]{ {\rm tr} \left[ \, {#1} \, \right] }

\newcommand{\deviatoric}[1]{ {\rm dev} \left[ \, {#1} \, \right] }
\newcommand{\divergence}[1]{ {\rm div} \left[ \, {#1} \, \right] }
\newcommand{\gradient}[1]{ {\rm \nabla} \left[ \, {#1} \, \right] }

\newcommand{\vect}[1]{\vec{#1}} 
\newcommand{\tensor}[1]{  { \boldsymbol{#1}} } 

\newcommand{\diffusivity}{\mbox{${\rm D} \mskip-8mu  | \,$}}

\newcommand{\mychi}{\mbox{$\chi \mskip-8mu  | $}}

\graphicspath{{./}{./figures/}}

\title{A novel two-mechanism model for all solid state Li-ion batteries: review and comparisons. }

\author[1]{L. Cabras}
\author[2]{V. Oancea}
\author[1]{A. Salvadori}

\affil[1]{Dipartimento di Ingegneria Meccanica e Industriale, Universit\`a di Brescia, Italy}
\affil[2]{Dassault Systemes Simulia Corp, United States}

\begin{document}

\maketitle

\begin{abstract}

In recent years there has been a major interest in developing all solid state batteries, for the sake of safety (issues due to toxic and flammable organic liquid electrolytes 
are well known) as well as of high energy density \cite{SCHNELL2018160,Zheng2018}. While liquid electrolyte cells constitute for now the vast majority of commercial cells, 
solid electrolyte batteries show great promise. 
In parallel to experimental investigations, computational simulations \cite{GrazioliEtAlCM2016} unveil the several 
physics that interconnect at different scales \cite{LiMonroeARCBM2020} during battery operations. 
We propose herein a new advanced model, which is multiscale compatible and fully three dimensional in nature.
Furthermore, we review some classical models, highlighting the conceptual evolutions that 
are ultimately collected in our work.
The model is validated against experimental evidences of  a Li/LiPON/LiCoO$_2$ thin film battery published in \cite{Danilovetal2011}.  

\end{abstract}



%

\section{Introduction}
\label{sec01}

Several case studies shown that conventional Li-ion batteries, based on porous electrodes and liquid electrolytes, are prone to chemo-mechanical degradation and affected by environmental and safety issues, because the toxic organic liquid electrolytes are flammable. All-solid-state batteries (ASSB), on the contrary, are based on solid electrolytes, which combine superior thermal and electrochemical stability and avoid hazardous liquid electrolyte leakage. This makes ASSB ideal devices to be deployed in medical implants and wireless sensor applications, and promising for implementing the green economy revolution \cite{Sun:2020tw}. 

Largely funded international projects \cite{Pasta_2020} and strategic action plans ( as for the European battery 2030+ initiative ) assess how the scientific community trusts and supports the development of next-generation storage systems that may meet the social quest for decarbonization. ASSBs are among the best candidates, provided that they achieve ultra-high-performances yet meeting sustainability and safety.
Among several challenges in the development of cells based on solid electrolytes, the two most crucial are to attain a high ionic conductivity at room temperature and to minimize the high resistances at the contact area between
active material and electrolyte \cite{Santhanagopalan2014}, either in composite electrolytes/electrodes morphologies \cite{BielefeldEtAlJPC2019} or in thin films (glass layers)\cite{Cao2014}. Experimental investigations are presently mainly focused on key materials and structures \cite{FanEtAlAEM2018,YuanACS2020,ZhengCSR2020} since ASSB suffer from significant chemo-mechanical problems \cite{ZhangEtAlNANOENERGY2020} as dendritic growth \cite{WALDMANN2018107,PorzEtAlAEM2017,SHISHVAN2020444}, various interfacial aging phenomena \cite{ChenEtAlAEM2021} and mechanical damage \cite{Tian2017}.

Side by side to experimental development, computational modeling are able to depict the several 
physics that interconnect at different scales, providing a profound understanding of processes and limiting factors. Several advanced mathematical models have been 
published and we do not aim here at a broad review, as done in \cite{GrazioliEtAlCM2016} or \cite{LiMonroeARCBM2020} among others. 

In this note we present a novel, fully three-dimensional and multiscale compatible (in the sense of \cite{SalvadoriEtAlJPS2015}), two-mechanisms advanced all-solid-state Li-ion battery model, leaving to a companion publication the detailed analysis of the solid electrolyte governing equations for a more readable and efficient organization. Besides this theoretical and numerical effort, we also review a set of few antecedent models.

It is always hard, if not even impossible, to categorize the whole bibliography on a subject and undoubtably identify those cornerstone papers that changed the flow of the subject itself. Perhaps Newman's theory might be assigned without presumptions in this class, for the electrochemistry of batteries. As in several aspects of human being, history is a succession of little advancements, in a collective way. In this spirit, we did not select the papers in this brief review as cornerstones and indeed we will be extremely grateful to readers that may pinpoint to other fundamental contributions. We selected three notable papers as they include ideas which make the backbone of the new model we are going to present in section \ref{subsec:novelformulation} and as such, papers with which outcomes from our new model shall be compared in terms of outcomes and principles.

The first paper we are going to describe in section \ref{subsec:Fabre} is a one-dimensional model of a Li/LiPON thin-film micro battery, developed by Fabre et al. \cite{Fabre2012}. The ionic transfer in the solid electrolyte is described by a single ion conduction model: as such, since the negative vacancies in the lattice are firmly held and do not flow, 
the concentration of lithium ions across the solid electrolyte $\rm Li^+$ is uniform and known a priori in view of the electroneutrality in the electrolyte, which holds almost everywhere as discussed in \cite{SalvadoriEtAlIJSS2015} (see also \cite{MykhaylovEtAlJMPS2019}). Because no concentration gradient drives the ionic motion, the focus of the model stands in finding the potential drop, governed essentially by Ohm's law.

The second contribution, which will be analyzed in section \ref{subsec:Lands}, is an advanced model framework for ASSB proposed by Landstorfer and coworkers in \cite{LandstorferEtAlPCCP2011}. As in \cite{Fabre2012},  they considered a single ion conduction model, enriching the description of the interfaces mechanisms, standing from a rigorous thermodynamic setting that ultimately leads to interface conditions of non Butler-Volmer type. Capacitance within interfaces was captured also in the third paper we will review herein, authored by Raijmakers and co-workers in 2020. Their fundamental contribution stands in a two-mechanism conduction model, in which both interstitial lithium and negative vacancies are allowed to move independently, thus creating a concentration gradient at steady state that resembles the liquid electrolytes distributions found for instance in \cite{SalvadoriEtAlJPS2015,SalvadoriEtAlJPS2015b}. 

In this note we revise and extend the vision of two-mechanism conduction proposed in \cite{RaijmakersEtAlEA2020}. In this regard, we split the positive ionic flux into two contributions, an interstitial one of the same kind of \cite{RaijmakersEtAlEA2020} and a hopping mechanism, capable of filling negative vacancies without fictitiously assuming that those have their own motility, which does not appear physically motivated. We also include the complex description of interfaces from \cite{RaijmakersEtAlEA2020}.

The paper is organized as follows. A brief review of models and their governing equations for all-solid-state batteries is proposed in section \ref{sec02}, where models are examined theoretically. An experimental benchmark, taken from \cite{Danilovetal2011}, is devised in section \ref{sec:benchmarks} in order to validate the novel model. The finite element solution scheme provides numerical approximations for the electric potential profiles, interface currents, fluxes and concentrations profiles. These quantities of interest were used to compare the new model against the antecedent published in \cite{RaijmakersEtAlEA2020}.

Concluding remarks complete this paper.

\section{A brief review of models and their governing equations for all-solid-state batteries}
\label{sec02}

\begin{figure}[!htb]
\centering
\includegraphics[width=120mm]{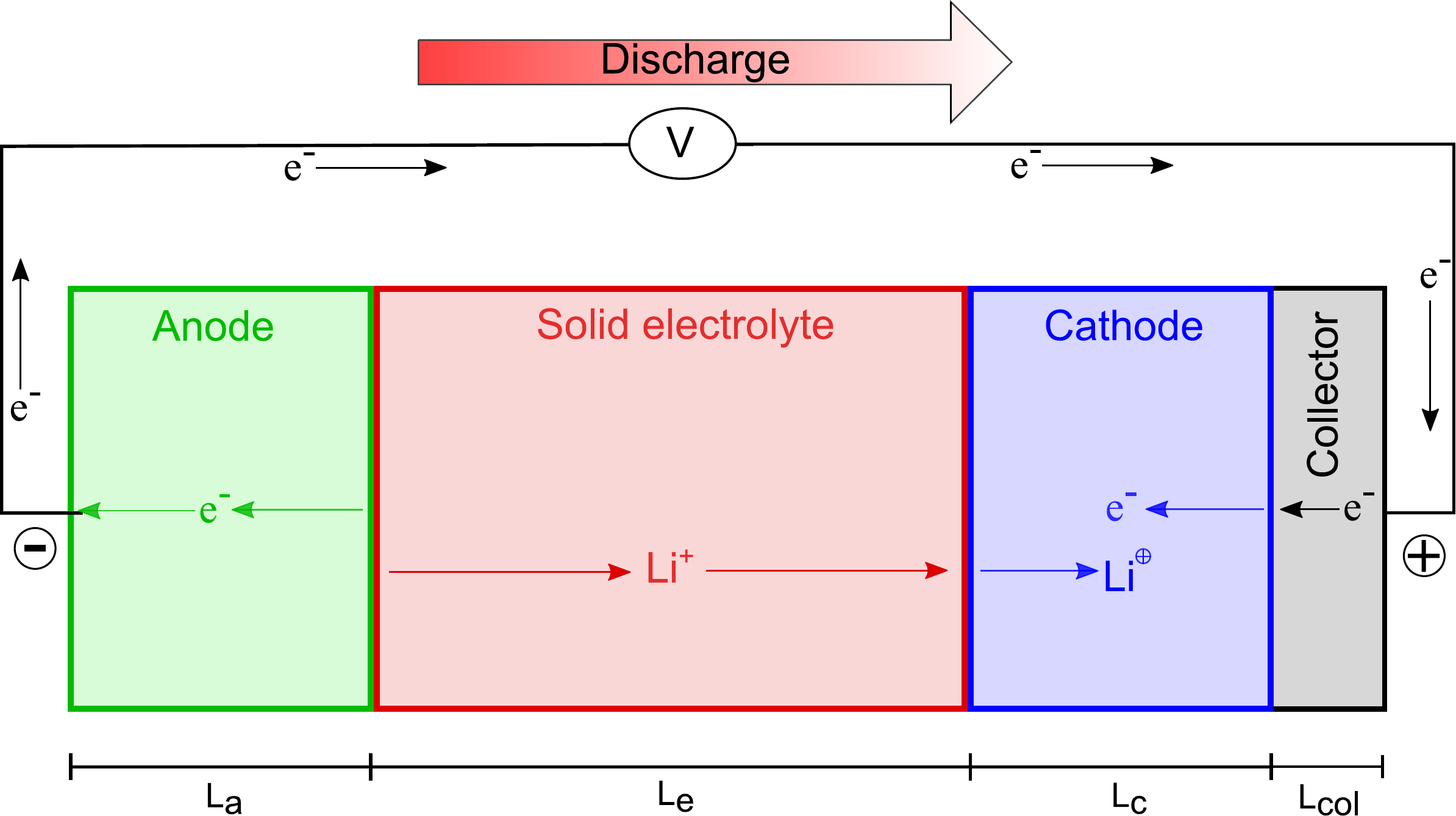}
\caption{ \em One dimensional model of a thin-film all-solid-state Li-ion battery (discharge conditions). The positive electrode is denoted with a $\oplus$-sign and the negative electrode side with a $\ominus$-sign. }
\label{fig:model}
\end{figure}

Different mathematical models, some of which reviewed in \cite{GrazioliEtAlCM2016,LiMonroeARCBM2020}, have been proposed and tailored to the battery microstructure, which simplistically can be categorized into two types: thin films and porous electrodes.
A broad set of models that include pseudo-2-D \cite{Doyle1995a,Fuller1994a}, multiscale \cite{SalvadoriEtAlJMPS2013,FrancoRCS2013} and fine-grained models \cite{LeeJPS2016} have been designed for electrodes made of porous materials, accounting for the different phases and attempting at capturing realism either in liquid \cite{FrancoEtAlCR2019} or solid electrolytes \cite{BielefeldEtAlJPC2019}. We will  restrict our focus on thin films, all-solid-state Lithium phosphorous oxy-nitride ``LiPON'' electrolytes. A planar geometry is generally accepted in thin films batteries, since the ratio between the lateral dimension and the thickness is large enough for the lateral dimension to be considered as infinite, see Fig.\ref{fig:model}.
Since furthermore electrodes and electrolyte can be well approximated by homogeneous planar materials, one-dimensional mathematical models are customary formulations for thin films batteries.

An all-solid-state electrochemical cell incorporates two electrodes and a solid electrolyte, as depicted schematically in Fig.\ref{fig:model}. Li-ions are extracted from the cathode (positive electrode) during charge and inserted back during discharge. The opposite holds for the anode (negative electrode). Assuming $\rm LiCoO_2$ (shortened in LCO) to be the positive electrode material, the basic electrochemical charge-transfer reaction writes
\begin{align}
{\rm LiCoO_2}
\underset{k_{-1}}{\overset{k_{1}}{\rightleftarrows}} 
{\rm Li}_{1-x}{\rm CoO_2} + x {\rm Li^+} + x {\rm e^-} \quad 0\leq x \leq \frac{1}{2}
\; .
\label{eq:a/e}
\end{align}
If lithium foil serves as the negative electrode material, deposition and extraction at the negative surface is described by the reaction
\begin{align}
{\rm Li}
\underset{k_{-2}}{\overset{k_{2}}{\rightleftarrows}}
{\rm Li^+ + e^-}
\; .
\label{eq:e/c}
\end{align}

\begin{figure}[!htb]
\centering
\includegraphics[width=150mm]{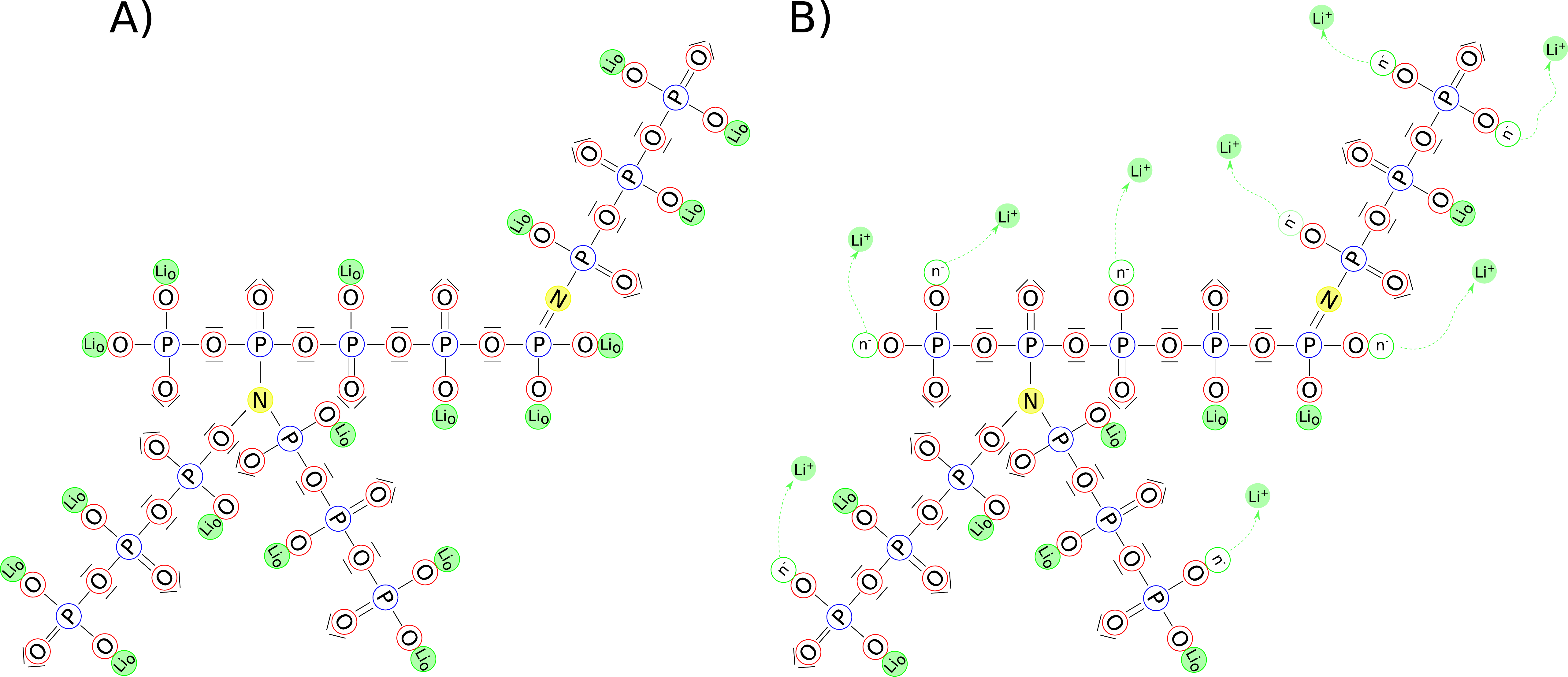}
\caption{\em LiPON matrix with triply- and doubly-coordinated nitrogen (a). Movements of charged particles towards the interstitial space and by means of particle hopping, representing the main ionic conductivity mechanisms in LiPON (b).}
\label{fig:ReticoloDanilov}
\end{figure}

The structure of the solid electrolyte is shown in Fig.\ref{fig:ReticoloDanilov}.  $\rm Li_0$ denotes the lithium (ionic) bound to the non-bridging oxygen atoms, $\rm Li^+$ is a lithium ion which has motion capabilities (either as transferred to the meta-stable interstitial state or hopping), and $\rm n^-$ is the uncompensated negative charge associated with a vacancy formed in the LiPON matrix at the place where lithium was originally located. 
The maximal concentration of host-sites, denoted with $c_0$, is determined by the stoichiometric composition of the electrolyte material. It is reached in the ideal case of absolute zero temperature, when all available host sites are fully filled with lithium ions and the ionic conductivity vanishes because all ions are immobile. In standard conditions, some of the Li ions are thermally excited and the chemical ionization reaction 
\begin{equation}
{\rm Li_0} \underset{k_b^{\rm ion}}{\overset{k_f^{\rm ion}}{\rightleftarrows}} {\rm Li^+ + n^-}
\label{eq:IonizationReaction}
\end{equation}
occurs, $k_f^{\rm ion}$ and $k_b^{\rm ion}$ being the forward and backward rate constants for the ionization and recombination reaction, respectively.
%
%

During charge the lithium ions cross the electrolyte and are reduced into metallic $\rm Li$ at the lithium foil surface; vice-versa during discharge. Mobile species in the electrolyte are therefore ions and predicting the behavior of electrochemical cells requires quantitative modeling of the kinetics of mobile ionic species. $\rm LiCoO_2$ electrode contains lithium oxide and the lithium exists as  ions as part of a lithium salt. We can thus state that $\rm Li$ intercalates in the cathode as ions as well, which will be denoted with $\rm Li^\oplus$ to stress its ionic nature ``shielded'' by its own electron, while distinguishing it from mobile charges $\rm Li^+$ in the electrolyte.

In the following, we will overview {\bluecom{three}} electro-chemical models proposed in the literature. Eventually, we will summarize a model discussed at large in a companion paper, to provide numerical comparisons in the rest of the paper.

\subsection{Charge/Discharge Simulation of an All-Solid-State Thin-Film Battery Using a One-Dimensional Model  \cite{Fabre2012}} 
\label{subsec:Fabre}

\begin{figure}[!htb]
\centering
\includegraphics[width=120mm]{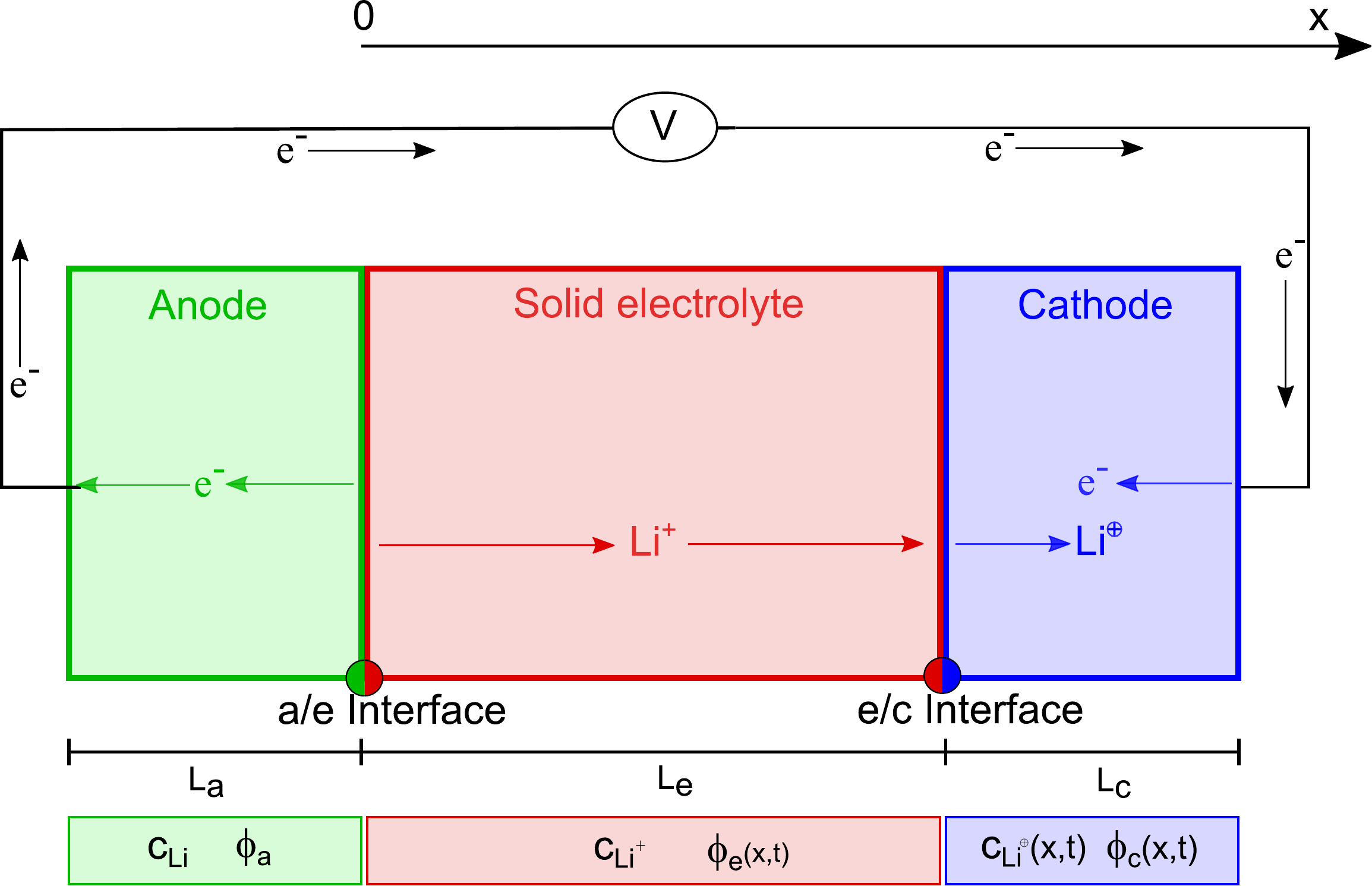}
\caption{ \em Representation of the 1-D model proposed in \cite{Fabre2012}. Schematic representation of an all-solid-state Li-ion battery under discharge conditions. }
\label{fig:model1}
\end{figure}

A one-dimensional model of a Li/LiPON thin-film micro battery was developed by Fabre et al. \cite{Fabre2012} and is depicted schematically in Fig.\ref{fig:model1}.  The $x$-axis, which points out the characteristic lengths of the model, too, is directed from the negative towards the positive electrode; its origin is located at the interface between the anode and the solid electrolyte.

Authors aimed at keeping their model as simple as possible, by introducing proper assumptions in order to come up with a reduced set of input parameters that can be measured from purposely designed experiments. The model is isothermal (no self-heating), 
neglects volume changes during charge/discharge and the active surface area, where redox processes occur, remains unaltered over cycling. The negative electrode is a metallic film of lithium with negligible Ohmic losses. The ionic transfer in the solid electrolyte is described by a single ion conduction model: as such, the concentration of lithium ions across the solid electrolyte $\rm Li^+$ is uniform. This property straightforwardly comes out as long as electro-neutrality approximation holds, a property of the governing equations largely discussed in \cite{SalvadoriEtAlIJSS2015,SalvadoriEtAlJPS2015,SalvadoriEtAlJPS2015b}.

Accordingly, the model traces lithium $\rm Li^\oplus$ diffusion and electron $e^{\!-}$ migration in the positive electrode, single ions $\rm Li^+$ migration in the solid electrolyte, and charge-transfer kinetics at the electrode/electrolyte interfaces. Three unknown fields are required, the concentration of lithium in the positive electrode $c_{\rm Li^\oplus}(x,t)$, the electric potentials $\phi_e(x,t)$ and $\phi_c(x,t)$ in the solid electrolyte and in cathode, respectively. 

\bigskip
\textbf{Governing equations} - 
The electric potentials in the positive electrode $\phi_{c}$ and in the solid electrolyte $\phi_{e}$ are related to the current densities $\vect{i}_{c}(x,t)$ and $\vect{i}_{e}(x,t)$ by means of Ohm's law, through the electrical $k_c$ and ionic $k_e$ conductivities, respectively:
\begin{subequations}
\begin{align}
\label{eq:Current_E_C_model1a}
&\vect{i}_{e}(x,t)=-k_{e}\nabla \phi_{e}(x,t)  &0\leq x \leq L_e 
\; ,
\\
\label{eq:Current_E_C_model1b}
&\vect{i}_{c}(x,t)=-k_{c}\nabla \phi_{c}(x,t)  &L_e\leq x \leq L_e+L_c
\; .
\end{align}
The third governing equation in the positive electrode is a planar solid-state diffusion equation, which describes the (neutral, in the sense detailed beforehand of shielded positive ion) transport of lithium $\rm Li^\oplus$  in the electrode:
\begin{align}
& \frac{\partial c_{\rm Li^\oplus}(x,t)}{\partial t}=\divergence{ \diffusivity_{{\rm Li}^\oplus} \, \nabla c_{\rm Li^\oplus}(x,t)}  & L_e\leq x \leq L_e+L_c \; ,
\label{eq:DiffusionCathode_model1}
\end{align}
\label{eq:FabreGoverningEquations}
\end{subequations}
with diffusion coefficient $\diffusivity_{{\rm Li}^\oplus}$. 
A concentration-dependent ionic diffusion coefficient was studied in \cite{Fabre2012}, which resulted in more accurate outcomes.
The two boundary conditions required for eq. \eqref{eq:DiffusionCathode_model1} are the lithium flux density at the reaction surface 
\begin{subequations}
\begin{align}
\frac{\partial c_{\rm Li^\oplus}(x,t)}{\partial x}&=\frac{-i_{e/c}(t)}{F \diffusivity_{{\rm Li}^\oplus}} \quad &x=L_e
\end{align}
and the zero-flux condition for lithium $\rm Li^\oplus$ at the electrode/collector interface:
\begin{align}
\frac{\partial c_{\rm Li^\oplus}(x,t)}{\partial x}&=0 \quad  &x= L_e+L_c
\end{align}
\label{eq:BoundaryConditions1_model1}
\end{subequations}
The potential is set arbitrarily to $\phi_e(0,t)=0$. Galvanostatic boundary conditions are finally imposed,
\begin{align}
\vect{i}_{c}(L_e+L_c,t)\cdot \vect{n}=i_{bat}(t),
\label{eq:BoundaryConditions_model1}
\end{align}
where $i_{bat}$ is the given galvanostatic current flowing across the 1D battery.

\bigskip
\textbf{Butler-Volmer equations} - 
Denoting with $\vect{n}$ the outward normal to the surface at boundary, conditions for $\phi_e$ and $\phi_c$, either at the anode-electrolyte ($a/e$) or at the electrolyte-cathode ($e/c$) interfaces
\begin{align}
\vect{i}_{e}(0,t)\cdot \vect{n}=-i_{a/e}(t) \; ; \qquad
\vect{i}_{e}(L_e,t)\cdot \vect{n}=-i_{e/c}(t) \; ; \qquad
\vect{i}_{c}(L_e,t)\cdot \vect{n}=i_{e/c}(t) \; ,  \qquad
\label{eq:InterfaceConditions_model1}
\end{align}
are modeled via Butler-Volmer equations in Fabre et al. \cite{Fabre2012}, in the form   
\begin{equation}
i_{s}(t)
=
i_{s}^0(t) 
\; 
\left(
\exp \left[ { \; \frac{\alpha F}{RT} \; \eta(t)} \right]
-
\exp \left[ {- \; \frac{(1-\alpha) F}{RT} \; \eta(t)} \right]
\right)
\label{eq:BV}
\end{equation}
with $s$ either $a/e$ or $e/c$ and $\alpha$ the so-called anodic (cathodic) charge transfer coefficient, usually both taken to be equal to $0.5$. The overpotential
 \begin{equation}
 \label{eq:overpotential}
 \eta(t)= \llbracket \, \phi \, \rrbracket -OCP 
 \end{equation}
is the difference between the jump $\llbracket \, \phi \, \rrbracket$ of the electric potential at the electrolyte/electrode interface (always defined as the electrode potential minus the electrolyte potential) 
and the open circuit potential (OCP) measured experimentally or calculated theoretically as in \cite{PurkayasthaMcMeekingCM2012}.

The exchange current $i^0_{a/e}$ is given by:
\begin{equation}
i^0_{a/e}(t)= \, F \, k_2 \; \left(c_{Li_+}^{sat}-c_{\rm Li^+}(0,t)\right)^{\alpha} \; c_{\rm Li^+}(0,t)^{1-\alpha},
\label{eq:BVanode1_model1}
\end{equation}
where $F$ is Faraday's constant, $k_2$ is the standard rate constant for the reaction \eqref{eq:a/e}, $c_{Li_+}^{sat}$ is the saturation concentration of lithium in the electrolyte.

The exchange current $i_{e/c}^0$ is given by:
\begin{equation}
i_{e/c}^0(t)=Fk_1\left(c_{\rm Li^\oplus}^{sat}-c_{\rm Li^\oplus}(0,t)\right)^{\alpha}c_{\rm Li^\oplus}(0,t)^{1-\alpha},
\label{eq:BVcathode1_model1}
\end{equation}
where $k_1$ is the standard rate constant for the reaction \eqref{eq:e/c} and $c_{Li_\oplus}^{sat}$ is the maximum concentration of lithium inside the cathode.

\subsection{An advanced model framework for solid electrolyte intercalation batteries \cite{LandstorferEtAlPCCP2011}} 
\label{subsec:Lands}

\begin{figure}[!htb]
\centering
\includegraphics[width=120mm]{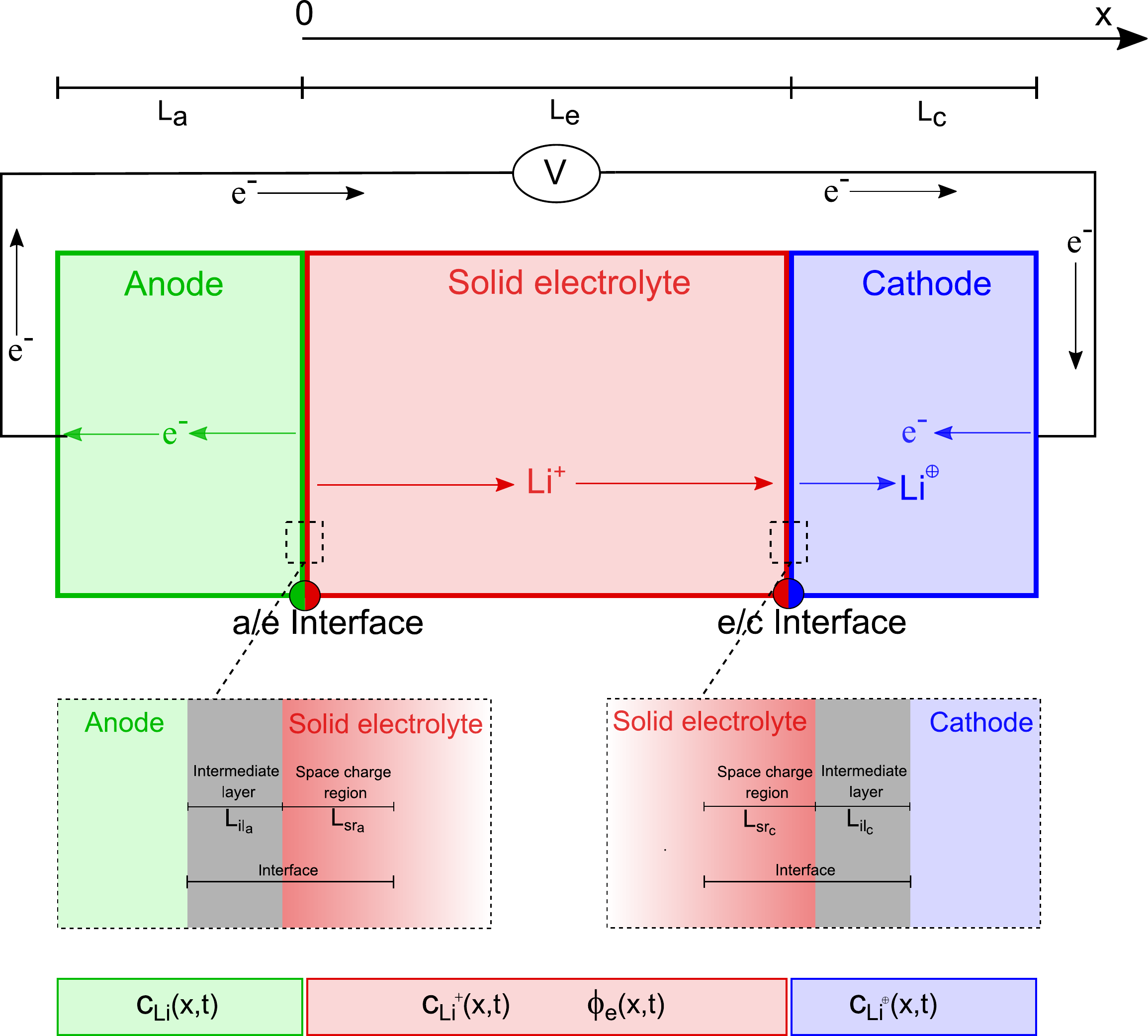}
\caption{\em Representation of the 1-D model proposed in \cite{LandstorferEtAlPCCP2011}. Schematic representation of an all-solid-state Li-ion battery under discharge conditions. On the bottom the schematic representation of the electrode/solid electrolyte interfaces is given, together with the unknown fields.}
\label{fig:model4}
\end{figure}

Landstorfer and coworkers studied in \cite{LandstorferEtAlPCCP2011} a non-porous electrode and a crystalline solid electrolyte. As in \cite{Fabre2012}, they assumed a solid electrolyte with just one mobile species ${\rm Li}^+$, with a uniform concentration of vacancies $c_{\rm n^-}$ that remains unaltered in time. The model entails a novel view of the electrode/solid electrolyte interface, which consists of an intermediate layer and a space charge region within the electrolyte. A visual representation of the model is given in Fig. \ref{fig:model4}.

\bigskip
\textbf{Governing equations} -
The electric potential $\phi_{c}$ is considered to be uniform within the electrodes, neglecting the ohmic loss \eqref{eq:Current_E_C_model1b}, while $\phi_{e}$ influences the ionic transport in the electrolyte through the constitutive law, which differs from the classical Nernst-Planck. 
Once ions randomly intercalate in the lattice structure of graphite or $\rm LiCoO_2$,
ionic transport in the electrodes is accounted for by a simple Fickian law diffusion, that is:
\begin{subequations}
\begin{align}
& \frac{\partial c_{\rm Li}(x,t)}{\partial t} = \divergence{\diffusivity_{\rm Li} \; \nabla c_{\rm Li}(x,t)}  
& -L_a\leq x \leq 0 \; ,
\\
& \frac{\partial c_{{\rm Li}^\oplus}(x,t)}{\partial t} = \divergence{\diffusivity_{{\rm Li}^\oplus}\nabla c_{{\rm Li}^\oplus}(x,t)} 
& L_e\leq x \leq L_e+L_c \; ,
\end{align}
\label{eq:model4_1}
\end{subequations}
where $\diffusivity_{\rm Li}$ and $\diffusivity_{{\rm Li}^\oplus}$ are the diffusion coefficients in the anode and in the cathode, respectively. 
Ionic motion in the solid electrolyte is ruled by non-equilibrium thermodynamics: the mass balance law relates the concentration of lithium $c_{{\rm Li}^+}(x,t)$ to the actual flux of cations $\vect{h}_{{\rm Li}^+}(x,t)$ 
\begin{align}
& \frac{\partial c_{{\rm Li}^+}(x,t)}{\partial t}=-\divergence{\vect{h}_{{\rm Li}^+}(x,t)}  &  L_e\leq x \leq L_e+L_c
\; .
\label{eq:model4_2}
\end{align}
Using the standard linear relationship of Onsager type between the flux $\vect{h}_{{\rm Li}^+}$ and the gradient of the electrochemical chemical potential ${\overline \mu}_{{\rm Li}^+}$, the Clausis-Duhem inequality is satisfied a priori and thermodynamics consistency is granted, as largely discussed in \cite{SalvadoriEtAlJPS2015,SalvadoriEtAlJPS2015b}. Classical algebra (see \cite{LandstorferEtAlPCCP2011} and also \cite{SalvadoriEtAlJMPS2018}) leads to
\begin{align}
&\vect{h}_{{\rm Li}^+}(x,t)= - \tensor{M}(x,t) \,  \nabla {\overline \mu}_{{\rm Li}^+}(x,t) 
& L_e\leq x \leq L_e+L_c \; ,
\label{eq:model4_3}
\end{align}
where
\begin{align}
\tensor{M}(x,t) = \frac{F \diffusivity_{{\rm Li}^+}}{RT} c_{{\rm Li}^+}(x,t) \; \mathds{1}
\label{eq:landsM}
\end{align}
is the expression take in  \cite{LandstorferEtAlPCCP2011}  for the mobility tensor and $\diffusivity_{{\rm Li}^+}$ is the diffusivity of ions within the solid electrolyte. Symbol $ \mathds{1}$ denotes the identity matrix. Note that eq. \eqref{eq:landsM} differs a little from the choice made in \cite{SalvadoriEtAlJMPS2018}. This remark also gives a justification for the different outcomes on the final form of the mass balance equation.

The chemical potential is the functional derivative with respect to the ionic concentration $c_{\rm Li^+}(x,t)$ of the Gibbs free energy (or Helmholtz free energy according to \cite{SalvadoriEtAlJPS2015,SalvadoriEtAlJPS2015b}).  The splitting of chemical and electrical potentials used in \cite{LandstorferEtAlPCCP2011} is classical, as it is the free energy of mobile guest atoms interacting with a host medium, described by a regular solution model \cite{AnandJMPS2012,DeHoffBook}.
The electrochemical potential was eventually derived as the sum of the chemical and electrostatic potential as follows,
{
\begin{subequations}
\begin{align}
&
{\overline \mu}_{{\rm Li}^+}(x,t) \! =
{\mu}_{{\rm Li}^+}(x,t) \!  + \!F\phi_e(x,t) 
&   \:L_e\!\leq\!x\!\leq\! L_e\!+\!L_c
\; ,
\label{eq:model4_4a}
\\
&
{ \mu}_{{\rm Li}^+}(x,t) \! =
\mu_{{\rm Li}^+}^0 
  +\! RT  \; \ln \frac{ \theta_{{\rm Li}^+}(x,t) }{ 1\!-\! \theta_{{\rm Li}^+}(x,t) } 
  +\! RT \; \mychi \left[ \; 1\!-\!2 \theta_{{\rm Li}^+}(x,t)\! \; \right]  
&   \:L_e\!\leq\!x\!\leq\! L_e\!+\!L_c
\; ,
\label{eq:model4_4b}
\end{align}
\label{eq:model4_4}
\end{subequations}
}
with $\theta_{{\rm Li}^+} = c_{{\rm Li}^+} / c^{sat}_{{\rm Li}^+}$.
The chemical potential in \eqref{eq:model4_4b} represents the entropy of mixing plus energetic interactions. The term $\mu_{{\rm Li}^+}^0 $ is the reference value of the chemical potential that specifies the free energy in the absence of interaction and entropic contributions. The real valued constant $\mychi$ in Eq. \eqref{eq:model4_4b}  - termed the exchange parameter \cite{ShellBook2015}  - characterizes the energy of interaction between mobile guest species and insertion sites. If all of the interactions between mobile species and sites are the same, then $\mychi=0$ and there is no energy of mixing: mixing is purely entropic. The contribution $RT \; \mychi \left[ \; 1\!-\!2 \theta_{{\rm Li}^+}(x,t)\! \; \right]  $, emanating from the excess Gibbs energy  \cite{SalvadoriEtAlJMPS2018}, may lead to phase segregation \cite{Bohn2013,dileoetalJMPS2014,Bower2015a}.
In view of the above, the mass flux to be inserted into the mass balance law \eqref{eq:model4_2} eventually holds:
{
\begin{align}
&
\vect{h}_{{\rm Li}^+}(x,t)
=
-
{\! \; \frac{\diffusivity_{{\rm Li}^+}}{RT} \left\{  \!\left(\frac{RT}{1\!-\!c_{{\rm Li}^+}}\!-\!2\mychi \; c_{{\rm Li}^+}\right)\!\nabla c_{{\rm Li}^+} \!+\!  c_{{\rm Li}^+} \, \nabla\phi_e \!  \right\}  } 
& L_e\!\leq\!x\!\leq\! L_e\!+\!L_c
\; .
\label{eq:model4_5}
\end{align}
}
The classical Nernst-Planck flux (see eq. \eqref{eq:FluxElec_model2} later on in the paper) can be attained if the dilute limit is assumed.

The whole electrolyte is thought as consisting of a {\em{space charge region}} and a bulk region, as shown in Fig.\ref{fig:model4}. In the bulk region electroneutrality is imposed a priori,  $c_{{\rm Li}^+}$ is defined by the (given and uniform) concentration of vacancies, and Ohm's law \eqref{eq:Current_E_C_model1a} allows recovering the electric potential. 
On the contrary, in proximity of the interfaces, where the concentrations of cations and anions differ, the electric potential is obtained by Gauss law, which provides, after constitutive prescriptions, the following Poisson equations:
\begin{subequations}
\begin{align}
&-\nabla^2\phi_{e}(x,t)=\frac{F}{\varepsilon_r\varepsilon_0}(c_{{\rm Li}^+}(x,t)-c_{\rm Li})    
& 0\!\leq\!x\!\leq\! L_{sr_a} 
\; ,
\\
&-\nabla^2\phi_{e}(x,t)=\frac{F}{\varepsilon_r\varepsilon_0}(c_{{\rm Li}^+}(x,t)-c_{{\rm Li}^\oplus})
& L_e\!-\!L_{sr_c}\!\leq\!x\!\leq\! L_e
\; ,
\label{eq:model4_6}
\end{align}
\end{subequations}
where $\varepsilon_r$ denotes the relative permittivity of the electrolyte and $\varepsilon_0$ the vacuum permittivity. 
Local electroneutrality is not enforced in the space charge region \cite{LandstorferEtAlPCCP2011},
rather a weak (i.e. global) electroneutrality 
\begin{align}
& F\int_V(c_{{\rm Li}^+}(x,t)-c_{\rm Li})dV=0
\; ,
& F\int_V(c_{{\rm Li}^+}(x,t)-c_{{\rm Li}^\oplus})dV=0 
\; ,
\label{eq:model4_7}
\end{align}
which allows for local deviations between cation and anion concentrations. 

\bigskip
\textbf{Modeling the intermediate layer as an interface} -
The two intermediate layers have been modeled as an interface between electrodes and the solid electrolyte, in terms of potential jumps (drops in \cite{LandstorferEtAlPCCP2011}) and flux continuity. By assuming perfect planar and infinite surfaces, the  two intermediate layers have been treated as plate capacitors. Authors borrowed from \cite{BonnefontEtAlJEC2001} the constitutive equation for potential jumps (defined as the potential at an electrode minus the one at the electrolyte) as
\begin{align}
&
 \llbracket \, \phi \, \rrbracket_{anode}
= \frac{\varepsilon_a}{C_a} \;  \nabla{\phi_e(0,t)}  \cdot \vec{n}
\; ,
&
 \llbracket \, \phi \, \rrbracket_{cathode}
= \frac{\varepsilon_c}{C_c} \;  \nabla{\phi_e(L_e,t)} \cdot \vec{n}
\; ,
\label{eq:model4_8}
\end{align}
with given surface capacitances $C_a$, $C_c$ and permittivities $\varepsilon_a$, $\varepsilon_c$.

Boundary conditions on fluxes at electrode - solid electrolyte interfaces resemble Butler-Volmer equations \eqref{eq:BV} in a form originally presented in \cite{Bazant2005} and named {generalized Frumkin-Butler-Volmer equations}, without making reference to the concept of overpotential \eqref{eq:overpotential}.  They read:
\begin{subequations}
\begin{align}
&
\vect{h}_{{\rm Li}^+}(0,t) \cdot \vec{n} 
=
-  \tilde{k}_{2} \; e^{\frac{\triangle G_{1}-\beta F(\phi_a-\phi_e(0,t))}{RT}}c_{\rm Li}(0,t)
+ \tilde{k}_{-2} \; e^{\frac{\triangle G_{-1}-(1-\beta)F(\phi_a-\phi_e(0,t)}{RT}}c_{{\rm Li}^+}(0,t)
\; ,
\label{eq:model4_10a}
\\ 
&
\vect{h}_{{\rm Li}^+}(L_e,t) \cdot \vec{n} 
=
-  \tilde{k}_{1} \; e^{\frac{\triangle G_{2}-\beta F(\phi_c-\phi_e(L_e,t))}{RT}}c_{{\rm Li}^\oplus}(L_e,t)
+ \tilde{k}_{-1} \; e^{\frac{\triangle G_{-2}-(1-\beta)F(\phi_c-\phi_e(L_e,t)}{RT}}c_{{\rm Li}^+}(L_e,t),
\label{eq:model4_10b}
\end{align}
\label{eq:model4_10}
\end{subequations}
where the reaction rate constants $k_1$ and $k_2$ have been defined in reactions \eqref{eq:a/e} and \eqref{eq:e/c} and are of Arrhenius type, $k_{n}=\tilde{k}_{n} \; e^{E_{n}/RT}$, $n=1,2$. The Gibbs energies of activation $\triangle G_{n}$ are further parameters of the model.

\subsection{An advanced all-solid-state Li-ion battery model \cite{RaijmakersEtAlEA2020}} 
\label{subsec:RaijmakersEtAlEA2020}

More complex one-dimensional mathematical models of a micro-battery $\rm Li/LiPON/LiCoO_2$ have been proposed in a series of publications from Notten's group \cite{Danilovetal2011,Danilov2016A,Danilov2016B}. In these studies, ionic transport in the solid electrolyte involves the ionization reaction \eqref{eq:IonizationReaction} of immobile, oxygen-bound lithium $\rm Li_0$ to mobile $\rm Li^+$ ions and negatively charged vacancies. Charge-transfer kinetics at both electrode/electrolyte interfaces, diffusion and migration of mobile lithium ions in the electrolyte ($\rm Li^+$) and positive electrode ($\rm Li^\oplus$) were accounted for. In their most recent work, \cite{RaijmakersEtAlEA2020}, additional features have been introduced, such as mixed ionic/electronic conductivity in the positive electrode, electrical double layers occurring at both electrode/electrolyte interfaces representing the space-charge separation phenomena that differ \cite{LandstorferEtAlPCCP2011}, variable ionic and electronic diffusion coefficients that depend on the lithium concentration inside the positive electrode.

\begin{figure}[!htb]
\centering
\includegraphics[width=130mm]{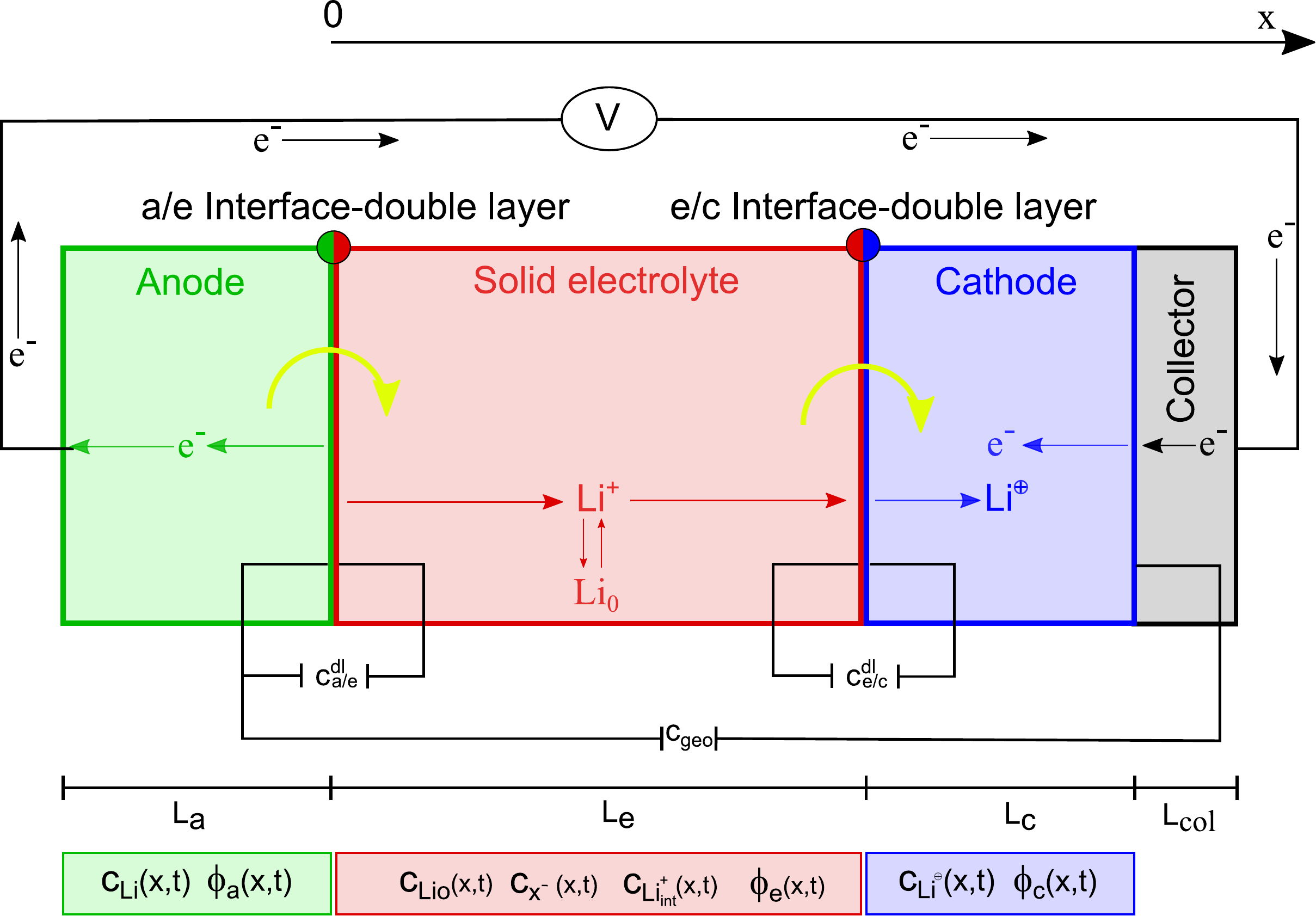}
\caption{\em A pictorial view of the 1-D model, proposed in \cite{RaijmakersEtAlEA2020}. The battery consists of a platinum current collector (Pt), onto which the positive electrode material is deposited, based on a $LCO$ chemistry, of a negative electrode made of metallic lithium $Li$ and of a $LiPON$ solid-state electrolyte. The thickness of the anode is denoted as $L_a$, the LiPON layer as $L_e$, the LCO layer as $L_c$, and that of the current collector as $L_{col}$. The (yellow) curved arrows indicate the charge-transfer reactions at both interfaces, while $i_{a/c}$ and $i_{e/c}$ are used to specify the charge-transfer current densities across both electrode/electrolyte interfaces. }
\label{fig:model2}
\end{figure}

The schematic representation of the battery, as proposed in \cite{RaijmakersEtAlEA2020},  under discharge conditions is shown in Fig. \ref{fig:model2}. The battery consists of a current collector, onto which the positive $\rm LiCoO_2$ electrode is deposited, of a metallic lithium $\rm Li$ foil as negative electrode and of a $\rm LiPON$ solid-state electrolyte.
As for \cite{Fabre2012} and \cite{LandstorferEtAlPCCP2011}, also this model is isothermal (no self-heating), electrodes are not porous and redox processes only occur at the interfaces between the electrolyte and the electrode layers; volume changes of the electrolyte during cycling are neglected and the active surface area does not change over cycling. 

As a distinctive feature of this class of models, the ionic transfer in the solid electrolyte is {\em{not}} described by a single ion conduction model and the concentration of lithium ions across the solid electrolyte is generally {\em{not}} uniform even though electro-neutrality approximation holds.
Denoting with $c_{\rm Li^+}$ the concentration of mobile ${\rm Li^+}$ ions, with $c_{\rm Li_0}$ the concentration of immobile lithium, with $c_{\rm n^-}$ the concentration of uncompensated negative charges, and with $\delta$ the fraction of Li at equilibrium, then the {\em{equilibrium}} concentration of the charge carriers is $c^{eq}_{\rm Li^+}=c^{eq}_{\rm n^-}=\delta c_0$ and the concentration of the remaining immobile lithium is  $c^{eq}_{\rm Li_0}=(1-\delta) c_0$. The overall rate of the charge carrier generation according to reaction \eqref{eq:IonizationReaction} is 
\begin{equation}
w=k_f^{\rm ion} \, c_{\rm Li_0}-k_b^{\rm ion} \, c_{\rm Li^+} \; c_{\rm n^-} \; .
\label{eq:w_reaction_rate}
\end{equation}
The ratio 
\begin{equation}
K_{\rm eq}^{\rm ion}  \! = \! k_f^{\rm ion}/k_b^{\rm ion}
\label{eq:equilibrium_constant_w}
\end{equation}
is the equilibrium constant of reaction \eqref{eq:IonizationReaction} and is related to the the fraction of Li at equilibrium $\delta$, see \cite{RaijmakersEtAlEA2020}.
The electro-neutrality approximation is fully exploited in \cite{RaijmakersEtAlEA2020}, implying that $$c_{\rm Li^+}(x,t)=c_{\rm n^-}(x,t) \; .$$ 

In addition to earlier models, two electrical {\em{double layer}} capacitances $C^{dl}_{a/e}$ and $C^{dl}_{e/c}$ and a {\em{geometric}} capacitance $C_{geo}$ were introduced in \cite{RaijmakersEtAlEA2020}. As in \cite{LandstorferEtAlPCCP2011}, double layer capacitances attempt at capturing the response of the space-charge very narrow layers, in liquid electrolytes termed after Stern, Go{\"u}y and Chapman, assumed to be electric capacitors. Whereas the concept resembles \cite{LandstorferEtAlPCCP2011}, the implementation is different. Capacitors in \cite{LandstorferEtAlPCCP2011} are described ``in series'', whereas in \cite{RaijmakersEtAlEA2020} ``in parallel'' (compare figures \ref{fig:model4} and \ref{fig:model2}). In view of this assumption, the current at electrode/electrolyte interfaces splits into two terms, the faradaic contribution that drives the reduction/oxidation charge-transfer reactions \eqref{eq:a/e}-\eqref{eq:e/c} ($ i_{s}^{ct} $) and the non-faradaic contribution that feeds the double layer ($ i_{s}^{dl} $), with $s$ either $a/e$ or $e/c$.

\bigskip
\textbf{Governing equations} -
During charging, the $\rm Li^+$ ions released from the positive surface must cross the solid electrolyte and are reduced into metallic Li at anode. Electrons, generated by the charge-transfer reactions \eqref{eq:a/e} and \eqref{eq:e/c}, must pass through the inner electronic circuit, with potential drops across the Li foils and the electronic collector that follow Ohm's law:
\begin{subequations}
\begin{align}
& \vect{i}_{a}(x,t)=-k_{a}\nabla \phi_{a}(x,t) 
& -L_a\leq x \leq 0
\; ,
\\
&\vect{i}_{col}(x,t)=-k_{col}\nabla \phi_{col}(x,t) 
& L_e+L_c\leq x \leq L_e+L_c+L_{col}
\; ,
\end{align}
\label{eq:Current_Collector_model2}
\end{subequations}
where $k_{col}$ and $k_{a}$ are the electrical conductivities in the collector and in the anode, respectively.

Transport of mass {\em{in the solid electrolyte}} is ruled by the mass continuity equation \eqref{eq:model4_2}, which shall be properly extended in order to account for the two ionic concentrations $c_{\rm Li^+}$ and $c_{\rm n^-}$ and to their fluxes. The mass balance equations proposed in \cite{RaijmakersEtAlEA2020}  to characterize the transport of $\rm Li^+$ and negatively charged vacancies\footnote{This description of transport of vacancies, in a form analogous to liquid electrolytes, appears to be questionable and is replaced by a different formulation in the novel approach to be presented in section \ref{subsec:novelformulation}.} read
\begin{subequations}
\begin{align}
&\frac{\partial c_{\rm Li^+}(x,t)}{\partial t}=-\divergence{\vect{h}_{\rm Li^+}(x,t)} + w(x,t) 
& 0\leq x \leq L_e
\; ,
\label{eq:MassEle_model2a}
\\
&\frac{\partial c_{\rm n^-}(x,t)}{\partial t}=-\divergence{\vect{h}_{\rm n^-}(x,t)} + w(x,t) 
& 0\leq x \leq L_e
\; ,
\label{eq:MassEle_model2b}
\end{align}
\label{eq:MassEle_model2}
\end{subequations}
where $\vect{h}_\alpha$ is the generic mass flux ($\alpha = \rm Li^+, n^-$), constitutively described by the Nernst-Planck law
\begin{align}
\vect{h}_\alpha(x,t)=-\diffusivity_\alpha \nabla c_\alpha(x,t)-\frac{z_\alpha F}{RT}\diffusivity_\alpha c_\alpha(x,t)\nabla\phi_e(x,t)
\; ,
\label{eq:FluxElec_model2}
\end{align}
$T$ is the absolute temperature, $z_\alpha$ is the valency of ion $\alpha$, typically $+1$ for $\rm Li^+$ cations and $-1$ for $\rm n^-$. 

The two mass balance equations \eqref{eq:MassEle_model2} contain the electric potential in view of eq. \eqref{eq:FluxElec_model2}. Coupling with an additional relation is mandatory, to model the migration process. The most common selection for such an additional relation in battery modeling is the electro-neutrality condition. 
By substituting eq. \eqref{eq:FluxElec_model2} into eq. \eqref{eq:MassEle_model2} and subtracting eq.\eqref{eq:MassEle_model2a} from eq.\eqref{eq:MassEle_model2b}, two independent partial differential equations finally arise:
\begin{subequations}
\begin{align}
&
\frac{\partial c_{\rm Li^+}(x,t)}{\partial t} = \divergence{\diffusivity_{\rm Li^+} \, \nabla c_{\rm Li^+}(x,t) +\frac{F \diffusivity_{\rm Li^+}}{RT}c_{\rm Li^+}(x,t)\nabla\phi_e(x,t)}+w(x,t)
\\
&
\divergence{\!(\diffusivity_{\rm n^-}-\diffusivity_{\rm Li^+})\nabla c_{\rm Li^+}(x,t)}\!-\!\divergence{(\diffusivity_{\rm n^-}\!+\!\diffusivity_{\rm Li^+})\frac{F}{RT}c_{\rm Li^+}(x,t)\nabla\phi_e(x,t)\!}\!=\!0.
\end{align}
\label{eq:MassEle1_model2}
\end{subequations}
To be solved they require the initial concentrations for $c_{\rm Li^+}$ 
\begin{align}
c_{\rm Li^+}(x,0)&=c^{eq}_{\rm Li^+}=\delta c_0
\label{eq:Init_model2}
\end{align}
and the Neumann conditions on fluxes at the left and right boundaries of the electrolyte
\begin{subequations}
\begin{align}
\vect{h}_{\rm Li^+}(0,t) \cdot \vect{n} &= - \frac{i^{ct}_{a/e}(t) + i_{a/e}^{dl}(t) }{F}  \\ 
\vect{h}_{\rm Li^+}(L_e,t) \cdot \vect{n} &= - \frac{i^{ct}_{e/c}(t) + i_{e/c}^{dl}(t) }{F}  \; .
\end{align}
\label{eq:BoundEle_model2}
\end{subequations}
The non-faradaic current (dis)charging the electrical double layers $i_\alpha^{dl}(t)$, can be defined in derivative form as 
\begin{equation}
i_\alpha^{dl}(t) = C_\alpha^{dl} \; \frac{\partial \llbracket \, \phi \, \rrbracket }{\partial t},
\label{eq:NonFaradaicCurrent}
\end{equation}
where the jump $\llbracket \, \phi \, \rrbracket$ of the electric potential at the electrolyte/electrode interface is always defined as the electrode potential minus the electrolyte potential and $\alpha ={a/e}, {e/c}$. Equation \eqref{eq:NonFaradaicCurrent} shall be compared with \eqref{eq:model4_8} in \cite{LandstorferEtAlPCCP2011}.

The faradaic current proposed in \cite{RaijmakersEtAlEA2020} emanates from charge transfer kinetics, in a form that extends Butler-Volmer equations \eqref{eq:BV} to the mass transfer-influenced conditions \cite{BardFaulknerBook}. 
\begin{subequations}
{
The expression of $i^{ct}_{a/e}$ at the metallic lithium electrode interface is:
\begin{equation}
 { i^{ct}_{a/e}} =  { i^{0}_{a/e}} 
 \;  
 \left( 
          \frac{c_{\rm Li}(0,t)}{\overline{c}_{\rm Li}}  
            \exp \left[ { \; \frac{\alpha_a F}{RT} \; \eta_a(t)} \right] 
            -
            \frac{c_{\rm Li^+}(0,t)}{\overline{c}_{\rm Li^+}}
            \exp \left[ {- \; \frac{(1-\alpha_a) F}{RT} \; \eta_a(t)} \right]
 \right) \; ,
\label{eq:BVanode_model2}
\end{equation}
where $\overline{c}_{\rm Li^+}$ is the average bulk concentration of species ${\rm Li^+}$,  $\overline{c}_{Li}$ is the bulk activity of the metallic Li, $\alpha_{a}$ is the charge transfer coefficient for reaction eq. \eqref{eq:a/e}, $\eta_{a}$ is the overpotential \eqref{eq:overpotential} of the charge transfer reaction at the negative electrode, and the exchange current $i_{a/e}^0$ is given by:
\begin{equation}
{ i^{0}_{a/e}} = \, F \, k_2 \; (\overline{c}_{{\rm Li^+}} )^{\alpha_a} \; (\overline{c}_{\rm Li})^{ 1-\alpha_a} \; ,
\label{eq:BVanode1_model2}
\end{equation}
with $k_2$ the standard rate constant for reaction eq. \eqref{eq:a/e}. The expression of $i_{e/c}$ at the positive electrode interface is given by:
\begin{equation}
i_{e/c}(t) = 
{ i^{0}_{e/c}}
 \;  
\left(
            \exp \left[ { \; \frac{\alpha_c F}{RT} \; \eta_c(t)} \right] 
   -
            \exp \left[ {- \; \frac{(1-\alpha_c) F}{RT} \; \eta_c(t)} \right]
\right) 
\; ,
\label{eq:BVcathode_model2}
\end{equation}
where $\alpha_{c}$ is the charge transfer coefficient for reaction eq.\eqref{eq:e/c}, $\eta_{c}$ is the overpotential \eqref{eq:overpotential} of the charge transfer reaction at the positive electrode, and the exchange current $i_{e/c}^0$ is given by:
\begin{equation}
i_{e/c}^0
=
\; F \; k_1 \; 
c_{{\rm Li}^\oplus}^{sat} 
\;
\left(1-\frac{\overline{c}_{{\rm Li}^\oplus}}{c_{Li^\oplus}^{sat}}\right)^{\alpha_{c}}
\; 
\left( \frac{\overline{c}_{{\rm Li}^\oplus} }{c_{Li^\oplus}^{sat}}  \right)^{1-\alpha_{c}}
\; 
\left( \, \overline{c}_{{\rm Li}^+} \, \right)^{\alpha_{c}}
\; ,
\label{eq:BVcathode1_model2}
\end{equation}
with $k_1$ the standard rate constant for reaction eq.\eqref{eq:e/c}. 
} 
\label{subeq:faradaic}
\end{subequations}
The reader may refer to \cite{RaijmakersEtAlEA2020} for further details on these equations and on the geometric capacitance.

\bigskip
{\em{In the electrode}} a mixed ionic/electronic conductivity is considered. The mass balance equations that characterize the transport of $\rm Li^\oplus$ ions and electrons ${\rm e^-}$ are therefore similar to eqs. \eqref{eq:MassEle_model2} for solid electrolyte and write
\begin{subequations}
\begin{align}
& \frac{\partial c_{\rm Li^\oplus}(x,t)}{\partial t}=-\divergence{\vect{h}_{{\rm Li}^\oplus}(x,t)} 
& L_e\leq x \leq L_e+L_c \; ,
\label{eq:MassCat_model2a}
\\
& \frac{\partial c_{\rm e^-}(x,t)}{\partial t}=-\divergence{\vect{h}_{\rm e^-}(x,t)} 
& L_e\leq x \leq L_e+L_c \; .
\label{eq:MassCat_model2b}
\end{align}
\label{eq:MassCat_model2}
\end{subequations}
The generic mass flux $\vect{h}_\alpha$, $\alpha = \{ {\rm Li^\oplus}, {\rm e^-} \}$ is constitutively described by the Nernst-Planck law \eqref{eq:FluxElec_model2}. This choice of independent motion of electrons and ionic intercalated lithium makes the governing equations different from Fabre's \cite{Fabre2012} equations ( \eqref{eq:DiffusionCathode_model1} to be compared with \eqref{eq:MassCat_model2a}) ).

Furthermore, the model in \cite{RaijmakersEtAlEA2020} accounts for an ionic concentration-dependent diffusion coefficients in the positive electrode, driven by the experimental evidence that solid-state diffusion depends strongly on the local electrochemical environment. 
The model further exploits the electro-neutrality approximation inside the cathode, implying $c_{\rm Li^\oplus}(x,t)=c_{\rm e^-}(x,t)$, and a space-time proportionality of the diffusion coefficients for ${\rm Li^\oplus}$ and ${\rm e^-}$. Classical mathematical passages allow to retrieve eq.  \eqref{eq:DiffusionCathode_model1} as the single PDE required to model the mass transport in the electrode, provided that the diffusivity $\diffusivity_{{\rm Li}^\oplus}$ is replaced by a suitable combination of electron and ionic diffusivities.

The initial concentration of the charge carrier at $t=0$, when no concentration profile developed yet, is equal to its equilibrium concentration 
\begin{align}
c_{\rm Li^\oplus}(x,0)&=c_{\rm Li^\oplus}^{eq}
\end{align}
Neglecting the role of the geometrical capacitance, the Neumann boundary conditions on fluxes at the left and right boundaries of the electrolyte read
\begin{subequations}
\begin{align}
\vect{h}_{{\rm Li}^\oplus}(L_e,t)\cdot\vect{n}&= \frac{i^{ct}_{e/c}(t) + i_{e/c}^{dl}(t) }{F} \\
\vect{h}_{{\rm Li}^\oplus}(L_e+L_c,t)\cdot\vect{n}&=0\\
\vect{h}_{\rm e^-}(L_e,t)\cdot\vect{n}&=0\\
\vect{h}_{\rm e^-}(L_e+L_c,t)\cdot\vect{n}&= - \frac{i_{bat}(t) }{F}\ 
\end{align}
\label{eq:BoundCat1model2}
\end{subequations}
with $i_{bat}(t)$ the given galvanostatic current flowing across the 1D battery, defined in eq. \eqref{eq:BoundaryConditions_model1}.

\subsection{A novel, two-mechanisms model for all-solid-state lithium-ion batteries}
\label{subsec:novelformulation}

%
\begin{figure}[!htb]
\begin{center}
\includegraphics[ width=12cm ]{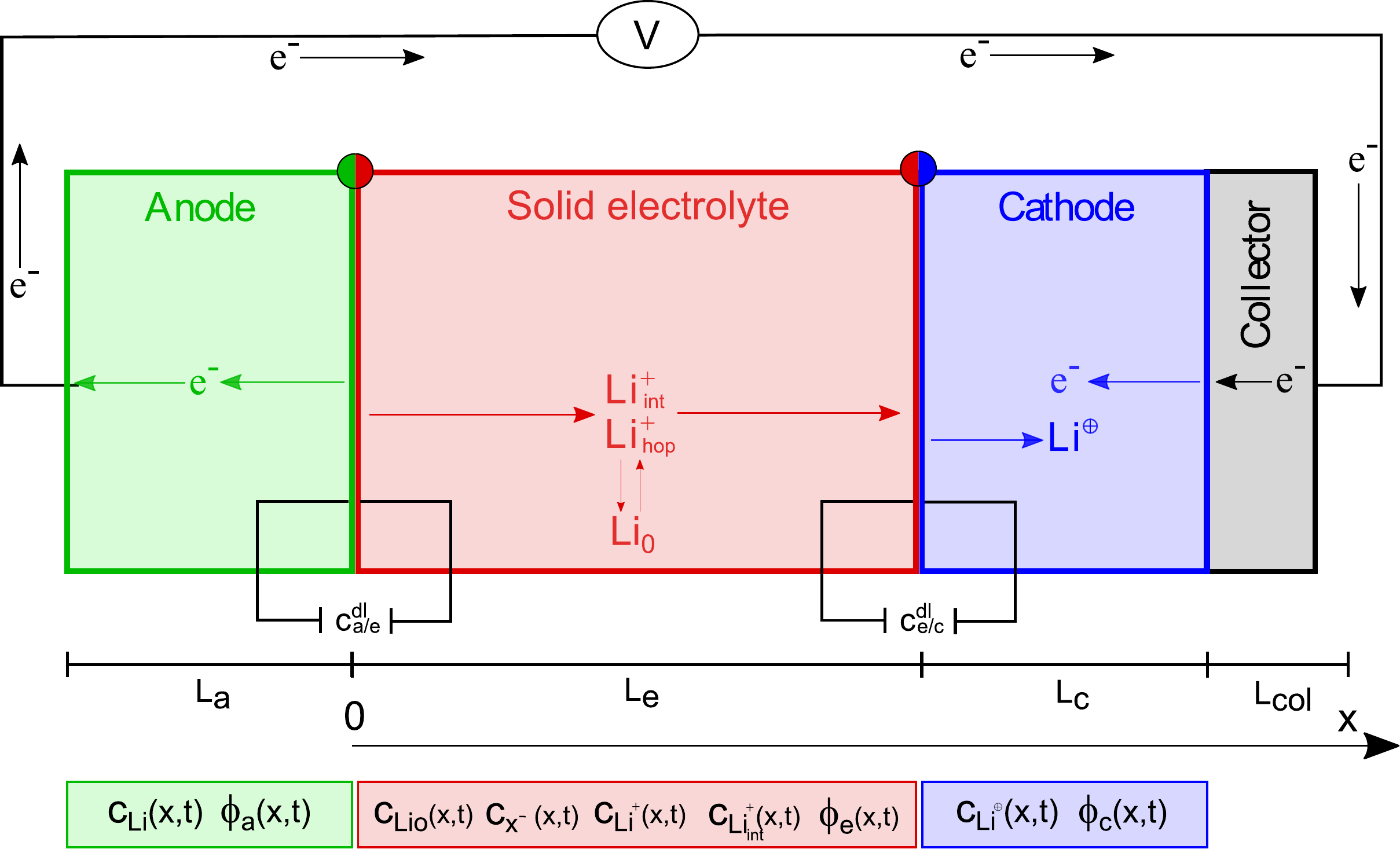}
\end{center}
\caption{\em A pictorial view of the novel 1-D battery model. Differences from \cite{RaijmakersEtAlEA2020} are highlighted, see also fig. \ref{fig:model2}.  }
\label{fig:GeneralBatteryScheme}
\end{figure}
%

Finally, a novel two-mechanisms model  \cite{CabrasEtAl2021a} is illustrated.
It is grounded in the thermo-mechanics of continua and advances  \cite{RaijmakersEtAlEA2020} in modeling the process of vacancies replenishment and in making it multi-scale-compatible, which appears to be relevant for composite cathodes \cite{BielefeldEtAlJPC2019}.
As described so well in  \cite{RaijmakersEtAlEA2020}, in standard conditions some of the Li ions are thermally excited and the chemical ionization reaction \eqref{eq:IonizationReaction} occurs, leaving behind uncompensated negative charges, associated with a vacancy in the LIPON matrix at the place formerly occupied by lithium.  In  \cite{RaijmakersEtAlEA2020} vacancies were depicted with the same conceptual formalism used for negative ions in liquid electrolytes, i.e. as able to move in the solid matter driven by an entropic brownian motion together with migration within an electric field (see eq. \eqref{eq:MassEle_model2b}). This picture is thermodynamically encapsulated in the constitutive law \eqref{eq:FluxElec_model2}.

In our novel formalism, we attempt at explicitly modeling the dynamic filling of vacancies by neighboring positions, which in turn create new vacancies. To this aim, we claim that after the chemical ionization reaction \eqref{eq:IonizationReaction} occurs, some ions, denoted henceforth with ${\rm Li}^+_{\rm hop}$, hop and fill neighboring vacancies, whereas the remaining $\rm Li^+$ ions move in a meta-stable interstitial state. This dynamic behavior is described by a further reaction, that converts part of the full amount of ions made available by reaction \eqref{eq:IonizationReaction} into hopping lithium with the ability to fill vacancies, leaving to the remaining ions the interstitial motion responsibility :
\begin{equation}
{\rm Li^+} \underset{k_b^{\rm hop}}{\overset{k_f^{\rm hop}}{\rightleftarrows}} {\rm Li}^+_{\rm hop}
\; ,
\label{eq:IonizationReaction2}
\end{equation}
where $k_f^{\rm hop}$ and $k_b^{\rm hop}$ are the rate constants for reaction \eqref{eq:IonizationReaction2}. When $k_f^{\rm hop} = 0$, no hopping mechanism is accounted for and the model restricts to a classical single ion conducting solid electrolyte \cite{Fabre2012}. 

In different words, reaction \eqref{eq:IonizationReaction} becomes the one which makes lithium ions capable to leave the host-sites and move within the complex amorphous LIPON structure, either by filling neighboring vacancies or to flow interstitially. The proportion of ions in these two mechanisms is governed by reaction \eqref{eq:IonizationReaction2}. In this formalism, positive ions are the only moving species, whereby the concentration of negatively charged vacancies is the outcome of the motion process and do not possess any intrinsic motility. In this sense, there is no direct flow $\vect{h}_{\rm n^-}(x,t)$ of negative charges, contrary to eq. (\ref{eq:MassEle_model2}b), and the local concentration of vacancies is altered merely by the chemical ionization reaction  eq. \eqref{eq:IonizationReaction2}.

\bigskip
\textbf{Governing equations} -
This conceptual picture {\em{in the electrolyte}} frames into the following set of mass balance equations, which characterize the immobile lithium $\rm Li_0$, the negative charges ${\rm n^-}$, the transport of the lithium ions ${\rm Li}^+_{\rm hop}$ that hop and the remaining $\rm Li^+$ that go interstitial:
\begin{subequations}
\begin{align}
&\frac{\partial c_{\rm Li_0}}{ \partial t}=-w                                     &0\leq x \leq L_e \; ,
\\
\label{eq:MassEle_model1b}
&\frac{\partial c_{\rm n^-}}{ \partial t}=w                                       &0\leq x \leq L_e \; ,
\\
\label{eq:MassEle_model1c}
&\frac{\partial c_{\rm Li^+}}{ \partial t} + \divergence{\vect{h}_{\rm Li^+}} = w - y              &0\leq x \leq L_e \;,
\\
\label{eq:MassEle_model1d}
&\frac{\partial c_{{\rm Li}^+_{\rm hop}}}{ \partial t} + \divergence{\vect{h}_{{\rm Li}^+_{\rm hop}}} = y    &0\leq x \leq L_e.
\end{align}
\label{eq:MassEle_model1}
\end{subequations}
In eqs. \eqref{eq:MassEle_model1}: i) $w$ is the the overall rate of the charge carrier generation under general (dynamic) conditions for reaction \eqref{eq:IonizationReaction}, as defined in eq.\eqref{eq:w_reaction_rate}; ii) $y$ is the reaction rate of the conversion of mobile lithium into interstitial, i.e.
\begin{equation}
y=k_f^{\rm hop}c_{\rm Li^+}-k_b^{\rm hop}c_{{\rm Li}^+_{\rm hop}}
\label{eq:y_reaction_rate}
\end{equation}
according to reaction \eqref{eq:IonizationReaction2}. The ratio 
\begin{equation}
 K_{\rm eq}^{\rm hop}  \! = \! k_f^{\rm hop}/k_b^{\rm hop}
\label{eq:equilibrium_constant_y}
\end{equation}
is the equilibrium constant of reaction \eqref{eq:IonizationReaction2}. 

The mechanical response of the cell is ruled by the usual balance of forces:
\begin{equation}
\label{eq:stressbalance}
  \divergence{\bsigma} + \vect{b} =  \vect{0} 
  \; ,
\end{equation}
where $\bsigma$ is the Cauchy stress tensor, $\vect{b}$ is the body force per unit volume, and we have assumed inertia forces to be negligible.
The symmetry of the stress tensor follows from the balance of angular momentum \cite{GurtinFriedAnand}.

The chemo-thermo-elastic strain $\tensor{\varepsilon}$ is considered to be made up of two separate contributions: an elastic recoverable part after unloading $\tensor{\varepsilon}^{el}$ and a swelling contribution due to the insertion of species in the host material $\tensor{\varepsilon}^s$:
\begin{equation}
\label{eq:decompositionofstrains}
    \tensor{\varepsilon} = \tensor{\varepsilon}^{el} + \tensor{\varepsilon}^{s} 
    \; .
\end{equation}
The swelling contribution (  $\alpha = {\rm Li^+}, {\rm Li}^+_{\rm hop}$ )
\begin{equation}
\label{eq:swellingstrain}
    \tensor{\varepsilon}^{s} =  \sum_\alpha \; \omega_\alpha \, \left(  c_{\alpha} - c_{\alpha}^0 \right) \; \mathds{1} 
\end{equation}
is assumed to be volumetric and proportional to {the deviation $c_\alpha-c_\alpha^0$ from the reference concentration $c_\alpha^0$} by means of the chemical expansion coefficients $\omega_\alpha$ of species $\alpha$. They equal one third of the partial molar volumes at a given temperature.

The hopping mechanism is thermodynamically quite different from the interstitial motion, thus making recourse to the same thermodynamic description for both mechanisms might be questionable. We will elaborate this issue in future works. For now, as in \cite{LandstorferEtAlPCCP2011}, we derive constitutive laws from a rigorous thermodynamic setting. Inspired by \cite{Ganser2019} and \cite{Bower2015b,Bucci2014} yet restricting to small strains, elaborating the electromagnetic contribution in the Helmholtz free energy $\psi$ from our previous works   \cite{SalvadoriEtAlJPS2015,SalvadoriEtAlJPS2015b} and the phase segregation scheme from \cite{SalvadoriEtAlJMPS2018}, we extend
the chemical potential \eqref{eq:model4_4b} of species $\alpha = {\rm Li^+}, {\rm Li}^+_{\rm hop}$ as
\begin{equation}
    {\mu}_\alpha \, = 
      \mu_{\alpha}^0 
    + R T  \, \ln[ \frac {\theta_{\alpha} }{  1-\theta_{\alpha} }]
    + R T \, \mychi \left( 1-2 \theta_\alpha \right) 
    +  \frac{\partial  \psi_{el} }{\partial c_\alpha}  
     \; ,
\label{eq:chempotbeta}
\end{equation}
as detailed in \cite{SalvadoriEtAlJMPS2018}. A simple choice for the elastic part of the free energy density $  \psi_{el}$ in the small strain range is the usual quadratic form 
\begin{eqnarray}
   \psi_{el}(  \tensor{\varepsilon},  c_\alpha ) 
   =  \frac{1}{2} \,  K( c_\alpha ) \, \trace{ \tensor{\varepsilon} - \tensor{\varepsilon}^{s}  } ^2 
   +
   \; G( c_\alpha ) \; || \, \deviatoric{ \tensor{\varepsilon} - \tensor{\varepsilon}^{s} }    \, ||^2 
   \; , 
\label{eq:electrode:psimech}
\end{eqnarray}
where $K$, $G$ are the bulk and shear modulus respectively and they are made dependent on species concentrations.
The stress tensor $\tensor{\sigma}^e( \tensor{\varepsilon}, c_\alpha )$ descends from thermodynamic restrictions (see \cite{SalvadoriEtAlJMPS2018} for details and extension to temperature dependency)
\begin{eqnarray}
 \label{eq:sigma2}
 \tensor{\sigma}^e 
 &=& 
   \frac{\partial  \psi_{el} }{\partial \tensor{\varepsilon} }
   =
 2 \, G  \,  \deviatoric{ \tensor{\varepsilon} } +
   \, K  \,  \left( \, \trace{\bvarepsilon} \,   - 
   || \,  \tensor{\varepsilon}^{s} \, ||
    \;  \right)  \, \mathds{1}  
   \; .
\end{eqnarray}
Note that the derivative $\partial \psi_{el} / \partial c_\alpha$, in eq. \eqref{eq:chempotbeta} is the sum of two contributions
\begin{eqnarray}
   \frac{\partial \psi_{el}}{ \partial c_\alpha}
   =  - \omega_\alpha \, \trace{ \tensor{\sigma}^e  } +
   \frac{1}{2} \,  \frac{\partial K}{ \partial c_\alpha } \, \trace{ \tensor{\varepsilon} - \tensor{\varepsilon}^{s}  } ^2 
   +
   \; \frac{\partial G}{ \partial c_\alpha} \; || \, \deviatoric{\tensor{\varepsilon} - \tensor{\varepsilon}^{s}  }    \, ||^2 
   \; .
\label{eq:psielderivativecl}
\end{eqnarray}
The first emanates from the swelling part of the strain, and is present even if the material properties are independent on concentration of species. Assuming that material properties are such, 
Nernst-Planck equation \eqref{eq:FluxElec_model2} is extended as follows
\begin{align}
\label{eq:Nernst_Planck_extended}
\vect{h}_\alpha(x,t) =
&
- \diffusivity_\alpha   \;  \left[ 1- 2 \mychi \; \theta_\alpha \, ( 1 - \theta_\alpha ) \right]  \;  \nabla c_\alpha(x,t)
\\ & \nonumber
-   3 \, \tensor{M}(c_\alpha)  \;  K  \; \omega_\alpha \,  \left[
            3 \, \omega_\alpha   \, \nabla{ c_\alpha}  
   - \nabla{ \trace{ \tensor {\varepsilon} }  }
   \right] 
\\ & \nonumber
- \frac{\diffusivity_\alpha F}{RT} c_\alpha(x,t) \; \nabla\phi_e(x,t)
\; ,
\end{align}
with $\alpha = {\rm Li^+}, {\rm Li}^+_{\rm hop}$. The mobility tensor we selected reads ( $\theta_{\alpha} = c_{\alpha} / c_\alpha^{sat}$ )
\begin{equation}
\label{eq:isotropicmobility1}
    \tensor{M} ( c_{\alpha} ) = \frac{\diffusivity_\alpha }{RT} \, c_\alpha^{sat} \; \theta_{\alpha} \, \left( 1 -  \theta_{\alpha} \right) \; \mathds{1}
    \; ,
\end{equation}
{\em{accounts for saturation}} and in this differs from \eqref{eq:landsM}.
Energetic and entropic contributions in the constitutive law \eqref{eq:Nernst_Planck_extended} have been already described in eq. \eqref{eq:model4_5}. The mechanical contribution to the mass flux is driven by the chemical expansion coefficient, and derives from thermodynamic consistency.

\bigskip
Mass balance equations, after inserting \eqref{eq:Nernst_Planck_extended} into eqs. \eqref{eq:MassEle_model1c}-\eqref{eq:MassEle_model1d}, do not form a complete set, because ionic transport entails movement of mass as well as of charge. In order to build a multiscale compatible theory, the generally accepted electroneutrality assumption cannot be taken, since it prevents to impose the conservation of energy across the scales: this concept has been illustrated with great detail in \cite{SalvadoriEtAlJMPS2013,SalvadoriEtAlJPS2015} and will not be elaborated here further. Multiscale compatibility is granted by using Ampere's law (with Maxwell's correction in the realm of small strains)
\begin{align}
\label{eq:IncrementalGaussLaw}
     & \divergence{ - \varepsilon \; \gradient{  \frac{  \partial \phi_e }{\partial t} } +   F \, \left( \vect{h}_{\rm Li^+} + \vect{h}_{{\rm Li}^+_{\rm hop}} \right) }=0
     & \quad 0\leq x \leq L_e \; .
\end{align}
When multiscale is not invoked, the electroneutrality assumption 
\begin{equation}
c_{{\rm Li}^+}({x},t) + c_{{\rm Li}^+_{\rm hop}}({x},t)  = c_{\rm n^-}({x},t)
\; .
\label{eq:electroneutrality:newmodel}
\end{equation}
can be called for.

\bigskip
A widespread choice for the initial conditions for concentrations and electric potential enforces equilibrium conditions. They hold
\begin{subequations}
\begin{align}
&\phi_e(x,0)=0 \; ,\\
&c_{{\rm Li_0}}(x,0)=c_{{\rm Li_0}}^{eq}=(1-\delta)c_0 \; ,\\
&c_{{\rm n}^-}(x,0) =c_{{\rm n^-}}^{eq}=\delta c_0 \; ,\\
&c_{{\rm Li}^+}(x,0)=c^{eq}_{\rm Li^+}= \frac{ \delta c_0  } {1+ K_{\rm eq}^{\rm hop}  } = K_{\rm eq}^{\rm ion} \left( \frac{1}{\delta} -1 \right) \; , \\
&c_{{\rm Li}^+_{\rm hop}}(x,0)=c^{eq}_{{\rm Li}^+_{\rm hop}}= K_{\rm eq}^{\rm hop}  \;  \frac{ \delta c_0  } {1+ K_{\rm eq}^{\rm hop}  } = \delta \; c_0 + K_{\rm eq}^{\rm ion} \left(1- \frac{1}{\delta} \right)  \; .
\end{align}
\label{eq:initialconditionconc}
\end{subequations}

\bigskip
\textbf{Interface conditions} - Interface conditions for this advanced model are eqs. \eqref{eq:BoundEle_model2}, accounting for identity \eqref{eq:NonFaradaicCurrent} for the non-faradaic current contribution. The faradaic current emanates from charge transfer kinetics, as proposed in \cite{RaijmakersEtAlEA2020}. In view of the splitting of lithium flux in interstitial and hopping, constitutively specified by eq. \eqref{eq:Nernst_Planck_extended},
faradaic interface conditions split, too:
\begin{align}
&
i^{ct}_{a/e} =  { i^{ct}_{a/e}}_{\rm Li^+} + { i^{ct}_{a/e}}_{{\rm Li}^+_{\rm hop}} \; ,
&
i^{ct}_{e/c} =  { i^{ct}_{e/c}}_{\rm Li^+} +  { i^{ct}_{e/c}}_{{\rm Li}^+_{\rm hop}} 
\label{eq:BVtot_model3}
\end{align}
with the interstitial and hopping contributions to charge transfer current clearly identified. They can be inferred from Butler-Volmer eqs. \eqref{eq:BV} or \eqref{subeq:faradaic}. The exchange current read:
\begin{subequations}
\begin{align}
&
{ i^{0}_{a/e}}_{\rm Li^+}  = \, F \, k_2 \; (\overline{c}_{{\rm Li^+}} )^{\alpha} \; (\overline{c}_{\rm Li})^{ 1-\alpha} \; ,
&
{ i^{0}_{e/c}}_{\rm Li^+}  = F \; k_1 \; c_{\rm Li^\oplus}^{sat} \; 
\left(
1-
\frac{\overline{c}_{\rm Li^\oplus}}{c_{{\rm Li}^\oplus}^{sat}}
\right)^{\alpha} 
\; 
\left(
\frac{\overline{c}_{\rm Li^\oplus}}{c_{{\rm Li}^\oplus}^{sat}}
\right)
^{1-\alpha}
\; 
(\overline{c}_{{\rm Li^+}})^{\alpha}
\\
&
{ i^{0}_{a/e}}_{{\rm Li}^+_{\rm hop}}   
= 
{ i^{0}_{a/e}}_{\rm Li^+}  
\;
\left(
\;
\frac{
\overline{c}_{{\rm Li}^+_{\rm hop}} 
}
{
\overline{c}_{{\rm Li}^+}  
}
\right)^{\alpha}
\; ,
&
{ i^{0}_{e/c}}_{{\rm Li}^+_{\rm hop}}   
= 
{ i^{0}_{e/c}}_{\rm Li^+}  
\;
\left(
\;
\frac{
\overline{c}_{{\rm Li}^+_{\rm hop}} 
}
{
\overline{c}_{{\rm Li}^+}  
}
\right)^{\alpha}
\; .
\end{align}
\label{eq:BV_model3}
\end{subequations}
where $\overline{c}_{\rm Li^+}$, $\overline{c}_{{\rm Li^+}_{\rm hop}}$ are average bulk concentrations of species ${\rm Li^+}$ and ${\rm Li^+}_{\rm hop}$, respectively. Note that, differently from \cite{RaijmakersEtAlEA2020}, those averages are not time independent.
Lacking more clear understanding, we assume that the non faradaic current $i_\alpha^{dl}(t) $ as in eq. \eqref{eq:NonFaradaicCurrent} is proportional to the faradaic splitting, i.e.
\begin{subequations}
\begin{align}
\label{eq:non_faradaic_a}
& & i_{s}^{dl} =  { i^{dl}_{s}}_{\rm Li^+} + { i^{dl}_{s}}_{{\rm Li}^+_{\rm hop}} \; , &\\
& { i^{dl}_{s}}_{\rm Li^+}  = \frac{{ i^{ct}_{s}}_{\rm Li^+} }{ i^{ct}_{s}} \;  c_s^{dl} \; \frac{\partial \llbracket \, \phi \, \rrbracket }{\partial t}   \; , 
& & { i^{dl}_{s}}_{{\rm Li}^+_{\rm hop}}   = \frac{{ i^{ct}_{s}}_{{\rm Li}^+_{\rm hop}}  }{ i^{ct}_{s}} \;  c_s^{dl} \; \frac{\partial \llbracket \, \phi \, \rrbracket }{\partial t}  \; ,
\label{eq:non_faradaic_b}
\end{align}
\label{eq:non_faradaic}
\end{subequations}
where the jump $\llbracket \, \phi \, \rrbracket$ of the electric potential at the electrolyte/electrode interface is always defined as the electrode potential minus the electrolyte potential and $s ={a/e}, {e/c}$.
 The Neumann conditions on fluxes at the left and right boundaries of the electrolyte eventually read:

\begin{subequations}
\begin{align}
& \vect{h}_{\rm Li^+}(0,t) \cdot \vect{n} = - ({  { i^{ct}_{a/e}}_{\rm Li^+} +  { i^{dl}_{a/e}}_{\rm Li^+} }) / {F}  \; ,
& \vect{h}_{\rm Li^+}(L_e,t) \cdot \vect{n} = - ( {  { i^{ct}_{e/c}}_{\rm Li^+} +  { i^{dl}_{e/c}}_{\rm Li^+} }) / {F}  \; ,
\\
& \vect{h}_{{{\rm Li}^+_{\rm hop}}}(0,t) \cdot \vect{n} = - ({  { i^{ct}_{a/e}}_{{{\rm Li}^+_{\rm hop}}} +  { i^{dl}_{a/e}}_{{{\rm Li}^+_{\rm hop}}} } ) / {F}  \; ,
& \vect{h}_{{{\rm Li}^+_{\rm hop}}}(L_e,t) \cdot \vect{n} = - ( {  { i^{ct}_{e/c}}_{{{\rm Li}^+_{\rm hop}}} +  { i^{dl}_{e/c}}_{{{\rm Li}^+_{\rm hop}}} }) / {F}  \; .
\end{align}
\end{subequations}
Continuity of displacements and normal tractions are interface conditions for the mechanical governing equation \eqref{eq:stressbalance}.
The electrodes governing equations and boundary conditions do not differ from section \ref{subsec:RaijmakersEtAlEA2020}.

\section{Benchmark comparison }
\label{sec:benchmarks}

The two models illustrated in sections \ref{subsec:RaijmakersEtAlEA2020} and \ref{subsec:novelformulation} are validated against the experimental outcomes described in \cite{Danilovetal2011}. That paper also provides values of material and geometrical parameters, which are collected in Table \ref{Tab:Input}. 

\begingroup
\setlength{\tabcolsep}{8pt} 
\renewcommand{\arraystretch}{1.3}
\begin{table}[!htb]
\scriptsize
\centering
\begin{tabular}{ c m{2.4cm} c p{7.90cm}}
\hlineB{3.0}
 \multicolumn{4}{c}{Input parameters} 
\\ \\
\hlineB{2.0}
 {\em Parameter}      & Value                         & Unit      & Description 
 \\
\hlineB{1.0}
\\
 $T$                          &  $298.5$                & $\rm K$         &Temperature\\
 $L_a$                        &  $0.50\cdot10^{-6}$     & $ \rm m$         &Thickness of the anode\\
 $L_e$                        &  $1.50\cdot10^{-6}$     & $\rm m$         &Thickness of the electrolyte\\
 $L_c$                        &  $0.32\cdot10^{-6}$     & $\rm m$         &Thickness of the cathode\\
 $L_{col}$                    &  $0.10\cdot10^{-6}$     & $\rm m$         &Thickness of the positive collector\\
 $A$                          &  $1.00\cdot10^{-4}$     & $\rm m^2$       &Geometrical surface area\\
 $c_{\rm Li^\oplus}^{sat}$        &  $2.34\cdot10^{4}$    & $\rm mol \; m^{-3}$    &Maximum concentration of $\rm Li^\oplus$ ions in the electrode\\
 $k_a$                        &  $1.08\cdot10^{7}$     &$\rm S \; m^{-1}$        &Electrical conductivities in the lithium anode\\
 $k_{col}$                        &  $10.0$     &$\rm S \; m^{-1}$        &Electrical conductivities in the current collector\\
 $k_{f}^{\rm ion}$                    &  $1.125\cdot10^{-5}$ ($\it 1.80\cdot10^{-5}$ )    &$\rm s^{-1}$          &Lithium ion generation reaction rate constant for eq.\eqref{eq:IonizationReaction}\\
 $k_{b}^{\rm ion}$                    &  $0.90\cdot10^{-8}$     & $ \rm m^3 \; mol^{-1} \, s^{-1}$ &Lithium ion recombination reaction rate constant for eq.\eqref{eq:IonizationReaction}\\
 $k_f^{\rm hop}$                    &  $8.10\cdot10^{-9}$ ($\it 1.69\cdot10^{-9}$ )    &$ \rm s^{-1}$          &Lithium ion generation reaction rate constant for eq.\eqref{eq:IonizationReaction2}\\
 $k_{b}^{\rm hop}$                    &  $0.90\cdot10^{-8}$     & $ \rm m^3 \; mol^{-1} \, s^{-1}$ &Lithium ion recombination reaction rate constant for eq.\eqref{eq:IonizationReaction2}\\
 $c_a^{dl}$                    &  $1.74\cdot10^{-4}$     & $\rm F \; m^{-2}$          &Double layer capacity per unit area of anode\\
 $c_c^{dl}$                    &  $5.30\cdot10^{-3}$     & $\rm F \; m^{-2}$          &Double layer capacity per unit area of cathode\\
 $\alpha_n$                   &   $0.6$                  & -           &Charge transfer coefficient for the negative electrode\\
 $\alpha_p$                   &  $0.6$                  & -           &Charge transfer coefficient for the positive electrode\\
 $\diffusivity_{\rm Li^+}$                   &  $5.10\cdot10^{-15}$    & $\rm m^2 \; s^{-1}$      &Diffusion coefficient for $\rm Li^+$ ions in the electrolyte\\
 $ \diffusivity_{{\rm Li}^+_{\rm hop}}$             &  $0.90\cdot10^{-15}$    & $\rm m^2 \; s^{-1}$      &Diffusion coefficient for ${\rm Li}^+_{\rm hop}$ ions in the electrolyte\\
 $\diffusivity_{{\rm Li}^\oplus}$              &  $1.76\cdot10^{-15}$    & $\rm m^2 \; s^{-1}$      &Diffusion coefficient for $\rm Li^\oplus$ ions in the cathode\\
 $k_{1}$                    &  $5.10\cdot10^{-6}$     &$\rm m^{2.5} \; mol^{-0.5} \; s^{-1}$          &Standard reaction rate constant for forward reaction in eq.\eqref{eq:e/c}\\
 $k_{2}$                    &  $1.09\cdot10^{-5}$     & $\rm m \; s^{-1}$ &Standard reaction rate constant for forward reaction in eq.\eqref{eq:a/e}\\
 $\delta$                     &  $0.18$                 &-            &Fraction of mobile ions in the electrolyte in equilibrium\\
 $c_0$                        &  $6.01\cdot10^{4}$      & $\rm mol \; m^{-3}$   &Maximal lithium concentration in the electrolyte\\
 $\varepsilon_r$              &  $2.25$                 &--          &Relative permittivity in the electrolyte
 \\
 \\
\hlineB{3.0}
\end{tabular}
\caption{\em Model parameters used during simulations}
\label{Tab:Input}
\end{table}
\endgroup

The positive electrode is a layer of $\rm LiCoO_2$ with thickness $\rm L_c = 0.32 \mu m$, deposited on a platinum substrate. A Lithium metal foil with thickness $\rm L_a=0.50 \mu m$ is used as negative electrode. The solid electrolyte is a one-micron-thick ($\rm L_e = 1.00 \mu m$) layer of LiPON. The surface area of the deposited electrodes is $A\!\!=10^{-4} {\rm m}^2$ and the theoretical storage capacity of the battery is $\rm10^{-5} Ah$. 

The electrochemical cell is subject to a galvanostatic process of discharge at different $C\!\!-\!\!rates$, under a temperature-controlled condition of $ \rm 25^o C$. The current corresponding to $C\!\!-\!rate = j$ is denoted with $I_{jC}$. For $C\!\!-\!\!rate\!\!=\!\!1$,  $I_{1C}$ amounts at $10^{-5} A$.  Initial and boundary conditions are made compatible with thermodynamic equilibrium at $t\!=\!0$, tuning the density current $i_{bat}(t)$ in time as:
\begin{equation}
i_{bat}(t)=(1-e^{-t}) \; \frac{I_{jC}}{A} \; ,
\label{eq:current}
\end{equation}
with $t$ in seconds. 
In view of \eqref{eq:current}, the concentrations of ions across all components of the battery at $t\!=\!0$ are uniform and at equilibrium, because no profiles developed yet. By enforcing the fraction of mobile lithium in the electrolyte $\delta=0.18$ and a maximum concentration of lithium host sites in the electrolyte  $c_0 = 60100\, \rm mol/m^3$, eq. \eqref{eq:initialconditionconc} provides
\begin{subequations}
\begin{align}
&c_{\rm Li}(x,0)=c_{\rm Li}^{\rm eq}=2.40\!\cdot\!10^{4} & {\rm mol/m^3}  & & -L_a\leq x\leq 0,\\
&c_{\rm Li_0}(x,0)= 4.93\!\cdot\!10^{4} & {\rm  mol/m^3 }  & & 0\leq x\leq L_e,\\
&c_{\rm n^-}(x,0)= 1.08\!\cdot\!10^{4} & {\rm  mol/m^3 } & & 0\leq x\leq L_e,\\
&c_{\rm Li^+}(x,0)= 5.68\!\cdot\!10^{3} & {\rm  mol/m^3 } & & 0\leq x\leq L_e,\\
&c_{{\rm Li}^+_{\rm hop}}(x,0) =  5.12\!\cdot\!10^{3} & {\rm  mol/m^3 } & & 0\leq x\leq L_e,\\
&c_{\rm Li^\oplus}(x,0)=c_{\rm Li^\oplus}^{\rm eq}=1.20\!\cdot\!10^{4} & {\rm  mol/m^3 } & & L_e\leq x\leq L_e+L_c
\; .
\end{align}
\label{eq:Bound_Simulation}
\end{subequations}
The electric potential at the interface between the anode and the solid electrolyte is fixed as:
\begin{equation}
\phi(0,t)=0\;[V] \quad \forall t.
\label{eq:Bound_Simulation1}
\end{equation}
All material parameters used in the simulation are listed in Table~\ref{Tab:Input}. According to eqs. \eqref{eq:equilibrium_constant_w} and \eqref{eq:equilibrium_constant_y}, the equilibrium constants read
\begin{align}
&
K_{\rm eq}^{\rm ion}   =  \frac{1.125\cdot10^{-5} }{ 0.90\cdot10^{-8} } = 1250
\; ,
&
K_{\rm eq}^{\rm hop}  =  \frac{8.10\cdot10^{-9} }{ 0.90\cdot10^{-8} } = 0.9 
\; .
\end{align}
throughout this benchmark comparison.

\begin{figure}[!htb]
\centering
\includegraphics[width=100mm]{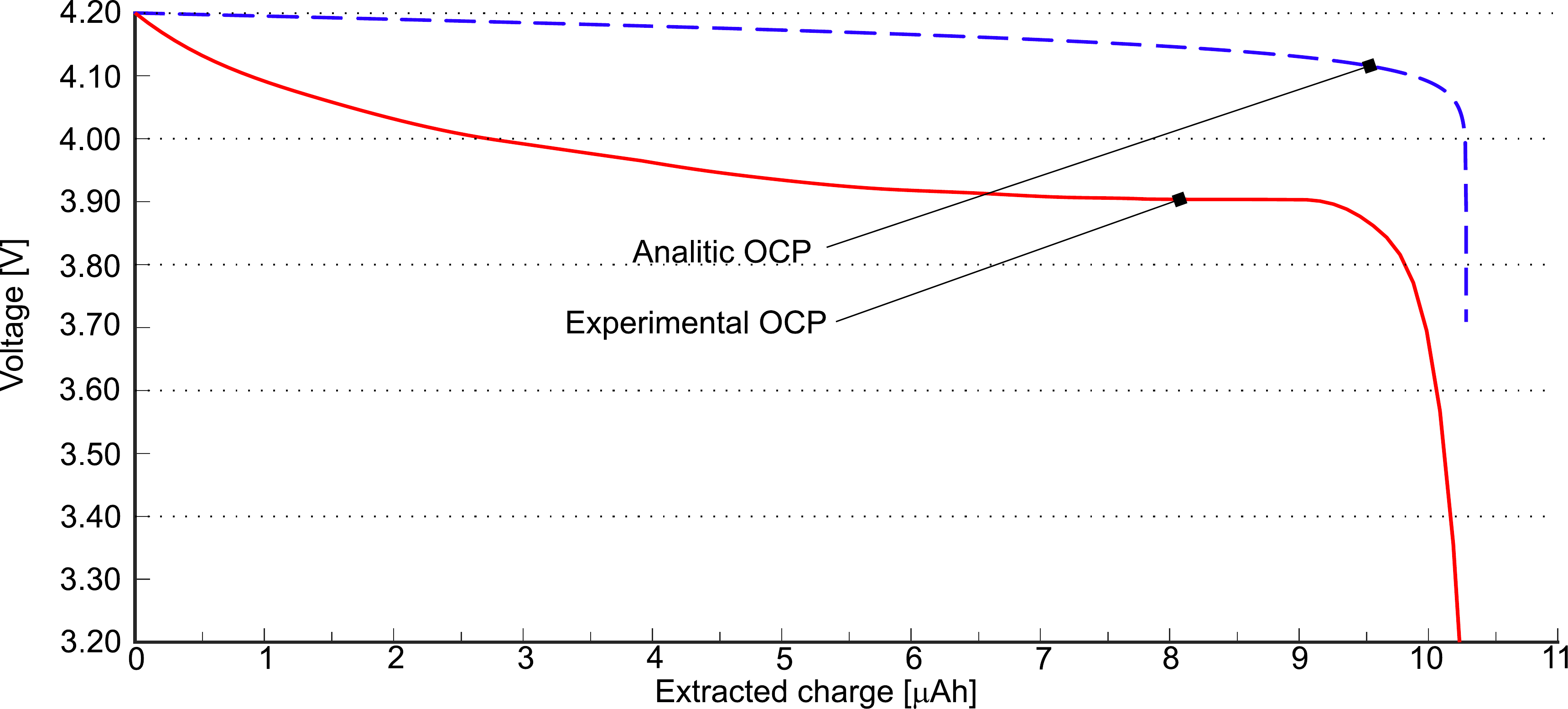}
\caption{\em Open circuit potential as a function of extracted charge. In red line the OCP obtained with experimental test and in blue line the OCP evaluated  as in \cite{PurkayasthaMcMeekingCM2012}.}
\label{fig:OCP}
\end{figure}

\begin{figure}[!htb]
\centering
    \begin{subfigure}[!htb]{.49\textwidth}
        \centering
        \includegraphics[width=8.0cm]{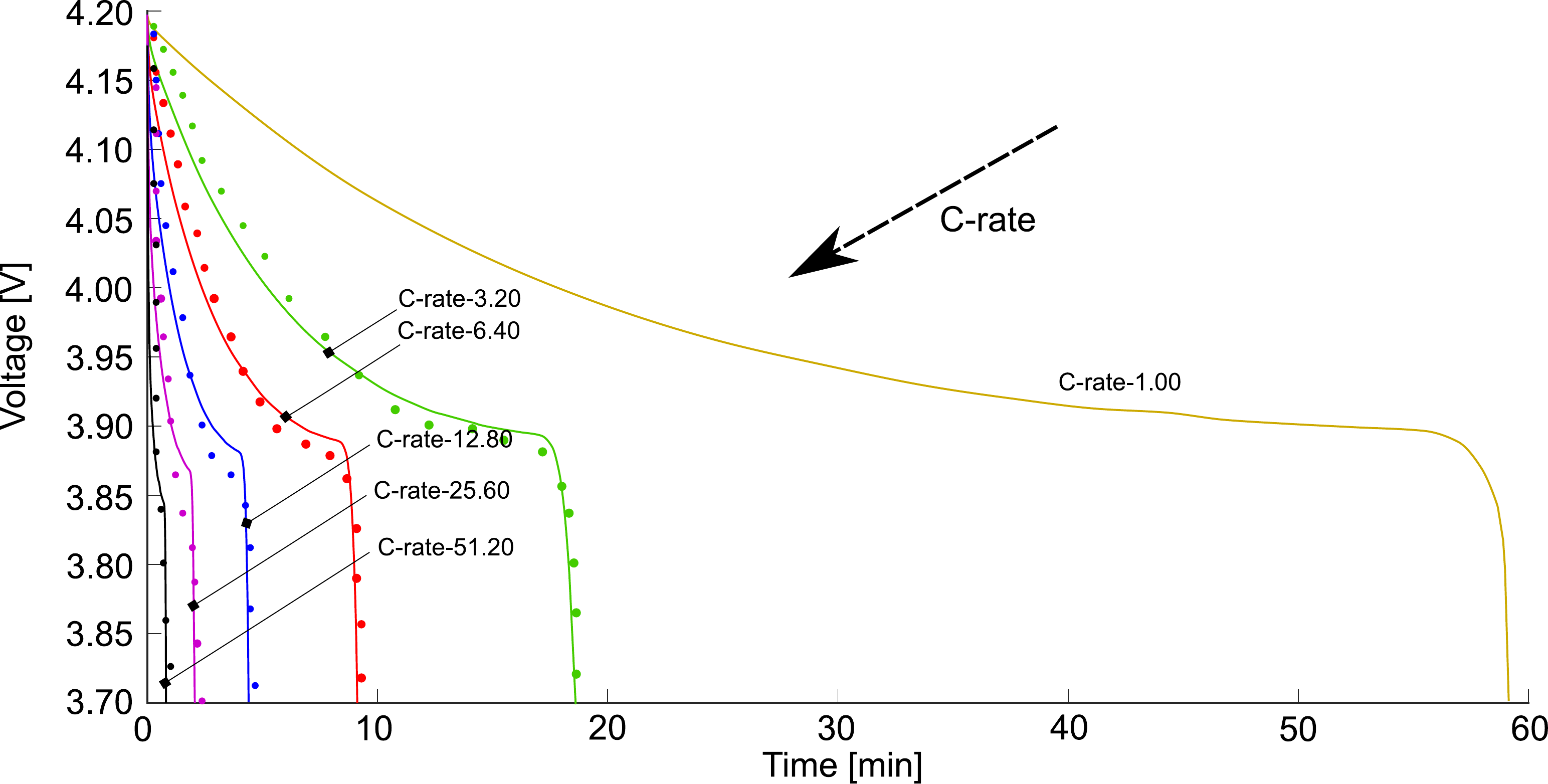}
        \caption{Voltage vs time }
    \end{subfigure}\hfill%
    \begin{subfigure}[!htb]{.49\textwidth}
        \centering
        \includegraphics[width=8.0cm]{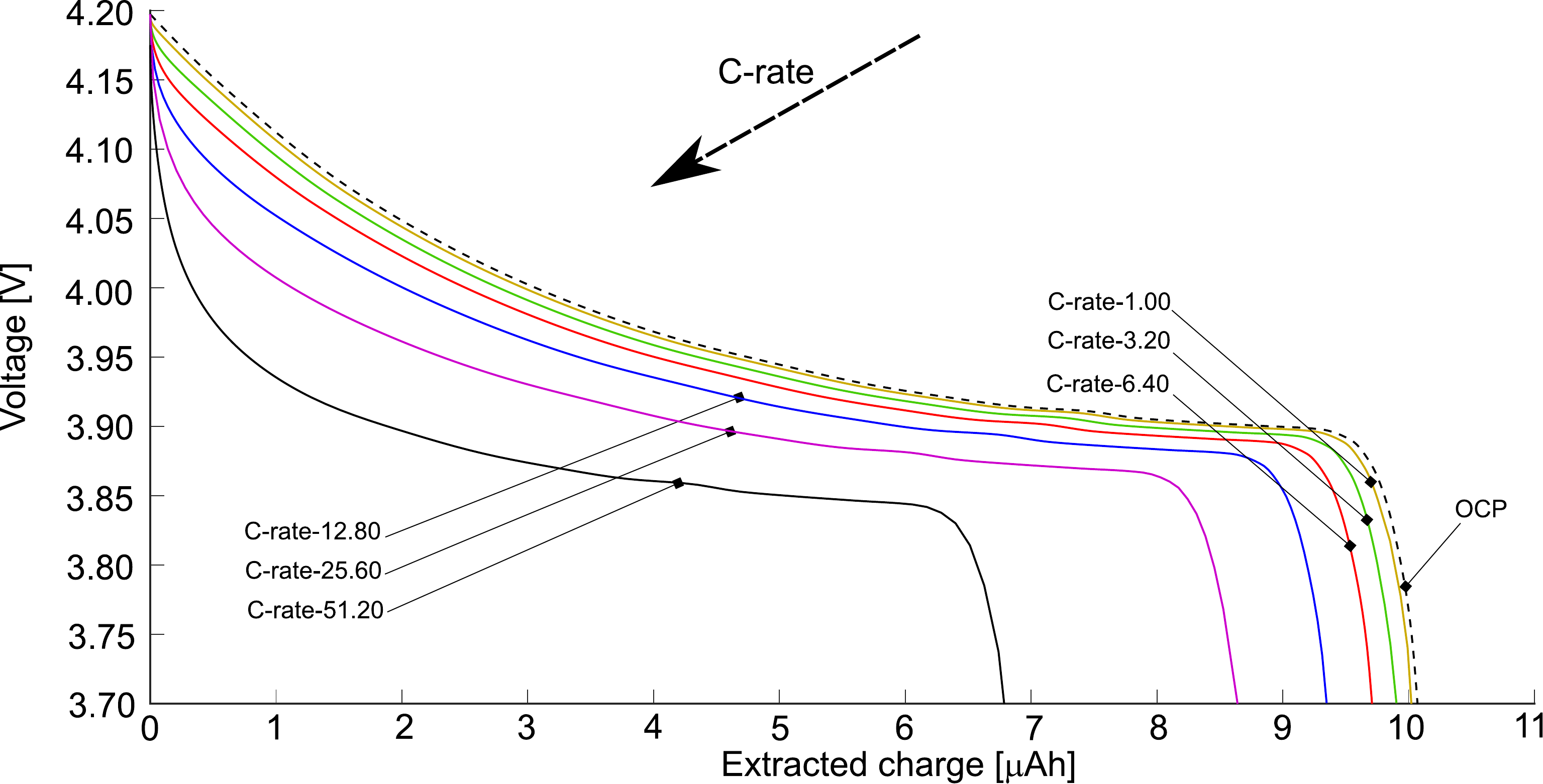}
        \caption{Voltage vs state of charge }
    \end{subfigure}\hfill%
%
\caption{\em Discharge curves as a function of time (a), and the extracted charge (b) for different $C\!\!-\!rates$. Colored lines correspond to different $C\!\!-\!rates$, whereas the dots identify the experimental values.}
\label{fig:discharge}
\end{figure}

\begin{figure}[!htb]
    \centering
    \begin{subfigure}[!htb]{.49\textwidth}
        \centering
        \includegraphics[width=8.0cm]{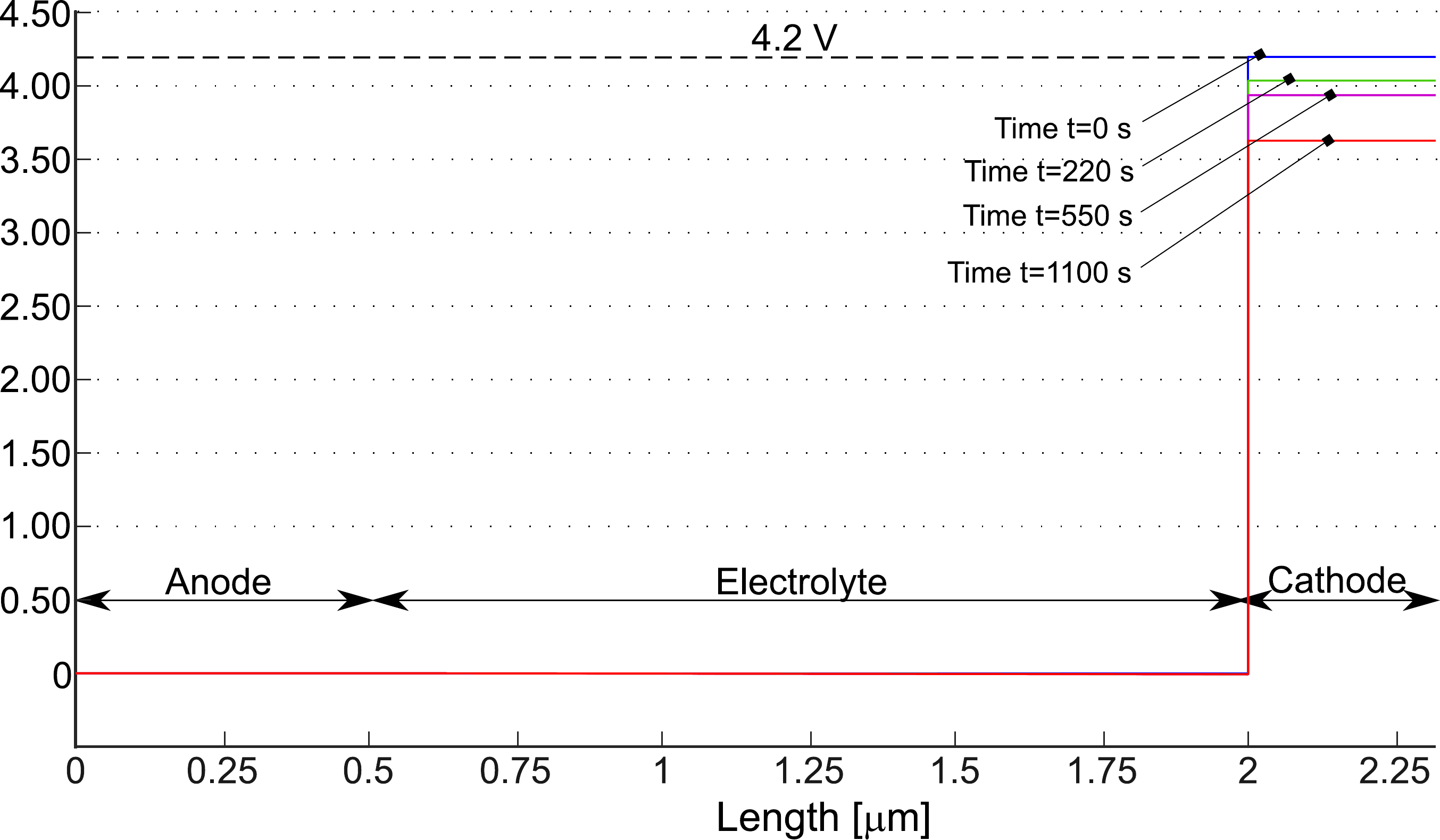}
        \caption{$C\!\!-\!\!rate=3.20$}
        \label{fig:ElectricPot_a}
    \end{subfigure}\hfill%
    \begin{subfigure}[!htb]{.49\textwidth}
       \centering
        \includegraphics[width=8.0cm]{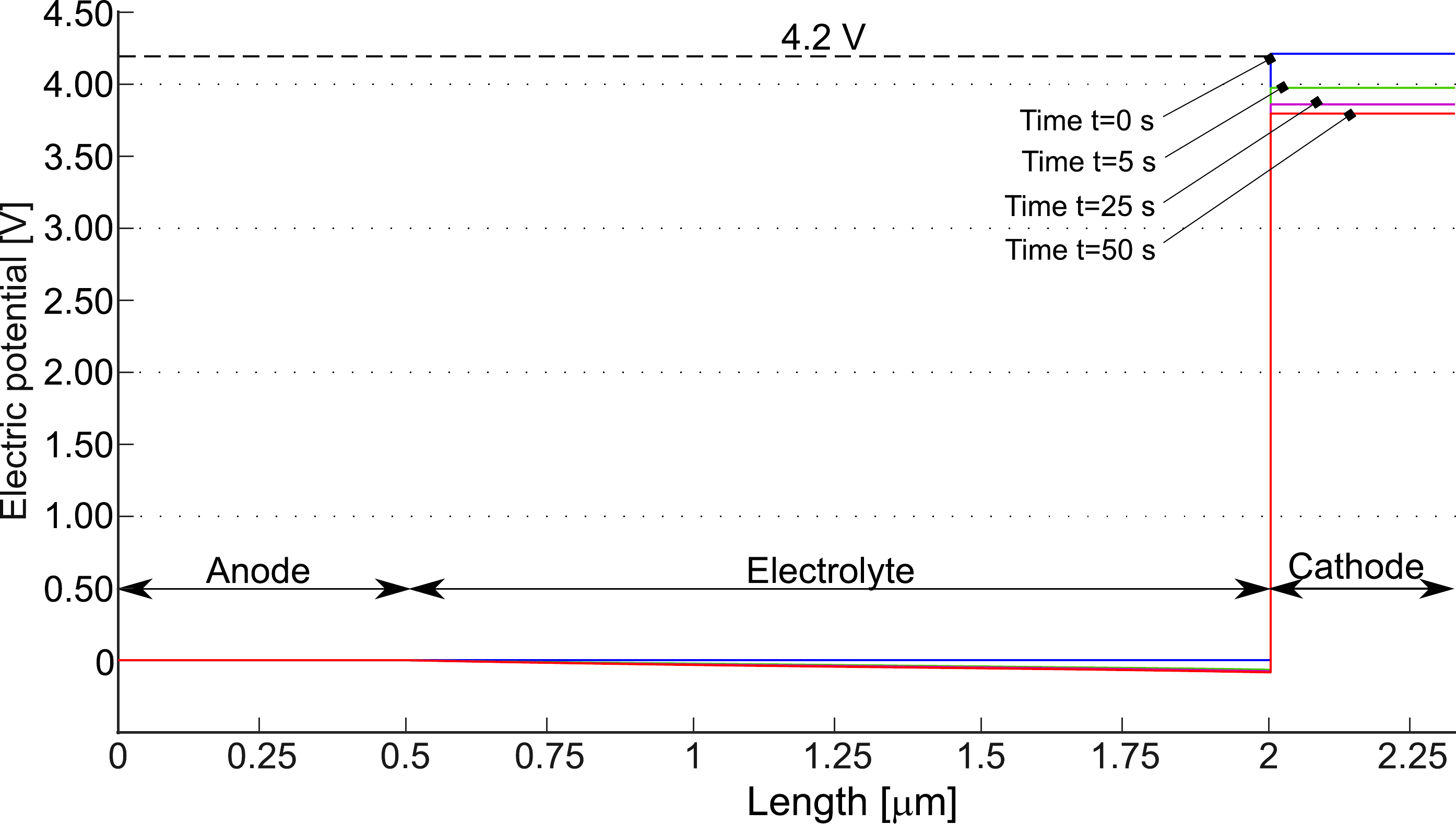}
        \caption{$C\!\!-\!\!rate=51.2$}
        \label{fig:ElectricPot_b}
    \end{subfigure}
    \\
    \begin{subfigure}[!htb]{.49\textwidth}
        \centering
        \includegraphics[width=8.0cm]{Carica-lunghezza-3_2}
        \caption{$C\!\!-\!\!rate=3.20$ for \cite{RaijmakersEtAlEA2020}}
        \label{fig:ElectricPot_c}
    \end{subfigure}\hfill%
    \begin{subfigure}[!htb]{.49\textwidth}
       \centering
        \includegraphics[width=8.0cm]{Carica-lunghezza-51_2}
        \caption{$C\!\!-\!\!rate=51.2$ for \cite{RaijmakersEtAlEA2020}}
        \label{fig:ElectricPot_d}
    \end{subfigure}
    \caption{\em The electric potential profile at different times for two different $C\!\!-\!\!rates$ for the new model ((a) and (b)) as well as for the model proposed in \cite{RaijmakersEtAlEA2020} ((c) and (d)).  At the initial time step, when the electric potential inside the single components of the battery is in equilibrium and no profile is developed. The last plotted time step corresponds to the instant when the concentration of lithium $\rm Li^\oplus$ inside the cathode reaches the saturation limit $c_{\rm Li^\oplus}^{\rm sat}$. 
}
    \label{fig:ElectricPot}
\end{figure}

\bigskip
\textbf{Solution schemes} 
-
The governing equations are numerically solved with the finite element method, with in house implementation of weak forms in the commercial numerical software Matlab.
The geometry and the unknown fields (see figs. \ref{fig:model2}, \ref{fig:GeneralBatteryScheme}) are discretized with $61$ linear elements; $1$ element is sufficient for the anode, since the lithium concentration is uniform ad the electric potential is linear;  the outcomes refer to a tessellation of $40$ finite elements covering the electrolyte and $20$ panels that discretize the cathode. In both cases, the mesh is refined near the electrode/electrolyte interface. The time marching is dealt with the backward Euler method, with fixed time increments of $ \Delta t\!=\!1.0 \, \rm s$.

The open circuit potential (OCP) used in the simulations is given in Fig. \ref{fig:OCP} as a function of the extracted charge. The OCP has been either reconstructed with splines stemming from experiments, the continuous line in Fig. \ref{fig:OCP}, or calculated analytically following the approach in \cite{PurkayasthaMcMeekingCM2012}, represented by a dashed curve in the figure. Numerical analysis will refer to both, but only the experimental OCP will be elaborated in the main text, while appendix \ref{app:anOCP} collects some outcomes from the analytical OCP.

The simulations account for a broad range of rates, from $1.0$ to $51.2$.
The corresponding experimental discharging curves have been plot with dots in Fig.\ref{fig:discharge}-a as a function of time, whereas Fig.\ref{fig:discharge}-b plots the same data as a function of the extracted charge: continuous lines correspond to simulations. Measurements and simulations agree well across the wide range of investigated $C\!\!-\!rates$. Obviously, the extracted charge decreases with increasing $C\!\!-\!rate$ due to the higher over-potentials. In fact lower discharge rates implies a lower rate of lithium insertion in the cathode and a more uniform Li distribution in the positive electrode.

\bigskip
\textbf{Electric potential profiles }
-
Figure \ref{fig:ElectricPot} depicts the electric potential $\phi(x)$ profile in the battery at different times for two different discharge rates, i.e. $C\!\!-\!rate\!=\!3.20$ (which theoretically allows to discharge the battery in $1125s$) and $C\!\!-\!rate\!=\!51.2$ (for which the battery in principle completes the discharge process in $70s$). Simulations quit when the concentration of lithium $\rm Li^\oplus$ inside the cathode reaches the saturation limit $c_{\rm Li^\oplus}^{sat}$, after $1085s$ and $50s$ respectively. Cathodic saturation is indeed the limit factor for the battery operation (see also \cite{MagriEtAl2021} for an extensive discussion on limiting factors in electrochemical cells, induced by materials and architectures).

At the initial time the electric potential discontinuity at the interfaces make Butler-Volmer currents vanishing. Based on the measured battery OCP, at full charge state $\Delta\phi\!=\!4.2 V$, as highlighted in Fig. \ref{fig:ElectricPot}. During the discharge, the potential drops in the anode and in the electrolyte. However, the major changes in the potential profile occur at interface between electrolyte and cathode: the battery voltage decreases significantly during discharge as depicted in Fig. \ref{fig:discharge} and it results almost uniform within the cathode. A similar behavior is observed for the model in \cite{RaijmakersEtAlEA2020}  ( see fig. \ref{fig:ElectricPot} (c) and (d)).

\begin{figure}[!htb]
    \centering
    \begin{subfigure}[!htb]{.245\textwidth}
        \centering
        \includegraphics[width=\textwidth]{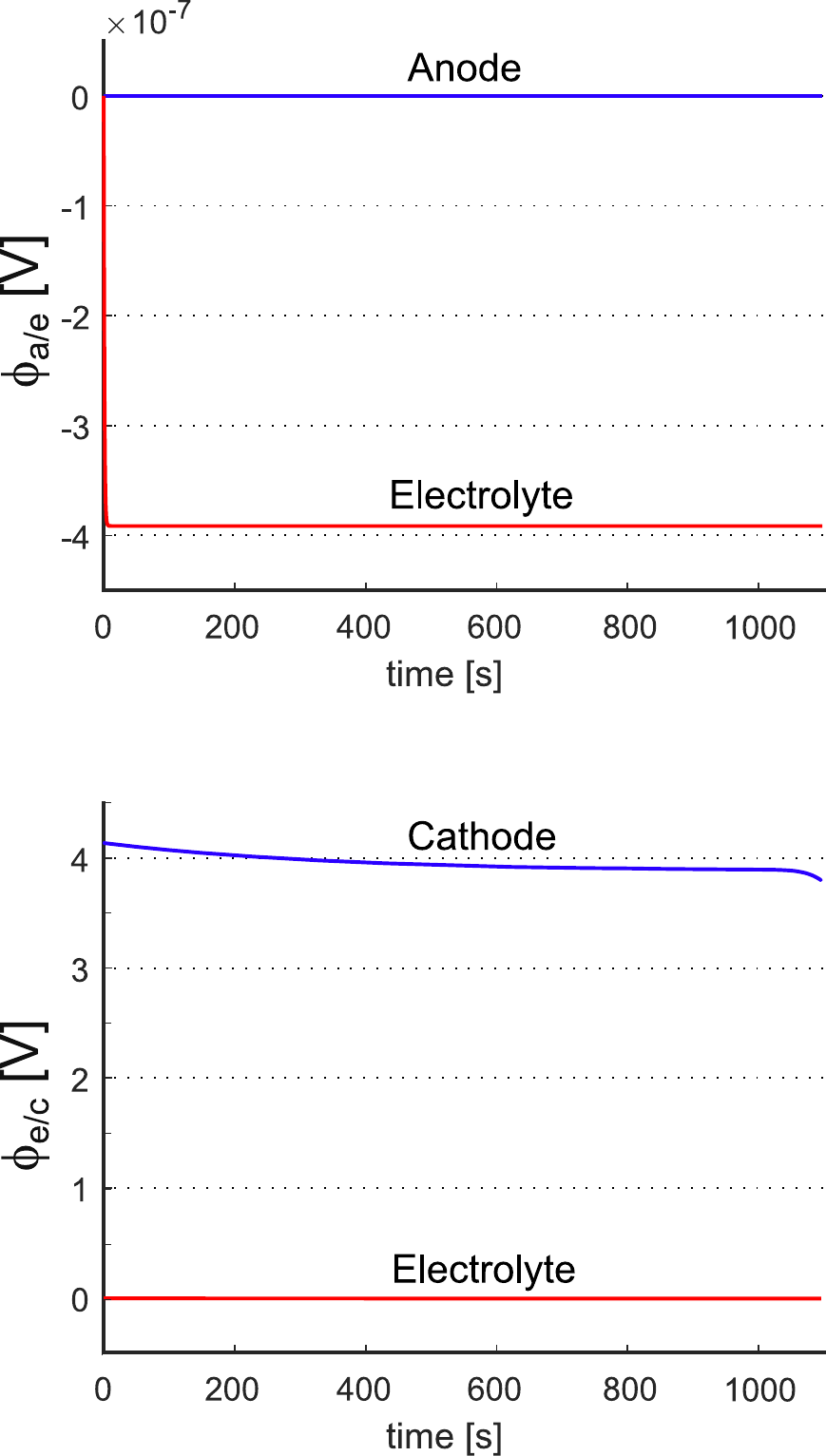}
        \caption{$C\!\!-\!\!rate=3.20$}
        \label{fig:Pot_a}
    \end{subfigure}\hfill%
    \begin{subfigure}[!htb]{.245\textwidth}
       \centering
        \includegraphics[width=\textwidth]{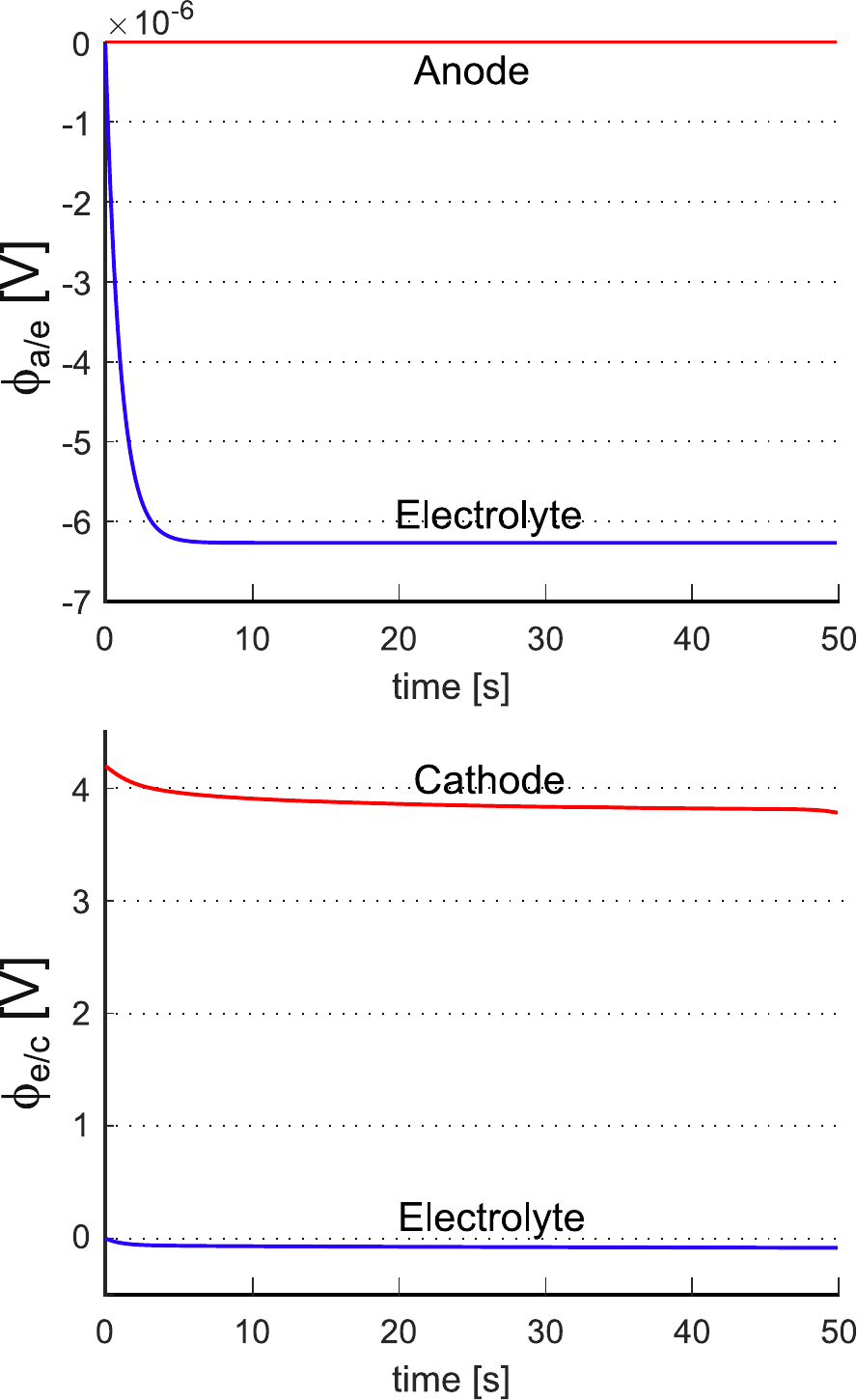}
        \caption{$C\!\!-\!\!rate=51.2$}
        \label{fig:Pot_b}
    \end{subfigure}
    \begin{subfigure}[!htb]{.245\textwidth}
        \centering
        \includegraphics[width=\textwidth]{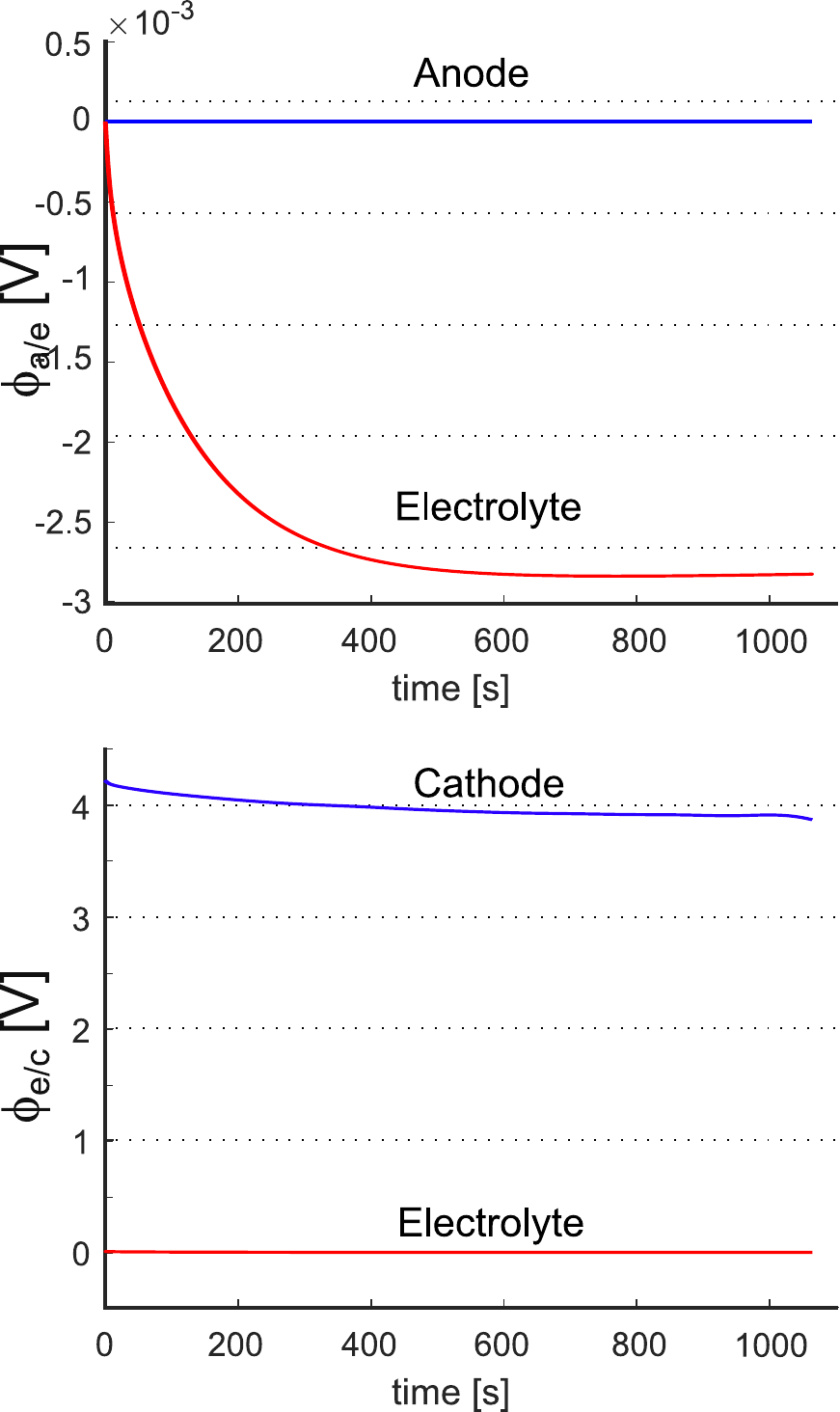}
        \caption{$C\!\!-\!\!rate=3.20$ for \cite{RaijmakersEtAlEA2020}}
        \label{fig:Pot_c}
    \end{subfigure}\hfill%
    \begin{subfigure}[!htb]{.245\textwidth}
       \centering
        \includegraphics[width=\textwidth]{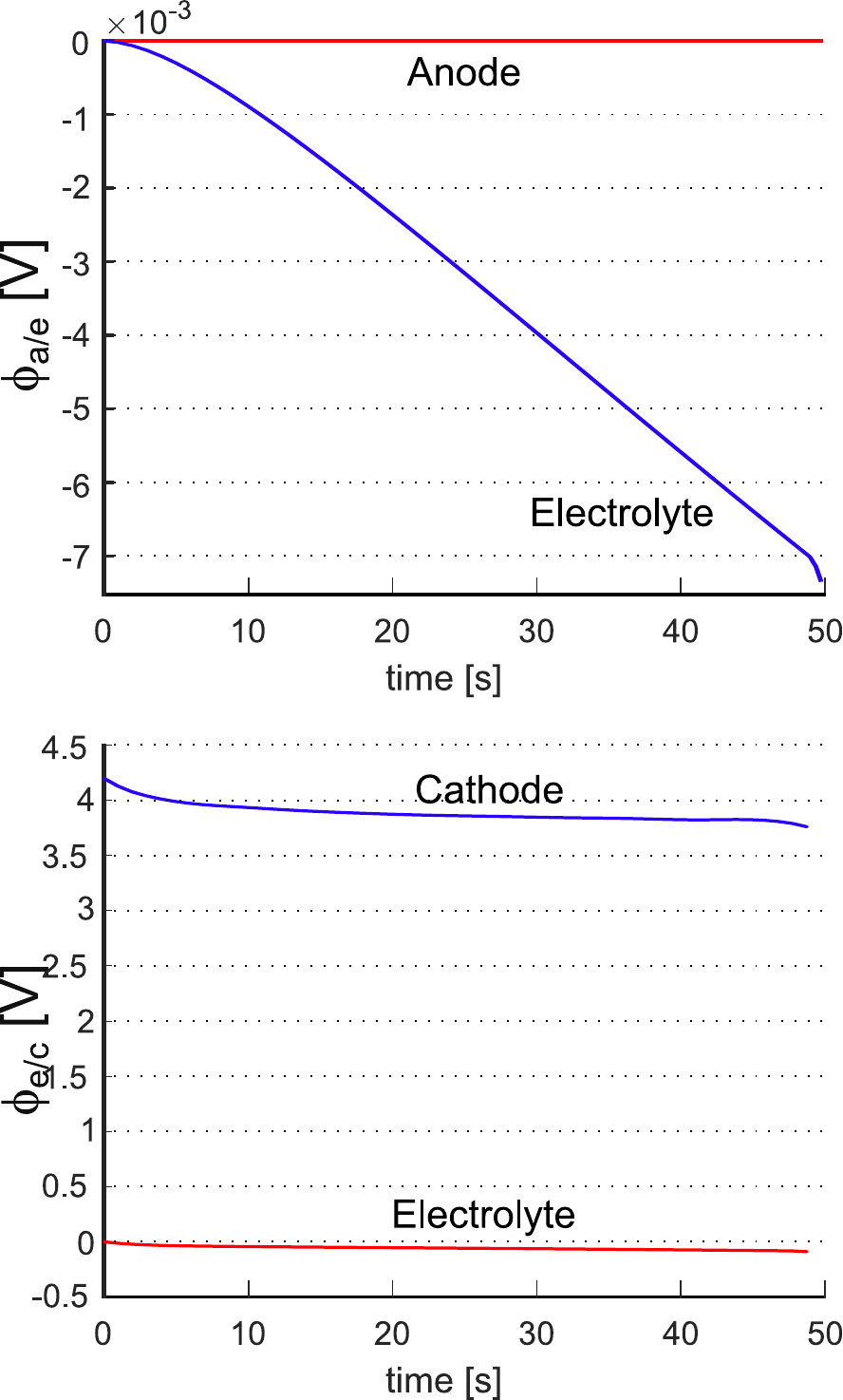}
        \caption{$C\!\!-\!\!rate=51.2$ for \cite{RaijmakersEtAlEA2020}}
        \label{fig:Pot_d}
    \end{subfigure}
    \caption{\em The electric potential in the interfaces with the electrodes as a function of time for two different $C\!\!-\!\!rates$ for the new model ((a) and (b)) as well as for the model proposed in \cite{RaijmakersEtAlEA2020} ((c) and (d)). For each interface both the values of the potential assumed inside the electrodes and inside the electrolyte are plotted.}
    \label{fig:ElectricPot_interfaces}
\end{figure}

The electric potential evolution in time at both interfaces is given in Fig.\ref{fig:ElectricPot_interfaces} and reflects the conclusions driven in Fig. \ref{fig:ElectricPot} 

\begin{figure}[!htb]
    \centering
    \begin{subfigure}{.495\textwidth}
        \centering
        \includegraphics[width=\textwidth]{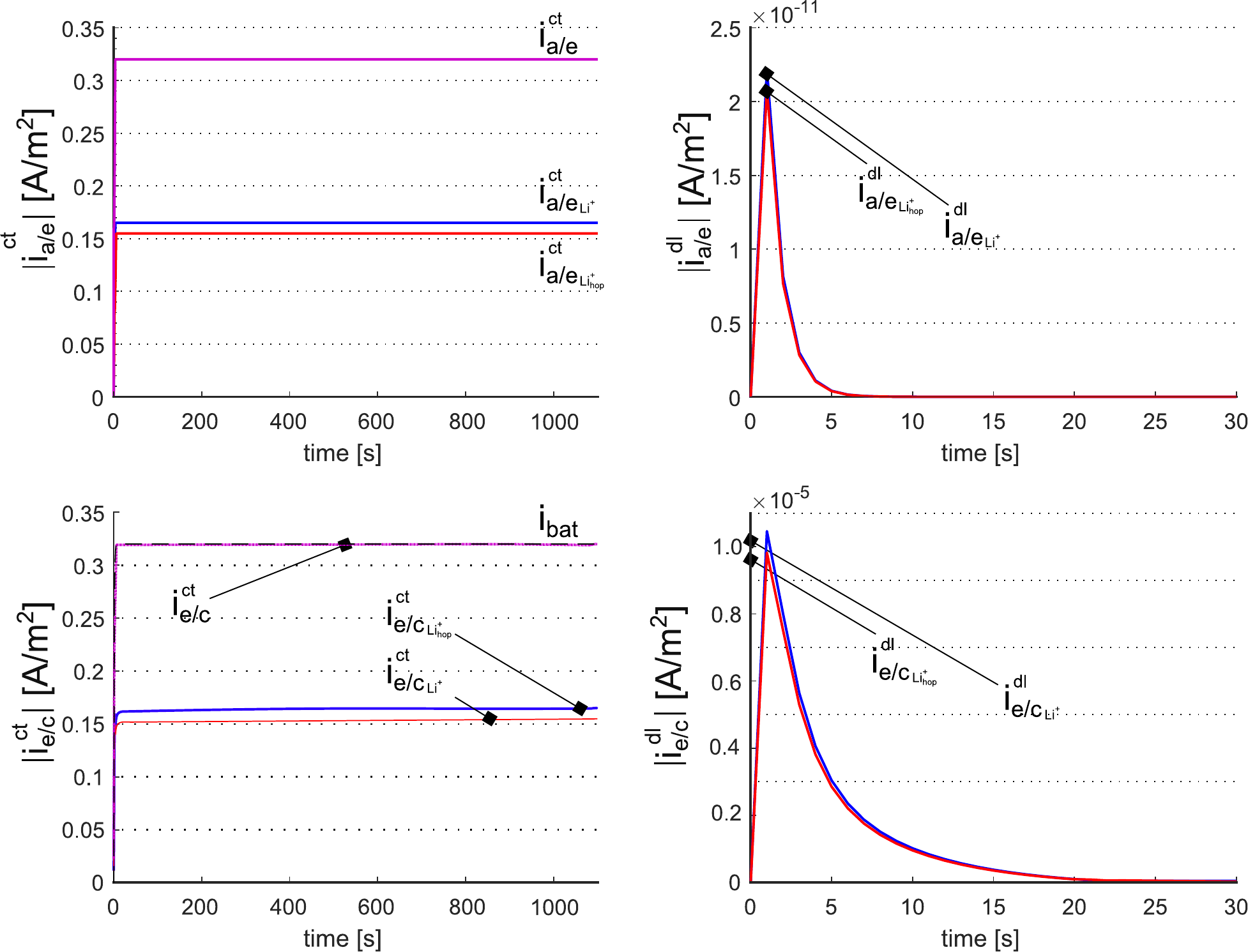}
        \caption{$C\!\!-\!\!rate=3.20$}
        \label{fig:ElectricCurr_a}
    \end{subfigure}
    \begin{subfigure}{.495\textwidth}
       \centering
        \includegraphics[width=\textwidth]{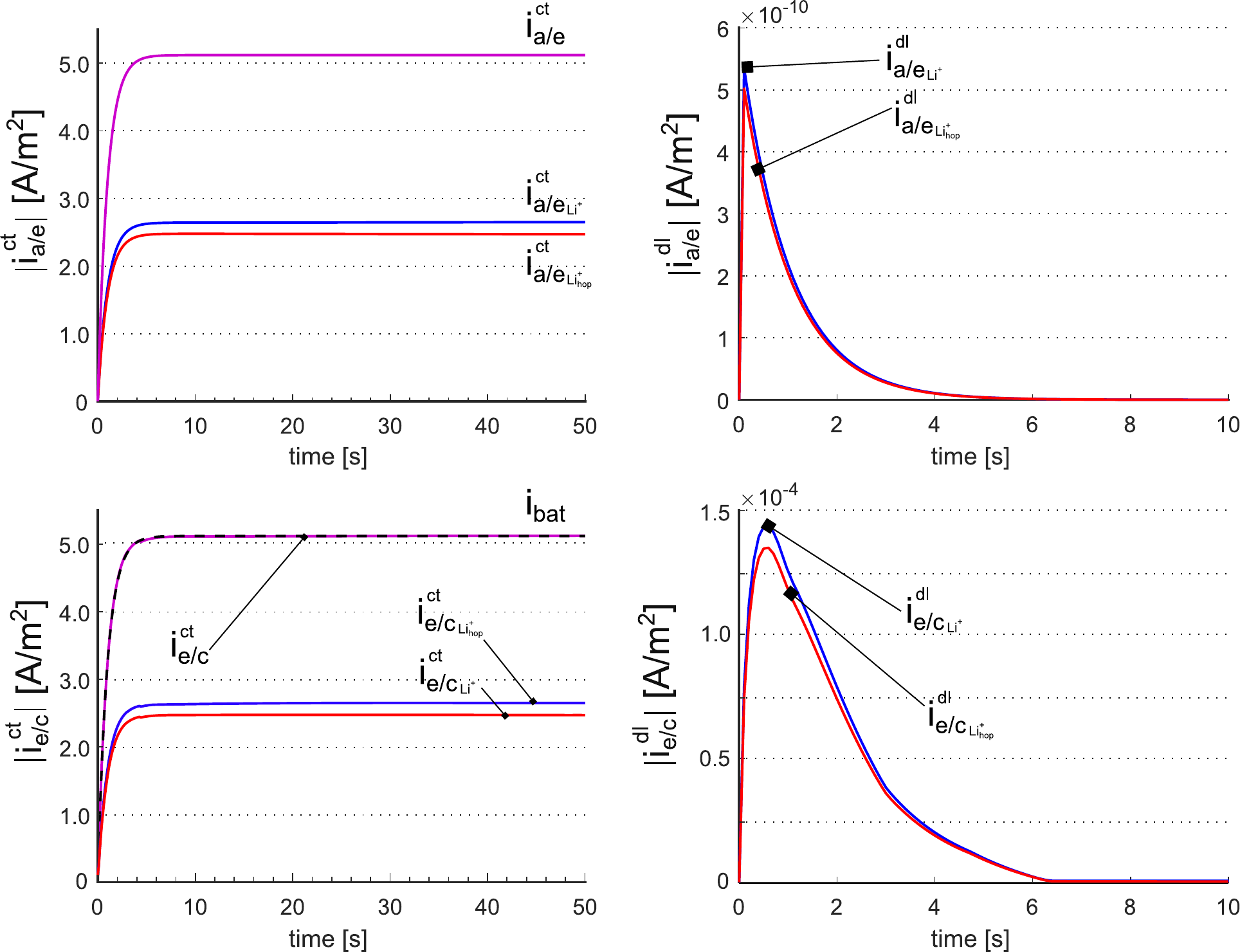}
        \caption{$C\!\!-\!\!rate=51.2$}
        \label{fig:ElectricCurr_b}
    \end{subfigure}
    \caption{\em Left (right) column: evolution in time of the charge transfer faradaic current $i^{ct}$ (double layer current $i^{dl}$), at the $a/e$ and $e/c$ interfaces at different C-rates.     }
    \label{fig:ElectricCurrents}
\end{figure}

\bigskip
\textbf{Interface currents }
-
The charge transfer faradaic current $i^{ct}$ and the double layer $i^{dl}$ currents flowing at the interfaces, described by eqs. \eqref{eq:BVtot_model3}-\eqref{eq:non_faradaic}, are shown in Fig.\ref{fig:ElectricCurrents} at different C-rates. The results account for the splitting of lithium flux in interstitial and hopping. 

The left column collects the evolution in time of the charge transfer faradaic current $i^{ct}$, at the $a/e$ and $e/c$ interfaces. At $t=0$ all currents vanish, in view of the thermodynamic equilibrium. The charge flow ramps up according to eq. \eqref{eq:current} and rapidly reaches a steady state, that is maintained in time. The hopping current ${ i^{ct}}_{{{\rm Li}^+_{\rm hop}}} $ is smaller than the corresponding interstitial at both interfaces, while their sum equals $i_{bat}(t)$ since $i^{dl}_{a/e}$ and $i^{dl}_{e/c}$ are vanishing after a short time.

The non-faradaic currents $i^{dl}_{a/e}$ and $i^{dl}_{e/c}$ are plot in the right columns of Fig.\ref{fig:ElectricCurrents} at different C-rates. The anodic double layer current is basically null throughout the discharge time, whereas the cathodic one reaches soon its maximum value, (still several order of magnitude smaller than  $i^{ct}$ ), and returns to zero at similar speed.

\begin{figure}[!htb]
    \centering
    \begin{subfigure}{.32\textwidth}
        \includegraphics[width=5.2cm]{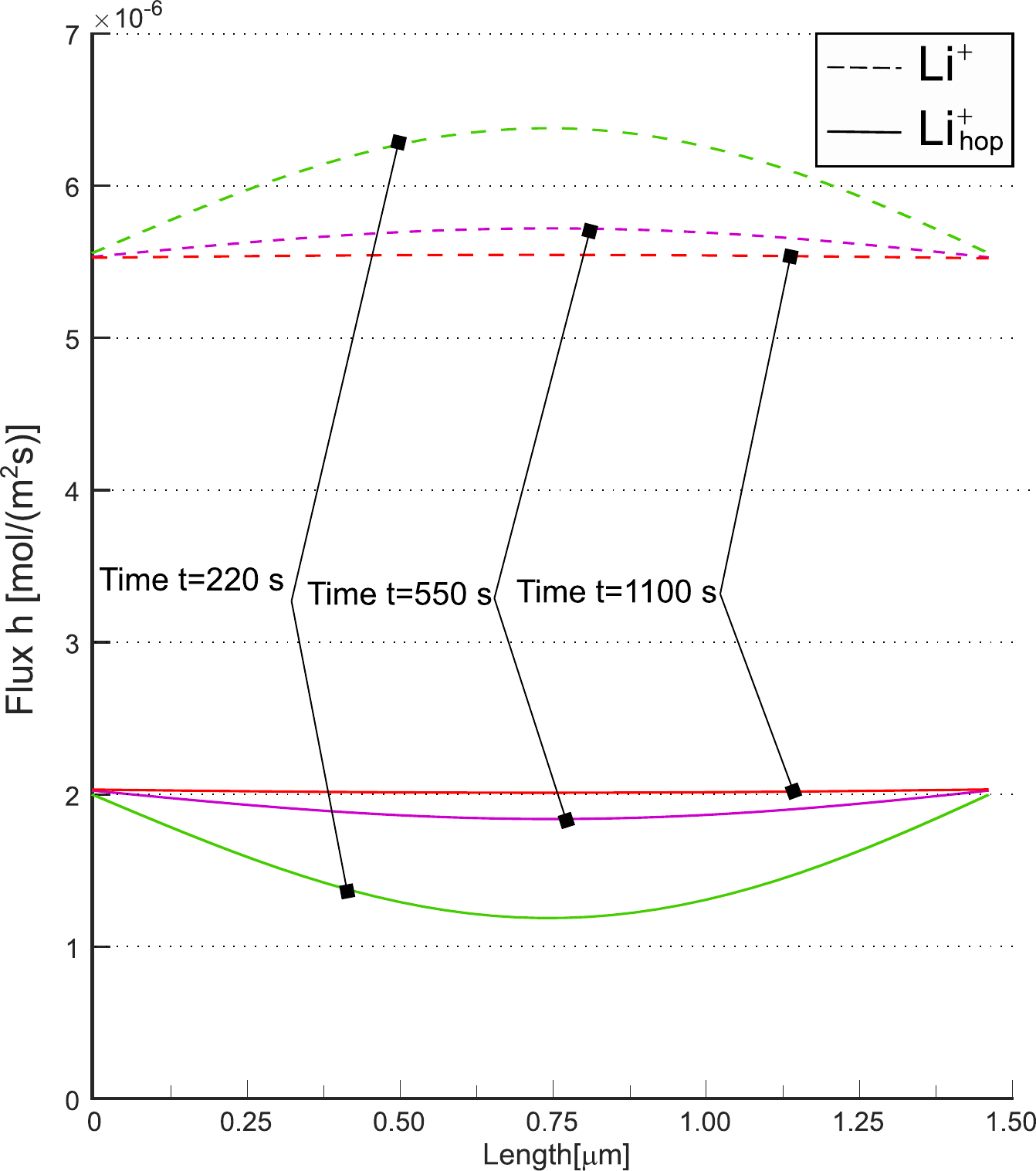}
        \caption{Total fluxes for  $C\!\!-\!rate\!=\!3.2$, $K_{\rm eq}^{\rm ion}   = 2000$ and $K_{\rm eq}^{\rm hop}  = 0.187 $. }
    \end{subfigure}\hfill%
        \begin{subfigure}{.32\textwidth}
        \centering
        \includegraphics[width=5.2cm]{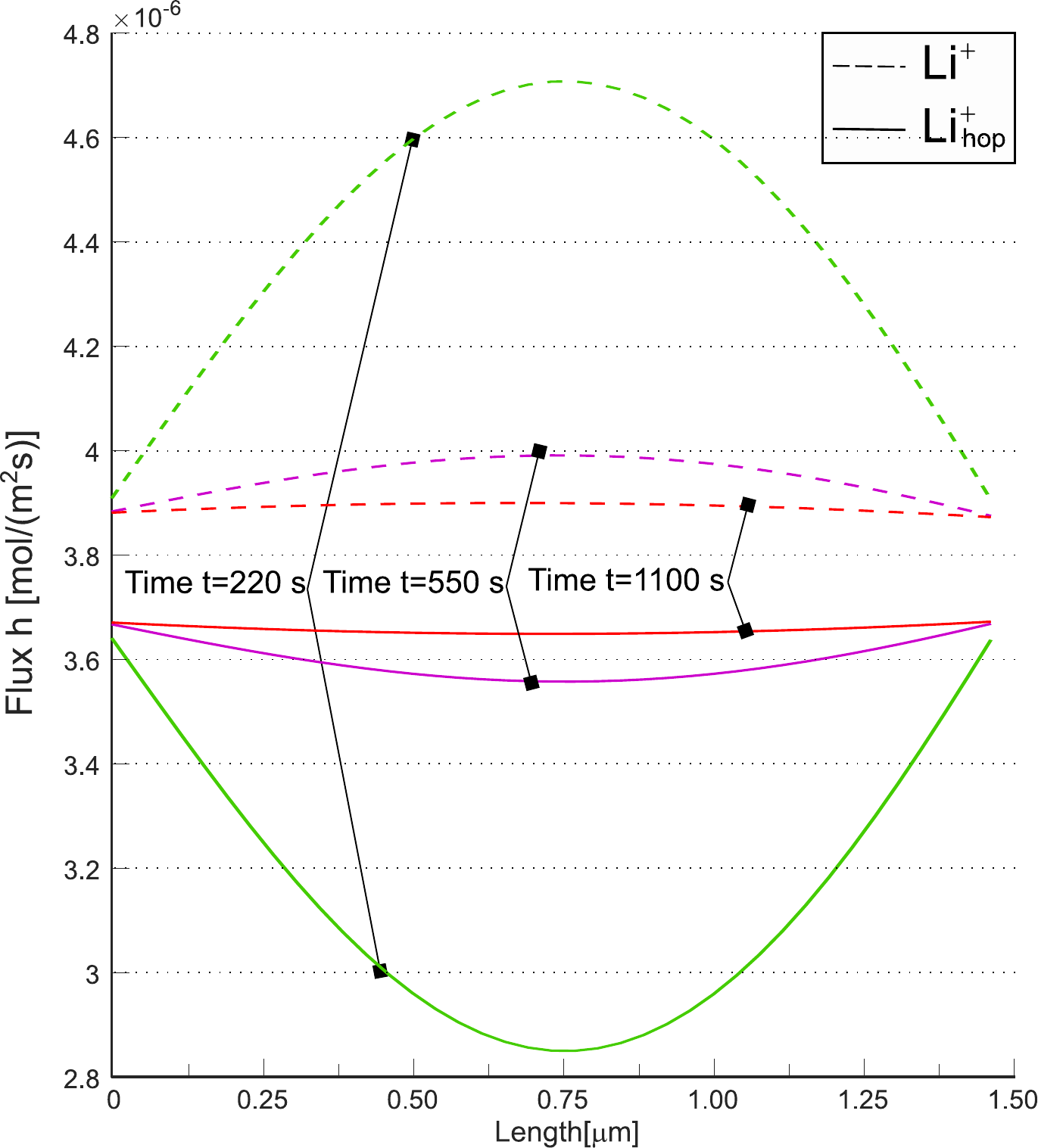}
        \caption{Total fluxes for $C\!\!-\!\!rate=3.2$, $K_{\rm eq}^{\rm ion}   = 1250$ and $K_{\rm eq}^{\rm hop}  = 0.9 $. }
    \end{subfigure}\hfill%
        \begin{subfigure}{.32\textwidth}
        \centering
        \includegraphics[width=5.2cm]{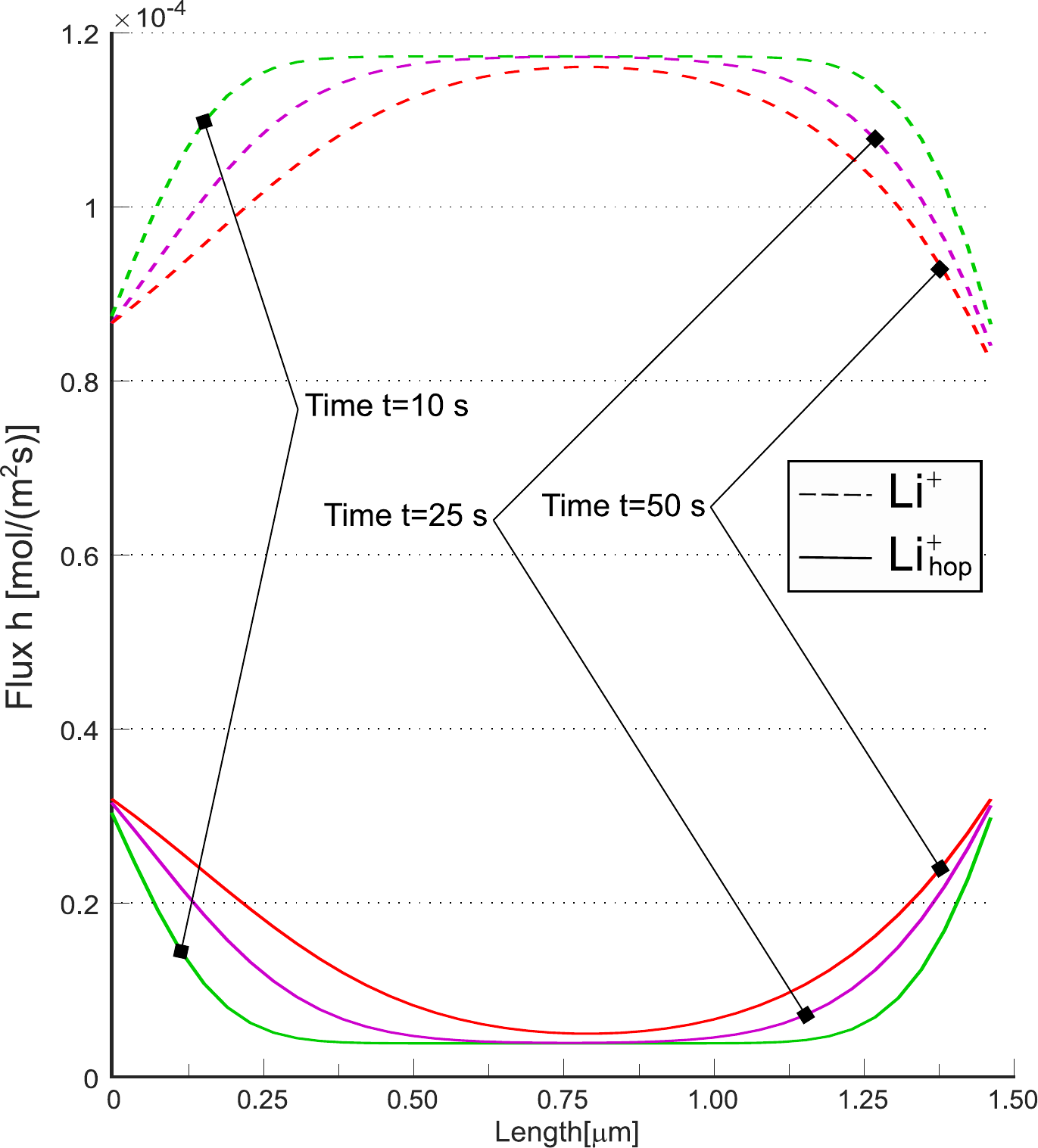}
        \caption{Total fluxes for $C\!\!-\!\!rate=51.2$, $K_{\rm eq}^{\rm ion}   = 1250$ and $K_{\rm eq}^{\rm hop}  = 0.9 $. }
    \end{subfigure}\hfill%
    \caption{\em Interstitial and hopping lithium fluxes inside the electrolyte at different C-rates and equilibrium constants }
    \label{fig:Fluxes}
\end{figure}

Currents and mass fluxes are related by the Faraday's constant $F$.
The mass fluxes of the two species of lithium across the electrolyte, $\vect{h}_{{\rm Li}^+}$ and $\vect{h}_{{\rm Li}^+_{\rm hop}}$ - evaluated according to eq. \eqref{eq:Nernst_Planck_extended}, are given in dashed and solid line in Fig. \ref{fig:Fluxes} respectively. 
Two $C-rates$ and two sets of equilibrium constants are considered. Figures \ref{fig:Fluxes}a and \ref{fig:Fluxes}b refer to the same C-rate (3.2) but to different equilibrium constants (specifically, fig.  \ref{fig:Fluxes}a kinetic constants are the ones in italic in Table \ref{Tab:Input}). Figures \ref{fig:Fluxes}b and \ref{fig:Fluxes}c share the same constants but refer to the different C-rates (3.2 vs 51.2). In all three cases, the fluxes at anode and cathode remains basically unchanged in time, confirming the evidence discussed in Fig.\ref{fig:ElectricCurrents}. The profile across the electrolyte however evolves towards a uniform profile, which corresponds to a steady state. 
In all cases both interstitial and hopping fluxes are positive, hence there is no counter-flux in any of the two mechanisms and lithium flows from the anode to the cathode in discharge.

Figure \ref{fig:Fluxes} highlights that the sum of the interstitial and hopping fluxes remains constant in space and time: this event is due to the electroneutrality. In fact, enforcing eq. \eqref{eq:electroneutrality:newmodel} in the linear combination of eqs. \eqref{eq:MassEle_model1b}-\eqref{eq:MassEle_model1d}, one easily gets
\begin{align}
\label{eq:electroneutrality:benchmark}
     & \divergence{  \vect{h}_{\rm Li^+} + \vect{h}_{{\rm Li}^+_{\rm hop}}  }=0
     &  0\leq x \leq L_e \; .
\end{align}
In turn, this implies that the total lithium $c_{\rm Li^+}^{tot} = c_{\rm Li^+} + c_{{\rm Li}^+_{\rm hop}} $ obeys the evolution equation
\begin{align}
	& \frac{\partial c_{\rm Li^+}^{tot}}{ \partial t} \; = w          
     	&  0\leq x \leq L_e 
	\; ,
\end{align}
thus highlighting the fundamental role of the ionization reaction \eqref{eq:IonizationReaction} in the charge/discharge process.

The hopping flux decreases from the electrodes toward the center of the electrolyte. Since the ionic flow goes from the anode to the cathode, this results in accumulation of hopping lithium at the anode and a depletion at the cathode. The interstitial lithium behaves in the opposite way and the same holds for vacancies that are not filled by the hopping lithium.

\begin{figure}[!htb]
    \centering
    \begin{subfigure}[!htb]{.49\textwidth}
        \centering
        \includegraphics[width=8cm]{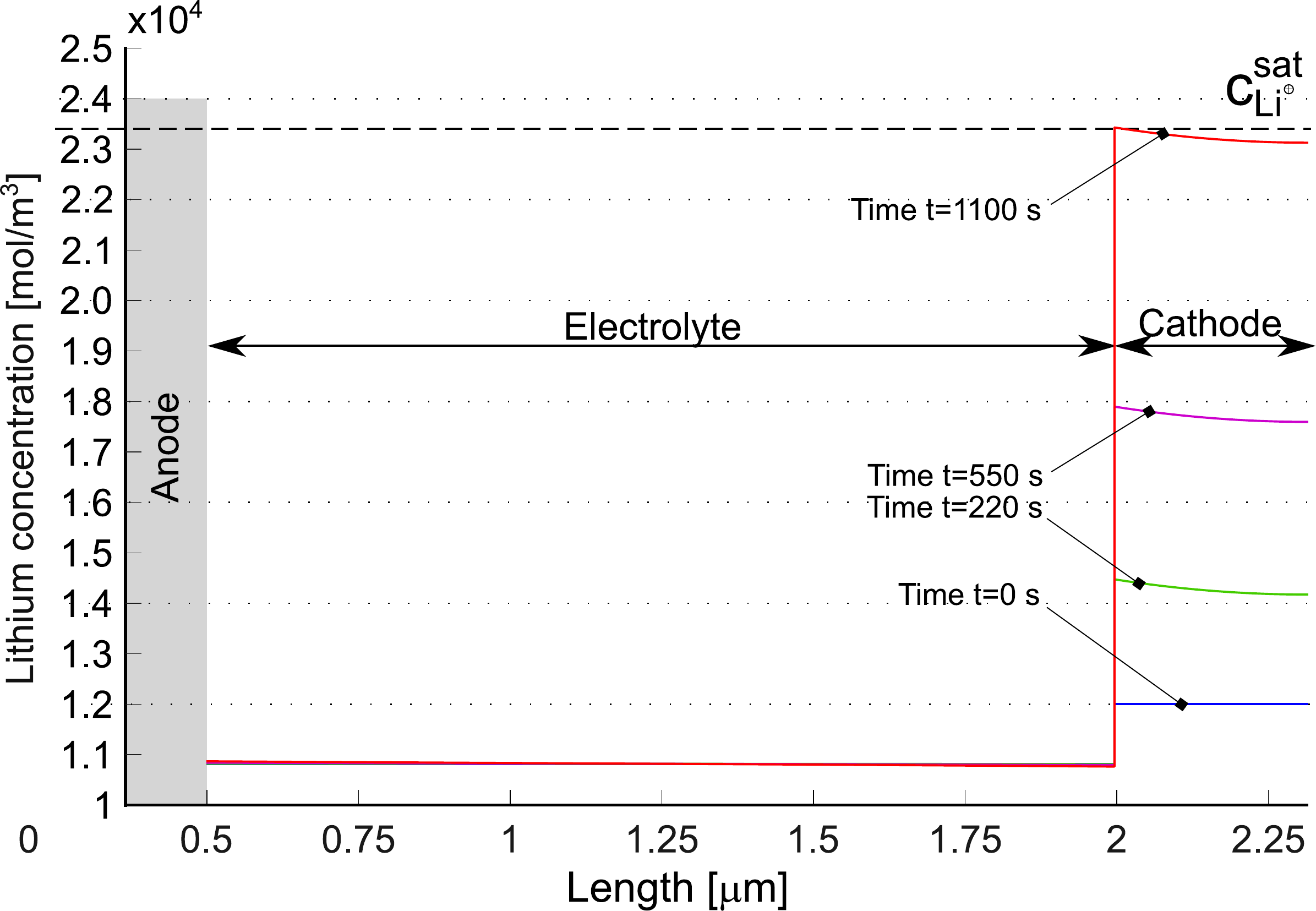}
        \caption{$C\!\!-\!\!rate=3.20$}
        \label{fig:ConcBatt_a}
    \end{subfigure}%
    \begin{subfigure}[!htb]{.49\textwidth}
       \centering
        \includegraphics[width=8cm]{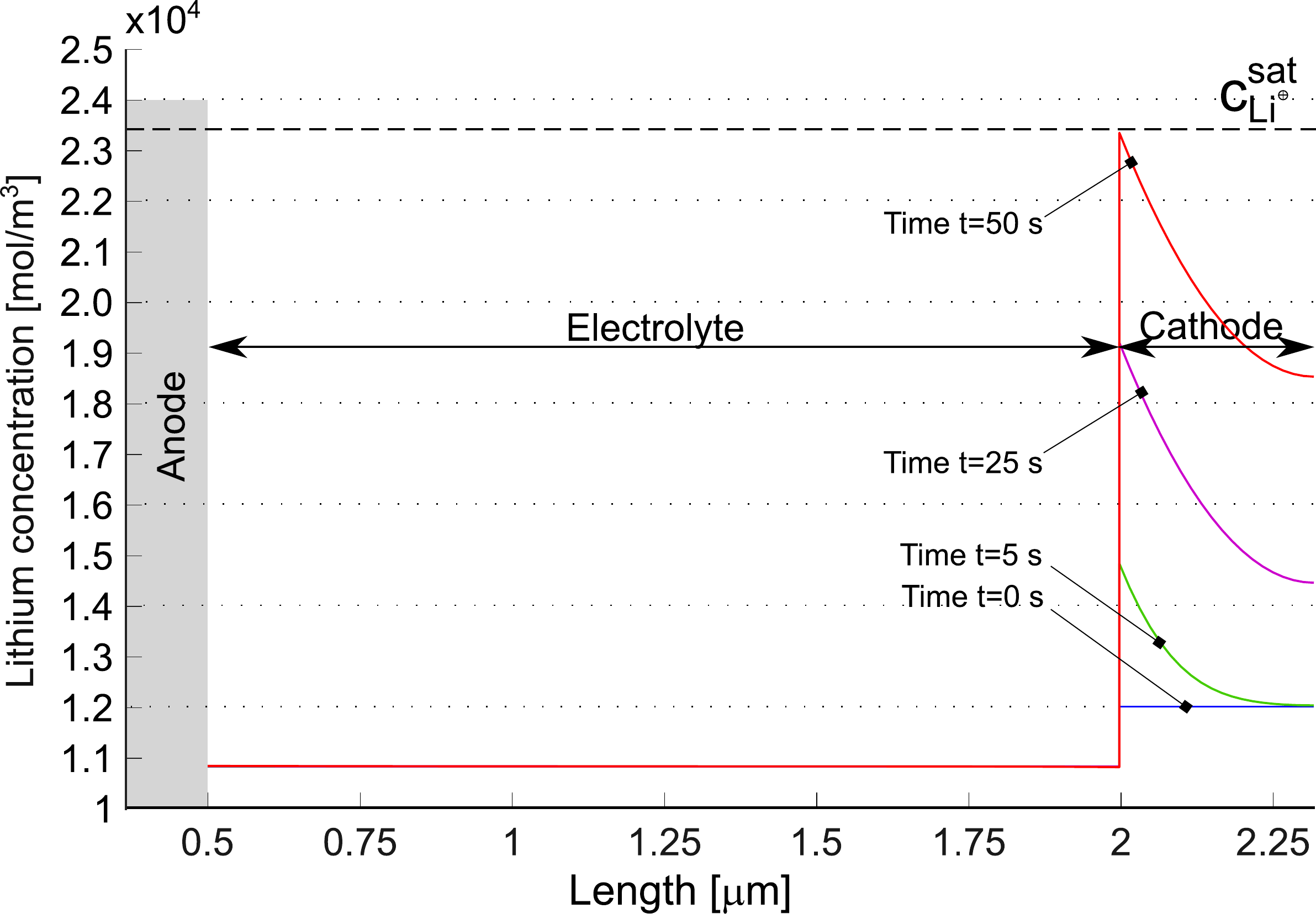}
        \caption{$C\!\!-\!\!rate=51.2$}
        \label{fig:ConcBatt_b}
    \end{subfigure}
     \centering
    \begin{subfigure}[!htb]{.49\textwidth}
        \centering
        \includegraphics[width=8cm]{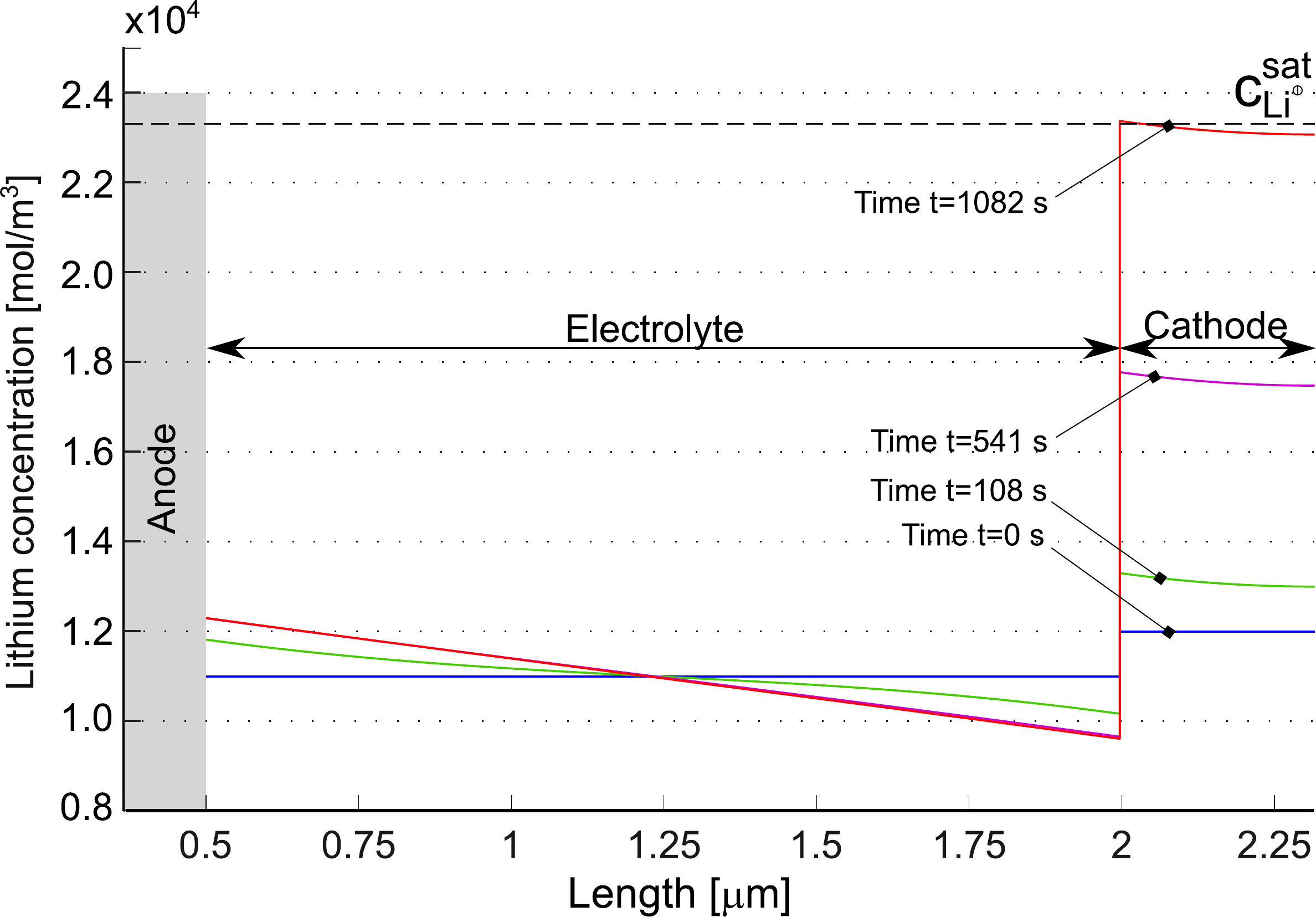}
        \caption{$C\!\!-\!\!rate=3.20$}
        \label{fig:ConcBatt_c}
    \end{subfigure}\hfill%
    \begin{subfigure}[!htb]{.49\textwidth}
       \centering
        \includegraphics[width=8cm]{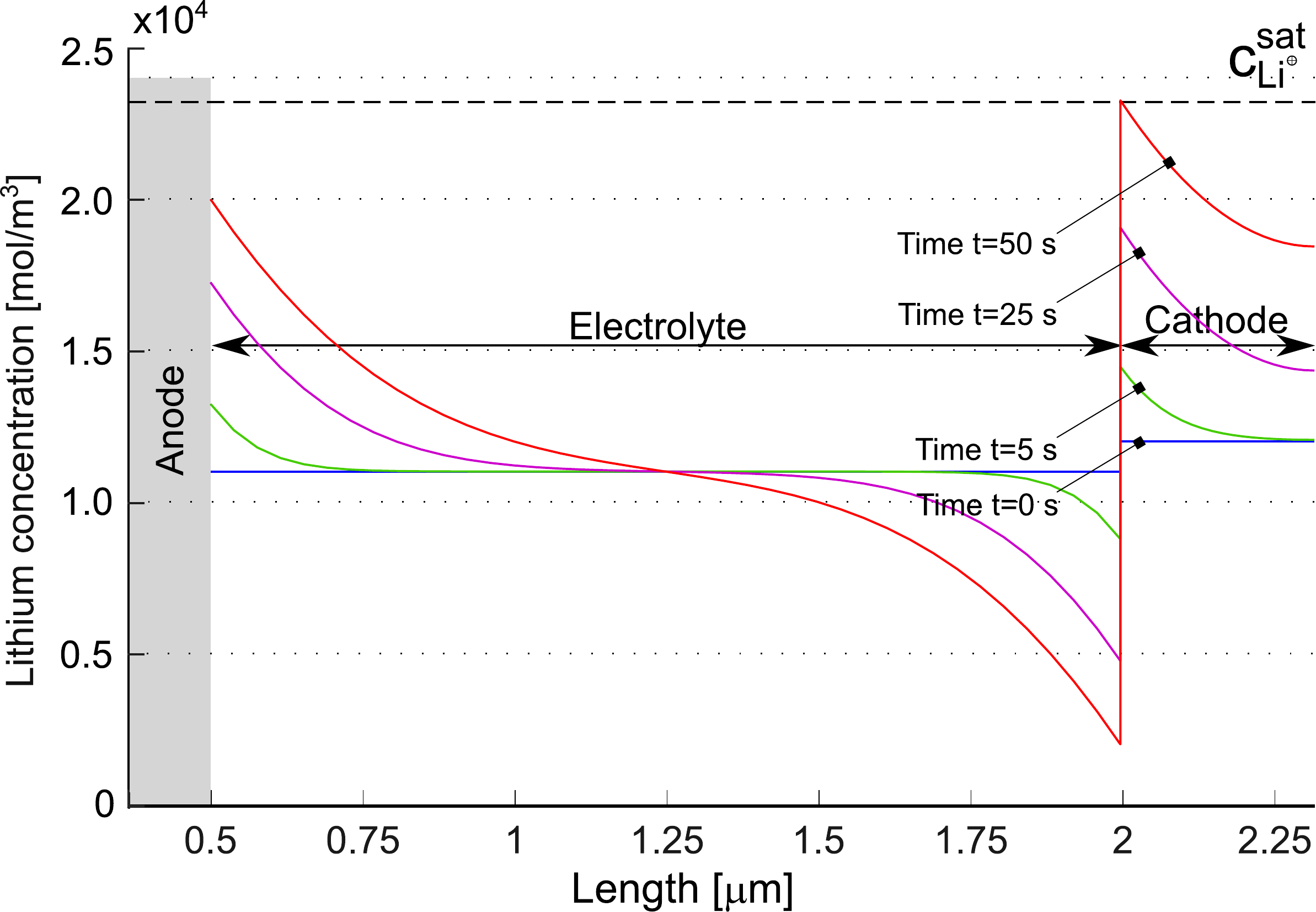}
        \caption{$C\!\!-\!\!rate=51.2$}
        \label{fig:ConcBatt_d}
        \end{subfigure}
    \caption{\em Lithium concentration profiles inside the battery  at different time steps for two different $C\!\!-\!\!rates$ for the new model ((a) and (b)) as well as for the model proposed in \cite{RaijmakersEtAlEA2020} ((c) and (d)).  The blue lines correspond to the initial time step, when the lithium concentrations inside the single components of the battery are in equilibrium and no profile is developed. The final time step, in red line, corresponds to the instant when the battery is considered discharged since the concentration of lithium $\rm Li^\oplus$ inside the cathode reaches the saturation limit $c_{\rm Li^\oplus}^{\rm sat}$. }
    \label{fig:ConcBatt}
\end{figure}

\bigskip
\textbf{Concentrations profiles }
-
Figure \ref{fig:ConcBatt} depicts the evolution of lithium concentration $c_{\rm Li^\oplus}(x)$ in the cathode and in the solid electrolyte for the new model ((a) and (b)) as well as for the model proposed in \cite{RaijmakersEtAlEA2020} ((c) and (d)). Since two ionic concentrations are concurrently present in the electrolyte, only their sum $(c_{\rm Li^+}\!+\!c_{{\rm Li}^+_{\rm hop}})$ has been plotted in fig. \ref{fig:ConcBatt}(a) and (b).

The anode, made of a metallic foil of lithium, is unaffected by the lithiation/delithiation processes and considered as an unlimited reservoir of lithium. The lithium ions intercalate inside the cathode and accumulates near at the electrolyte/cathode interface. The discharge process ends when lithium in cathode reaches its saturation limit $c_{\rm Li^\oplus}^{sat}=23400 \rm mol \, m^{-3}$. Saturation of the interface electrolyte/cathode is thus the limiting factor for the performance of this battery.

In \cite{RaijmakersEtAlEA2020}, the concentration of $(c_{\rm Li^+}\!+\!c_{{\rm Li}^+_{\rm hop}})$ ions in the electrolyte, initially uniform, increases near the anode interface, while decreasing at the cathode interface. This ``liquid electrolyte'' kind of behavior is justified since in \cite{RaijmakersEtAlEA2020} negative charges are allowed to move. In the novel formulation, negative charges are filled by hopping lithium, which in turn is allowed to intercalate. For this sake, the time evolution of negative charges and total lithium is driven by che ionization reaction rate $w$, as already discussed, and is very small in figures \ref{fig:ConcBatt}(a) and (b). 

\begin{figure}[!htb]
    \centering
    \begin{subfigure}[!htb]{.99\textwidth}
        \centering
        \includegraphics[width=15cm]{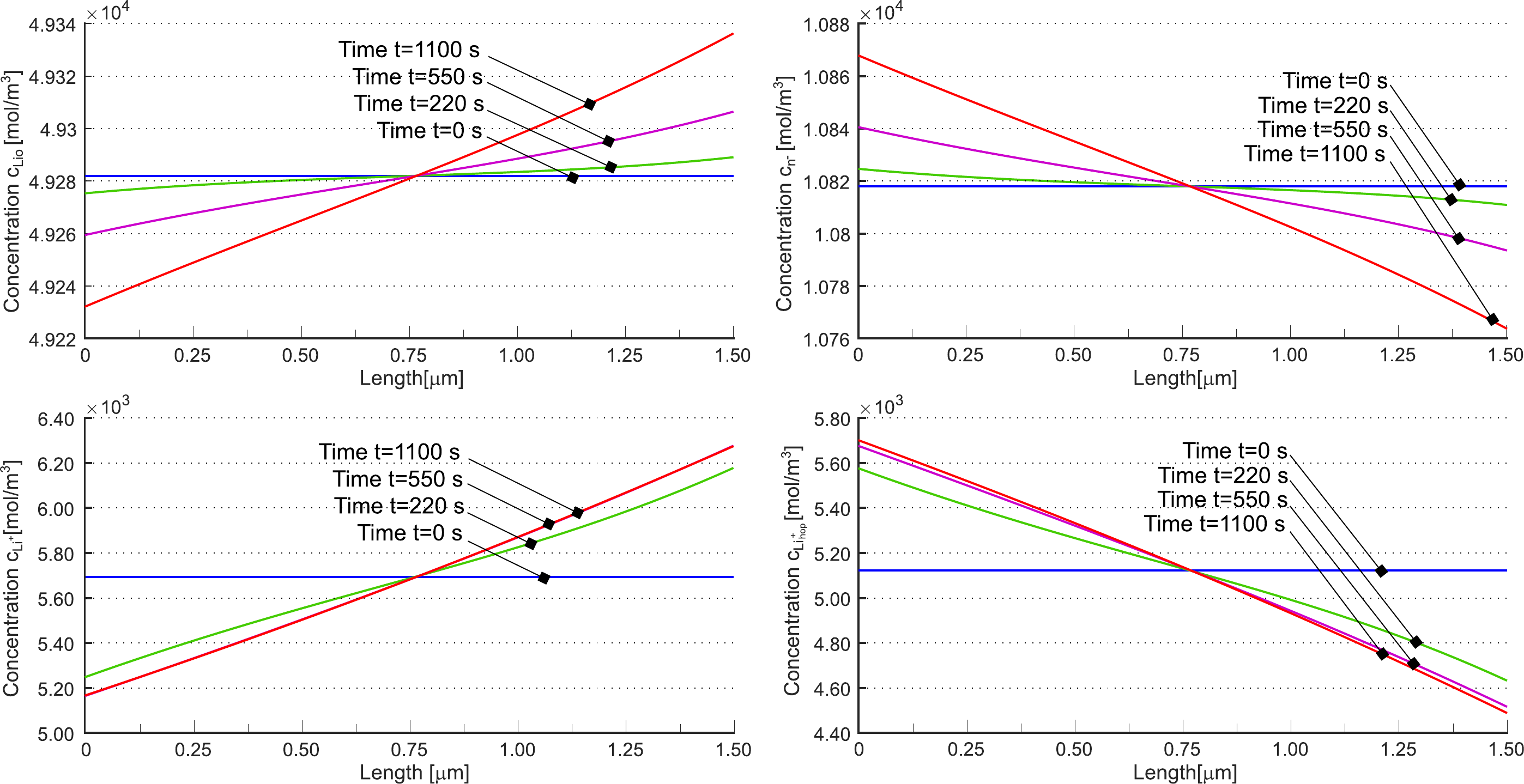}
        \caption{$C\!\!-\!\!rate=3.20$}
        \label{fig:ConcBatt1_a}
    \end{subfigure}\hfill%
    \begin{subfigure}[!htb]{.99\textwidth}
       \centering
        \includegraphics[width=15cm]{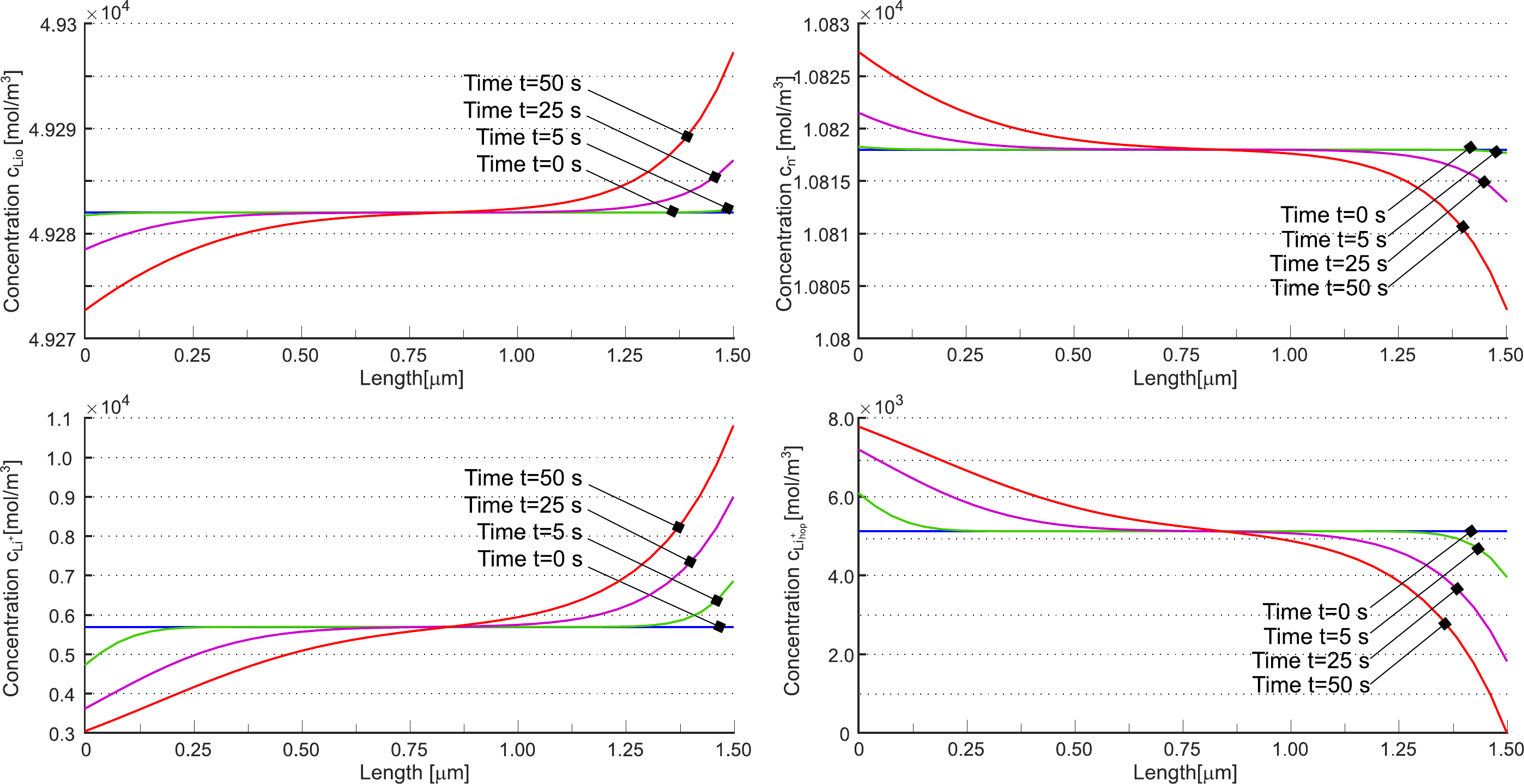}
        \caption{$C\!\!-\!\!rate=51.2$}
        \label{fig:ConcBatt1_b}
    \end{subfigure}
    \caption{\em Lithium concentration profiles of the different species $\rm c_{LiO}$, $\rm c_{n^-}$, $\rm c_{Li^+}$ and $\rm c_{Li^+_{hop}}$ inside the solid electrolyte at different time steps for a $C\!\!-\!rate\!=\!3.2$ (a) and a $C\!\!-\!rate\!=\!51.2$ (b). }
    \label{fig:ConcBatt1}
\end{figure}

Concentrations $\rm c_{LiO}$, $\rm c_{n^-}$, $\rm c_{Li^+}$ and $\rm c_{Li^+_{hop}}$ are plotted separately in Fig.\ref{fig:ConcBatt1}. As clearly visible, the concentration of hopping lithium resembles qualitatively the cationic behavior in \cite{RaijmakersEtAlEA2020}, at different C-rates. The number of uncompensated negative vacancies, balanced by the interstitial lithium, is higher at the cathode in discharge. Note that, in view of electroneutrality and of the assumption of no negative charges flow, the total concentration of lithium cannot vanish, differently from liquid electrolyte and from \cite{RaijmakersEtAlEA2020} (compare figs. \ref{fig:ConcBatt} (b) and (d) ) . Nonetheless, the contributions $\rm c_{Li^+}$ and $\rm c_{Li^+_{hop}}$ can separately become zero (see fig. \ref{fig:ConcBatt1} (d) ). The consequences of this event are yet unclear.

The concentration of $\rm Li^\oplus$ in the cathode increases with the discharge time, in agreement with the Li-intercalation reaction of Eq. \eqref{eq:e/c}. The higher the C-rate, the steeper the expected concentration gradient, hence diffusion in the electrode may become an issue at high C-rates. At $51.2$, this phenomena is clearly visible in fig. \ref{fig:ConcBatt} (b). Similar behavior occur  in \cite{RaijmakersEtAlEA2020} - see fig. \ref{fig:ConcBatt} (d).

\begin{figure}[!htb]
    \centering
    \begin{subfigure}[!htb]{.49\textwidth}
        \centering
        \includegraphics[width=6cm]{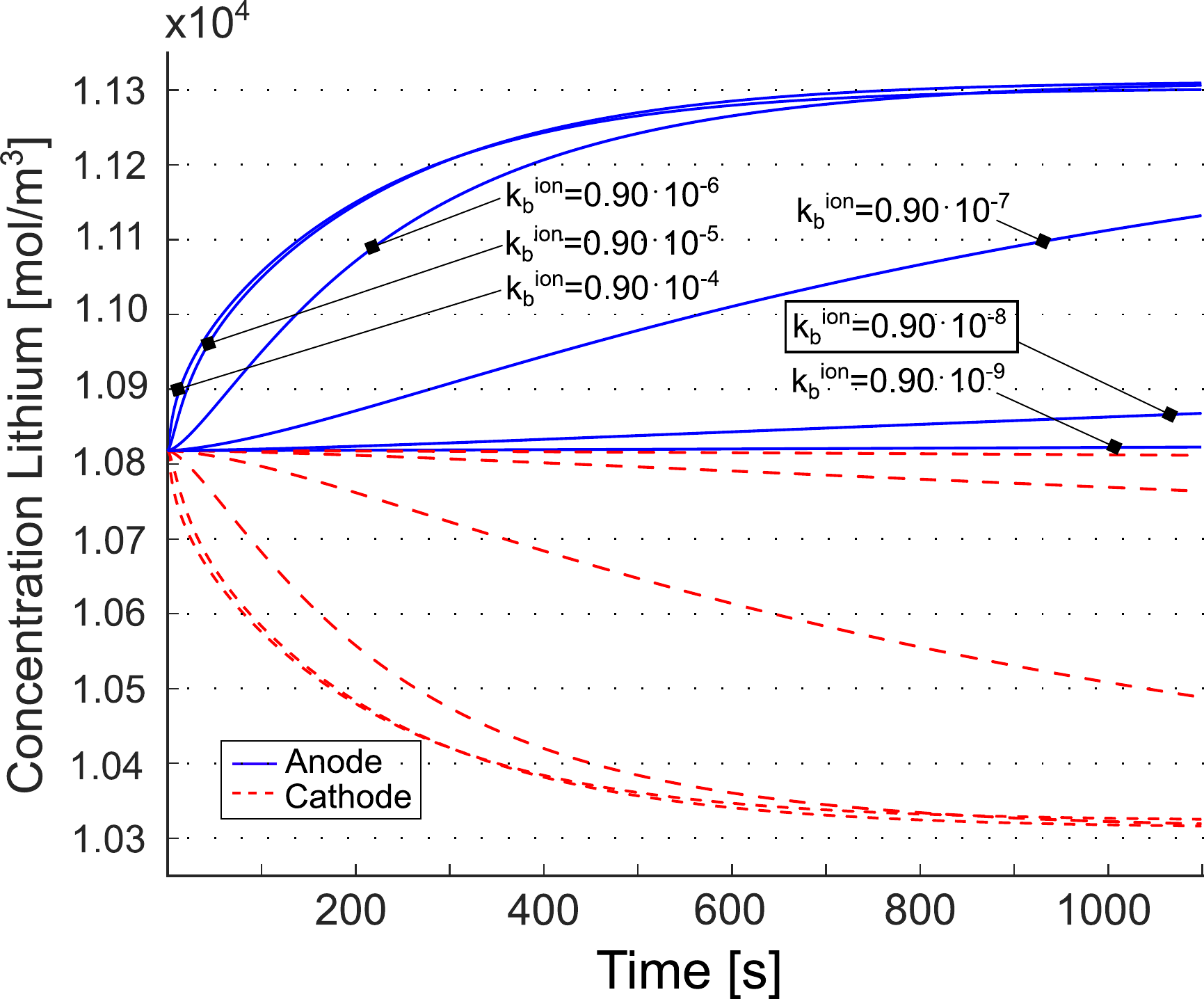} 
        \caption{$C\!\!-\!\!rate=3.20$}
        \label{fig:ConcBattElectrode_a}
    \end{subfigure}
    \begin{subfigure}[!htb]{.49\textwidth}
       \centering
        \includegraphics[width=6cm]{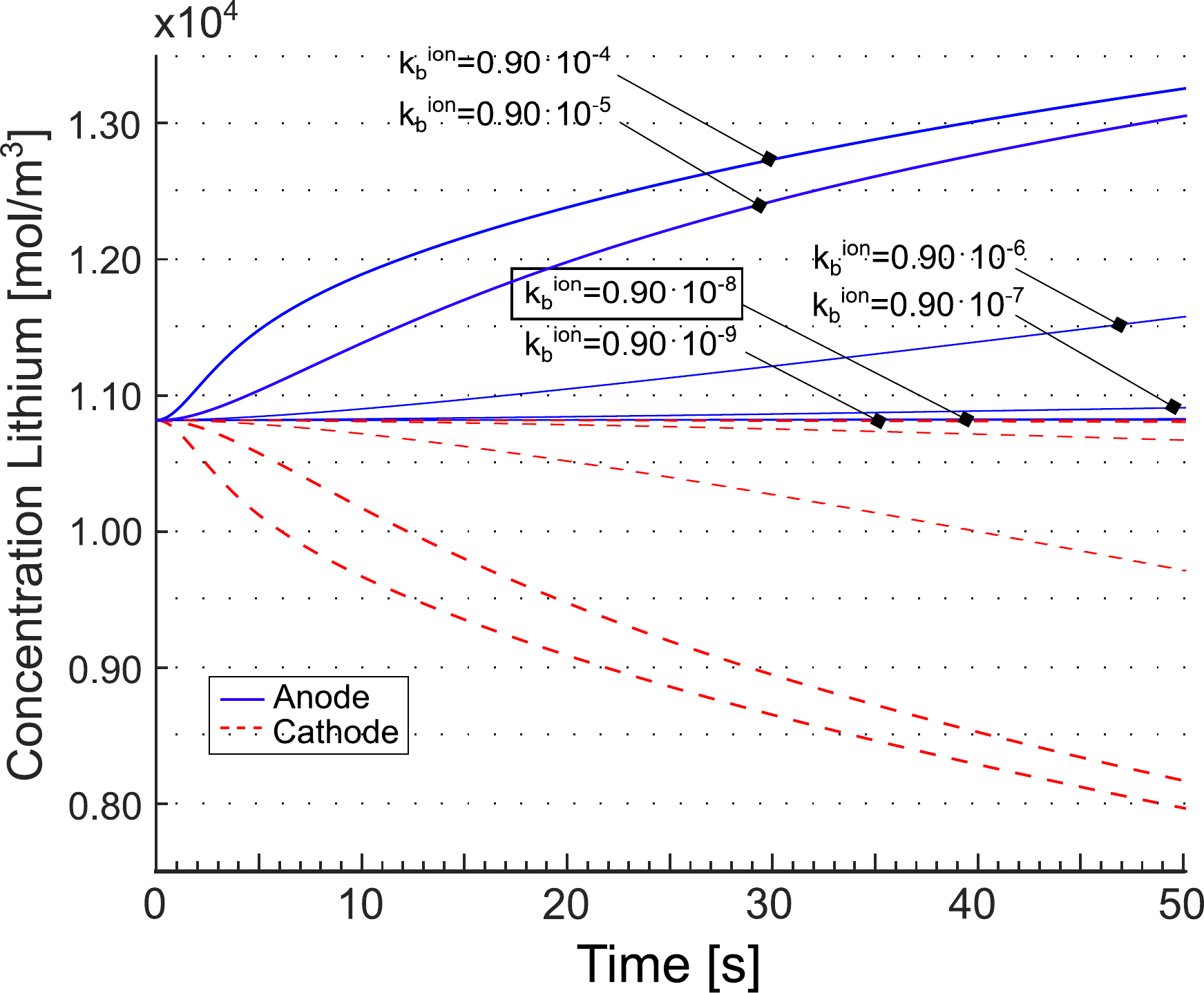} 
        \caption{$C\!\!-\!\!rate=51.2$}
        \label{fig:ConcBattElectrode_b}
    \end{subfigure}
    \centering
    \begin{subfigure}[!htb]{.49\textwidth}
        \centering
        \includegraphics[width=6cm]{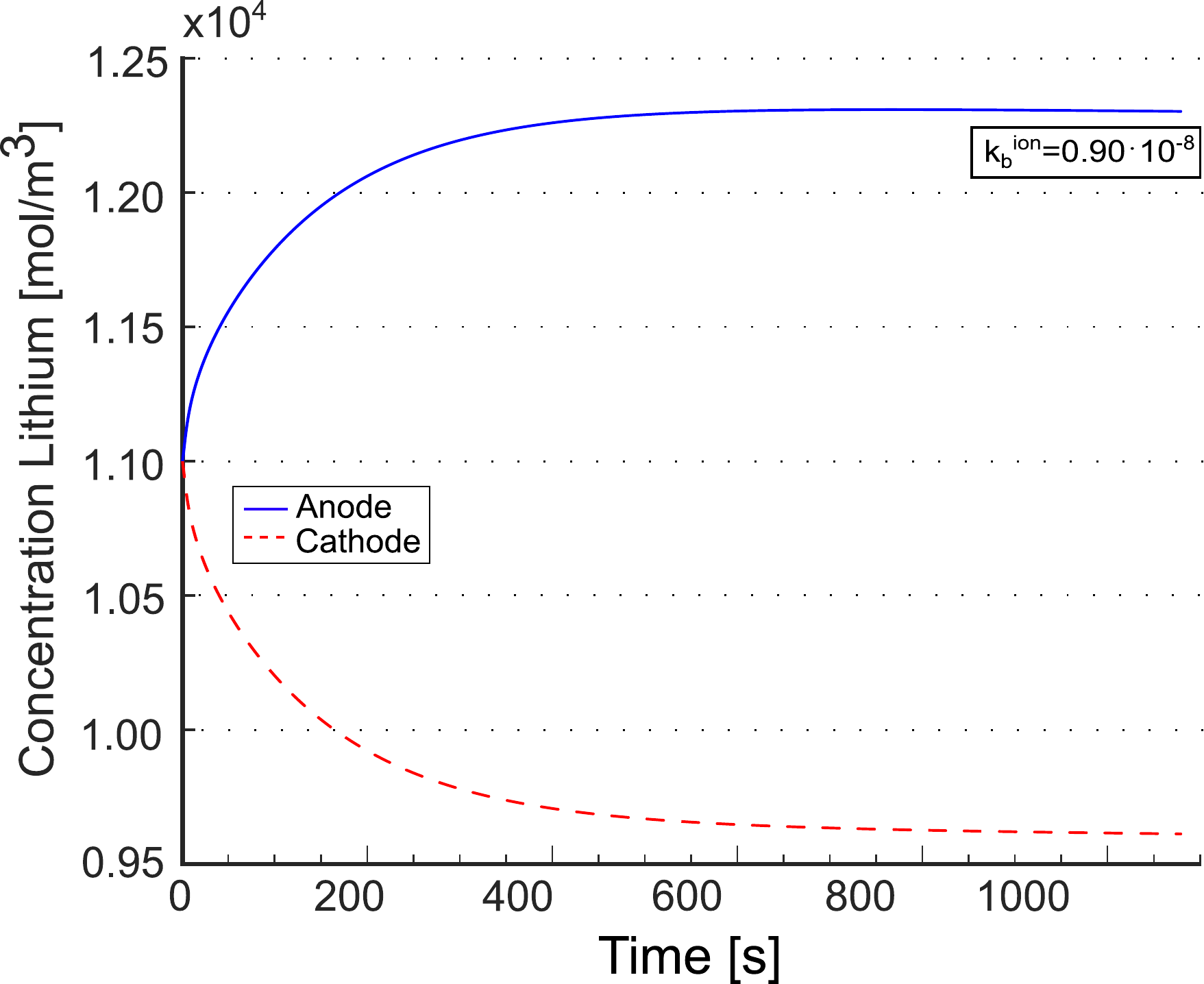}
        \caption{$C\!\!-\!\!rate=3.20$}
        \label{fig:ConcBattElectrode_c}
    \end{subfigure}
    \begin{subfigure}[!htb]{.49\textwidth}
       \centering
        \includegraphics[width=6cm]{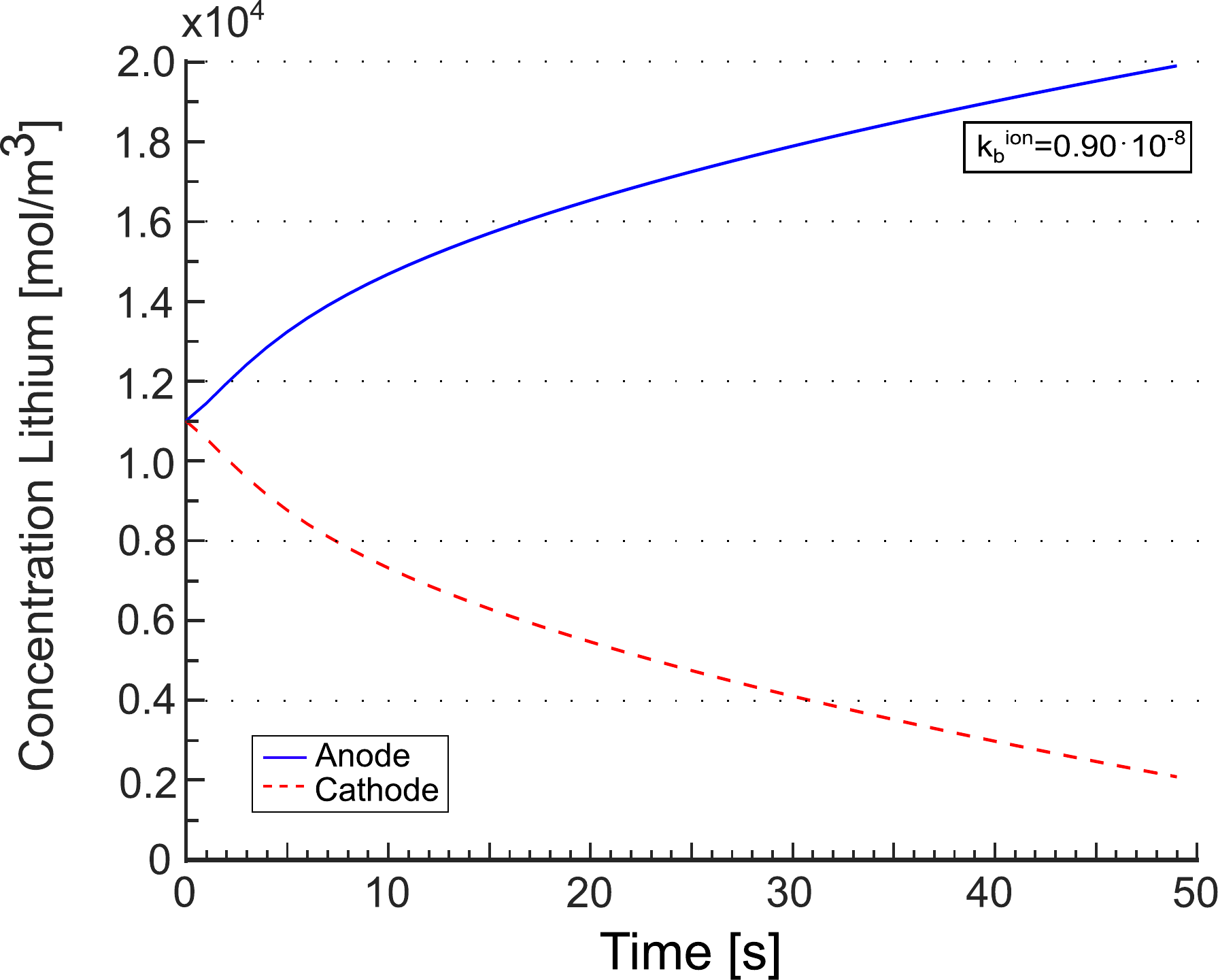}
        \caption{$C\!\!-\!\!rate=51.2$}
        \label{fig:ConcBattElectrode_d}
    \end{subfigure}
    \caption{\em Lithium concentration at the interfaces with the electrodes for two different $C\!\!-\!\!rates$ for the new model ((a) and (b)) as well as for the model proposed in \cite{RaijmakersEtAlEA2020} ((c) and (d)). The blue lines correspond to the values measured at the interface anode/electrolyte and the red line corresponds to the interface electrolyte/cathode. Note that in figures (a) and (b) the initial tangent vanishes: this is explained by the resulting electroneutrality coupled to the equilibrium condition $w=0$ imposed at $t=0$ in eq. \eqref{eq:MassEle_model1b}. }
    \label{fig:ConcBattElectrode}
\end{figure}

The total lithium concentration is plotted in fig. \ref{fig:ConcBattElectrode} for the novel ((a) and (b)) as well as for the model in \cite{RaijmakersEtAlEA2020} ((c) and (d)). For the novel model the tangent at $t=0$ vanishes: this is explained by the resulting electroneutrality coupled to the equilibrium condition $w=0$ imposed at $t=0$ in eq. \eqref{eq:MassEle_model1b}. The concentrations at the two electrodes evolve according to mass balance equations \eqref{eq:MassEle_model1c} and \eqref{eq:MassEle_model1d}.
As noticed in fig. \ref{fig:Fluxes}, the fluxes tend in time to achieve a uniform value across the electrolyte, i.e. the divergence term becomes less and less important and the  evolution of concentrations turns out to be driven by the evolution of $w$. It is thus expected that, at the same equilibrium constant, increasing the reaction constant (i.e. making the reaction faster) would allow reaching the steady state in due time. This is numerically confirmed in Fig. \ref{fig:ConcBattElectrode} (a) and (b), where the higher the $k_f^{\rm ion}$ and $k_b^{\rm ion}$, keeping $K_{\rm eq}^{\rm ion}  = 1250$, the sooner the concentration plateau is reached.  For the model in \cite{RaijmakersEtAlEA2020} the concentration response is similar to liquid electrolytes \cite{SalvadoriEtAlJPS2015,SalvadoriEtAlJPS2015b}, as expected. Note that steady state in that case is obtained only at low C-rates.

\section{Conclusion}
\label{sec06}

A novel model of ASSB has been compared, theoretically and computationally, to a few antecedents, in terms of a few quantities of interest such as interface currents, the electric potential, fluxes and concentrations profiles. While the four models dealt within the paper are presented at higher degree of complexity, they share some common assumptions. They are all thin-film based: the first model encapsulates this assumption in the constitutive equations, while the others are in principle fully three-dimensional and hence applicable also to composite electrode/electrolyte systems. All presented models are isothermal (no self-heating), although it is quite well known, at least in liquide electrolytes, that temperature affects the response of the battery \cite{LatzZauschBJN2015}.  All above assumptions shall be removed in order to replicate realistic behavior of batteries, especially with the aim of capturing the aging processes in ASSB that may preclude their own market success.

With battery cycling, the volume of electrodes changes due to repeated insertion and removal of Li atoms, causing disconnection and reduced contacts. Swelling is expected also in the solid electrolyte lattice, although this phenomena has certainly been less studied experimentally. In a recent work of Tian et al. \cite{Tian2017}, the effect of imperfect contact area was incorporated into a 1-D Newman battery model, by assuming the current and Li concentration will be localized at the contacted area of the interfaces. Among the presented publications, only the last one does not neglect volume changes during charge/discharge. In our simulations, though, the active surface area, where redox processes occur, were unaltered over cycling. Our future research plans are addressed to handle the mechanical problem in a three dimensional setting, with large strain mechanics as in \cite{Ganser2019, DiLeoIJSS2015a,Fathiannasab_2020,FATHIANNASAB2021229028}.

\vspace{6mm}
{\bf Acknowledgements}.  Authors express their gratitude to Prof. D. Danilov (TU/e, The Netherlands) and to Prof. P. McGinn (University of Notre Dame, USA) for sharing ideas and suggestions on modeling and on solid state batteries. 


\bibliographystyle{unsrt}
\bibliography{/Users/albertosalvadori/Bibliography/Bibliography}

\appendix

\section{Analytical OCP simulations}
\label{app:anOCP}

\begin{figure}[!htb]
\centering
\begin{tabular}{cc}
\includegraphics[width=80mm]{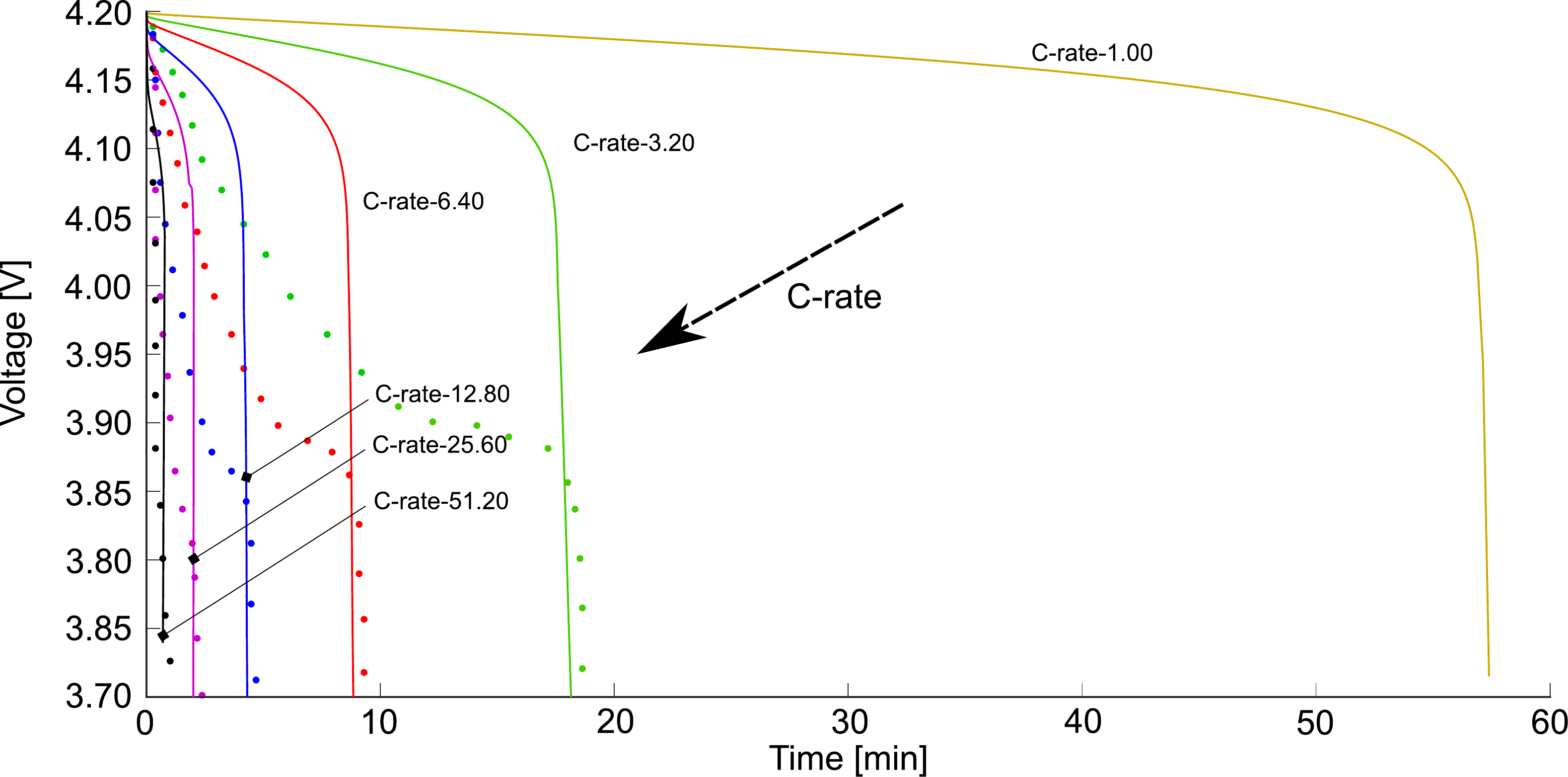} &\includegraphics[width=80mm]{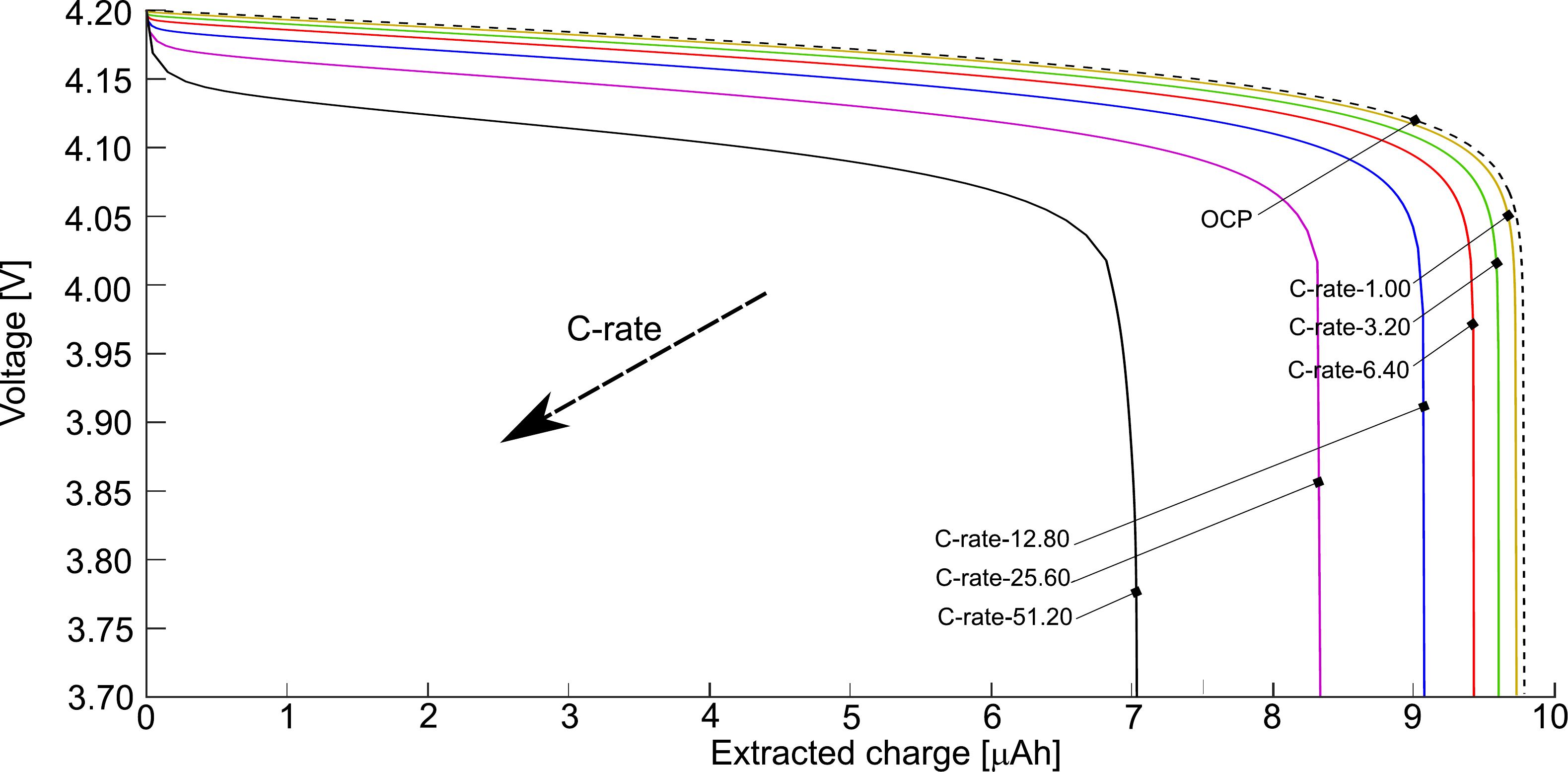}
\end{tabular}
\caption{\em Discharge curves as a function of time (a), and the extracted charge (b) for different $C\!\!-\!rates$, obtained following the analytical approach to evaluate the Ocp as given in \cite{PurkayasthaMcMeekingCM2012}.}
\label{fig:DischargeAnalytical}
\end{figure}

An approach to calculate the open circuit over-potential analytically is described in \cite{PurkayasthaMcMeekingCM2012}, based on the ideal chemical potential $\mu_{Li}$. The discharge curves as a function of the extracted charge, obtained following this analytical approach, are given in Fig.\ref{fig:DischargeAnalytical}.
The difference between the results obtained considering the OCP evaluated from experimental tests and from the analytical approach are not negligible for this kind of battery meaning the electro-chemical bahaviour must be described resorting to the experimental open circuit potential.

\section{Diffusion and migration fluxes}
\label{app:diffmigr}

\begin{figure}[!htb]
    \centering
    \begin{subfigure}[!htb]{.32\textwidth}
        \includegraphics[width=5.2cm]{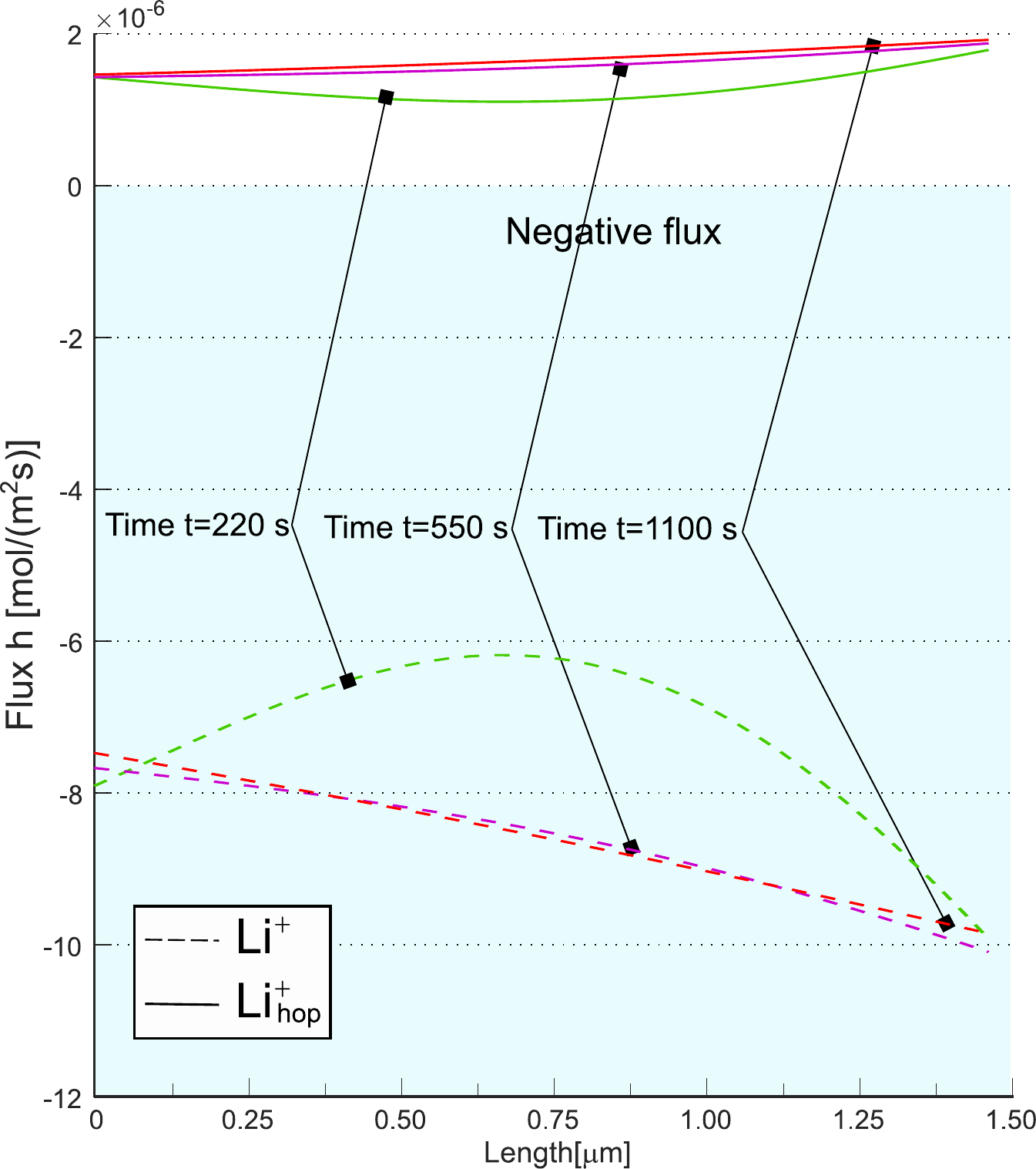}
        \caption{Diffusive}
        \label{fig:app:FluxBatt_a}
    \end{subfigure}\hfill%
        \begin{subfigure}[!htb]{.32\textwidth}
        \centering
        \includegraphics[width=5.2cm]{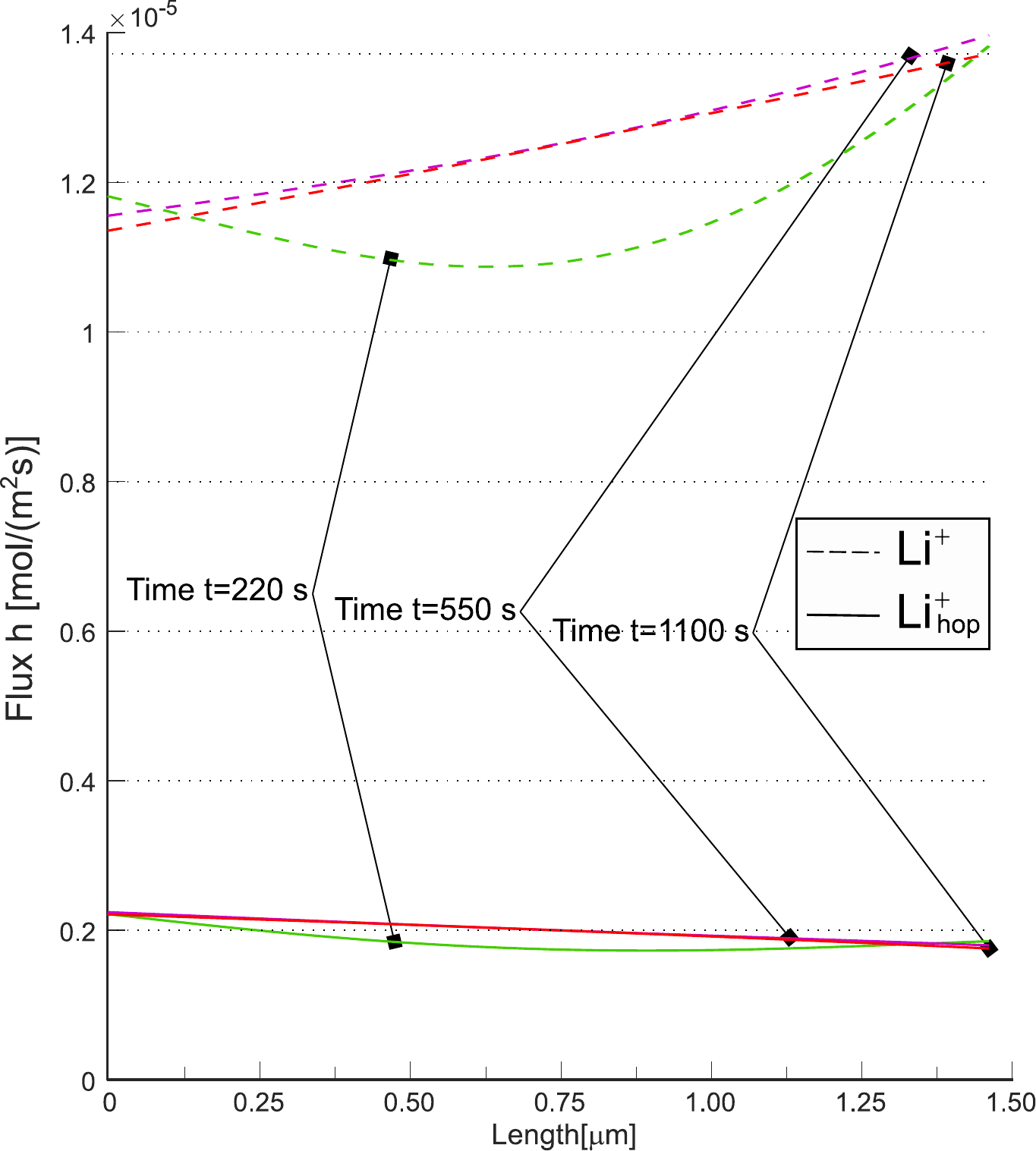}
        \caption{Migration}
        \label{fig:app:FluxBatt_b}
    \end{subfigure}\hfill%
        \begin{subfigure}[!htb]{.32\textwidth}
        \centering
        \includegraphics[width=5.2cm]{Flussi_3_2_3}
        \caption{Total fluxe}
        \label{fig:app:FluxBatt_c}
    \end{subfigure}
    \begin{subfigure}[!htb]{.32\textwidth}
       \centering
        \includegraphics[width=5.2cm]{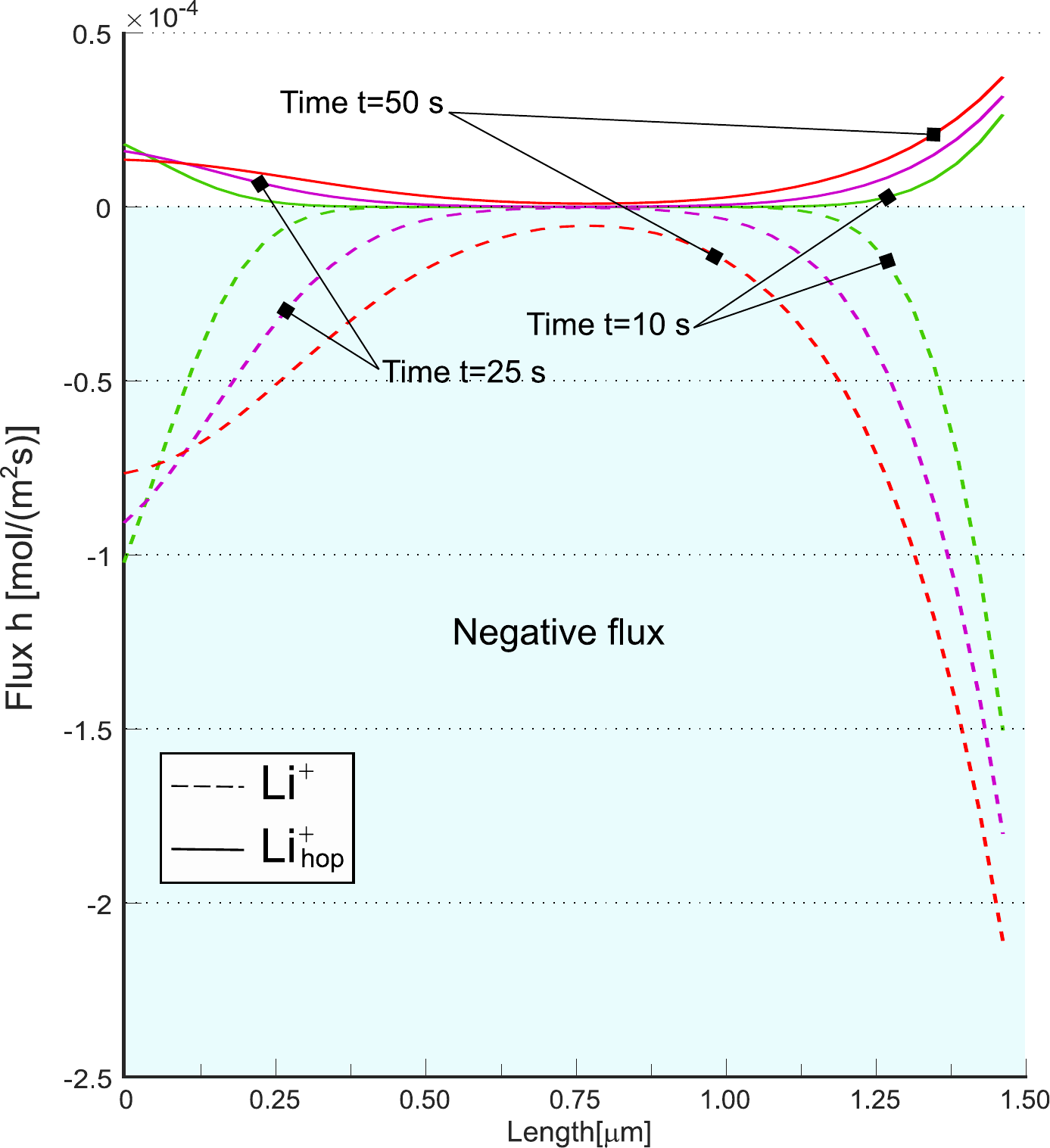}
        \caption{Diffusive}
        \label{fig:app:FluxBatt_d}
    \end{subfigure}
    \begin{subfigure}[!htb]{.32\textwidth}
       \centering
        \includegraphics[width=5.2cm]{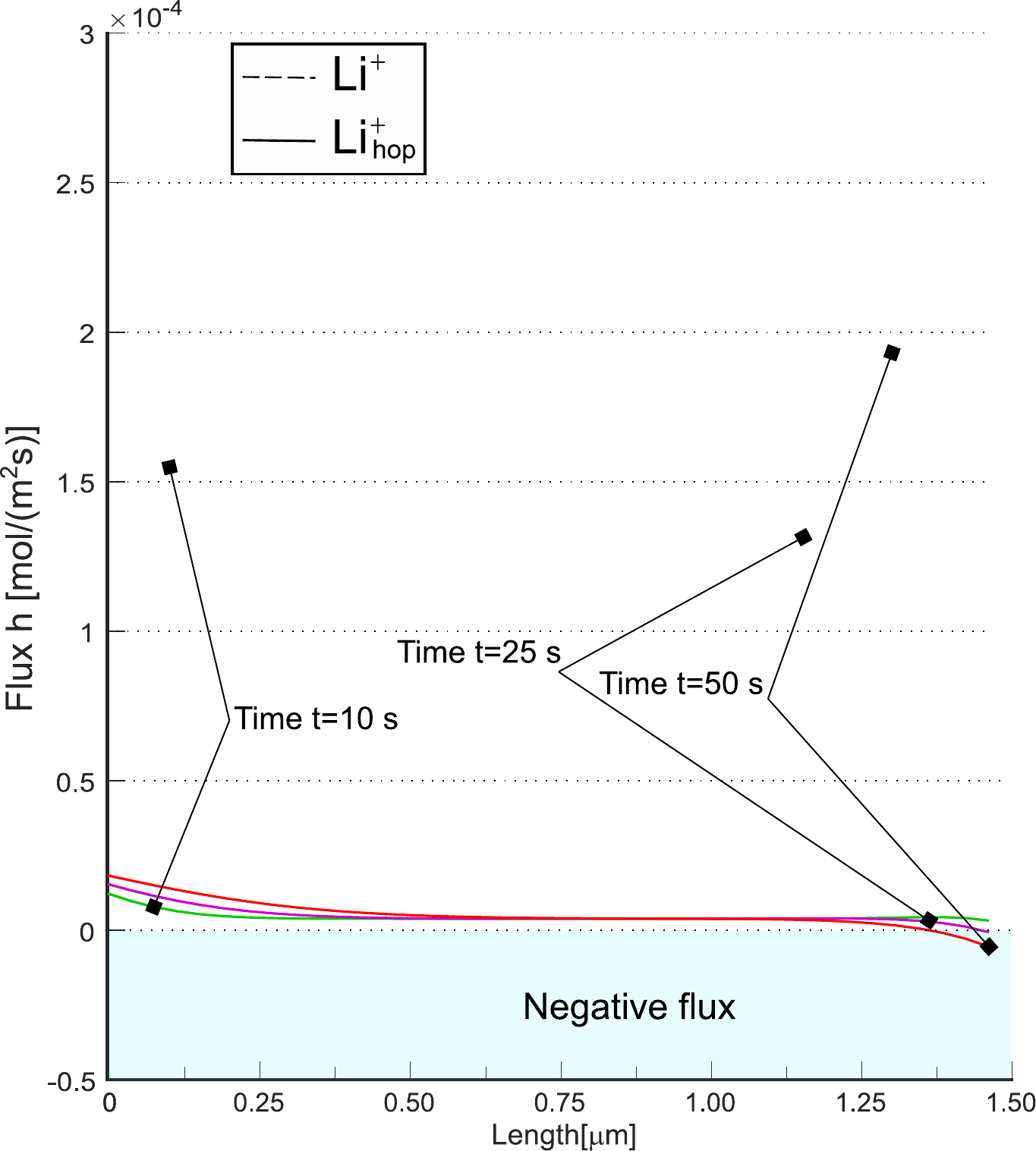}
        \caption{Migration}
        \label{fig:app:FluxBatt_e}
    \end{subfigure}
    \begin{subfigure}[!htb]{.32\textwidth}
       \centering
        \includegraphics[width=5.2cm]{Flussi_51_2_3}
        \caption{Total flux}
        \label{fig:app:FluxBatt_f}
    \end{subfigure}
    \caption{\em Lithium fluxes inside the electrolyte. (a)-(c) for $C\!\!-\!rate\!=\!3.2$, (d)-(f) for $C\!\!-\!rate\!=\!51.2$}
    \label{fig:app:Fluxes}
\end{figure}

In Fig.\ref{fig:app:Fluxes} the fluxes of the two species of lithium, $\rm Li^+$ and $\rm Li^+_{hop}$, inside the electrolyte are given in dashed and solid line respectively, for the two considered $C-rates$. The fluxes in Fig.\ref{fig:Fluxes}c,f have been calculated according to the Nernst-Plank equation \eqref{eq:model4_5}.  The total flux has been decomposed in  the diffusive part, emanating from from a concentration gradient Fig.\ref{fig:Fluxes}a,d, and in the migration term, stemming from the applied electric field Fig.\ref{fig:Fluxes}b,e.


\begin{figure}[!htb]
    \centering
    \begin{subfigure}[!htb]{.66\textwidth}
        \includegraphics[width=11cm]{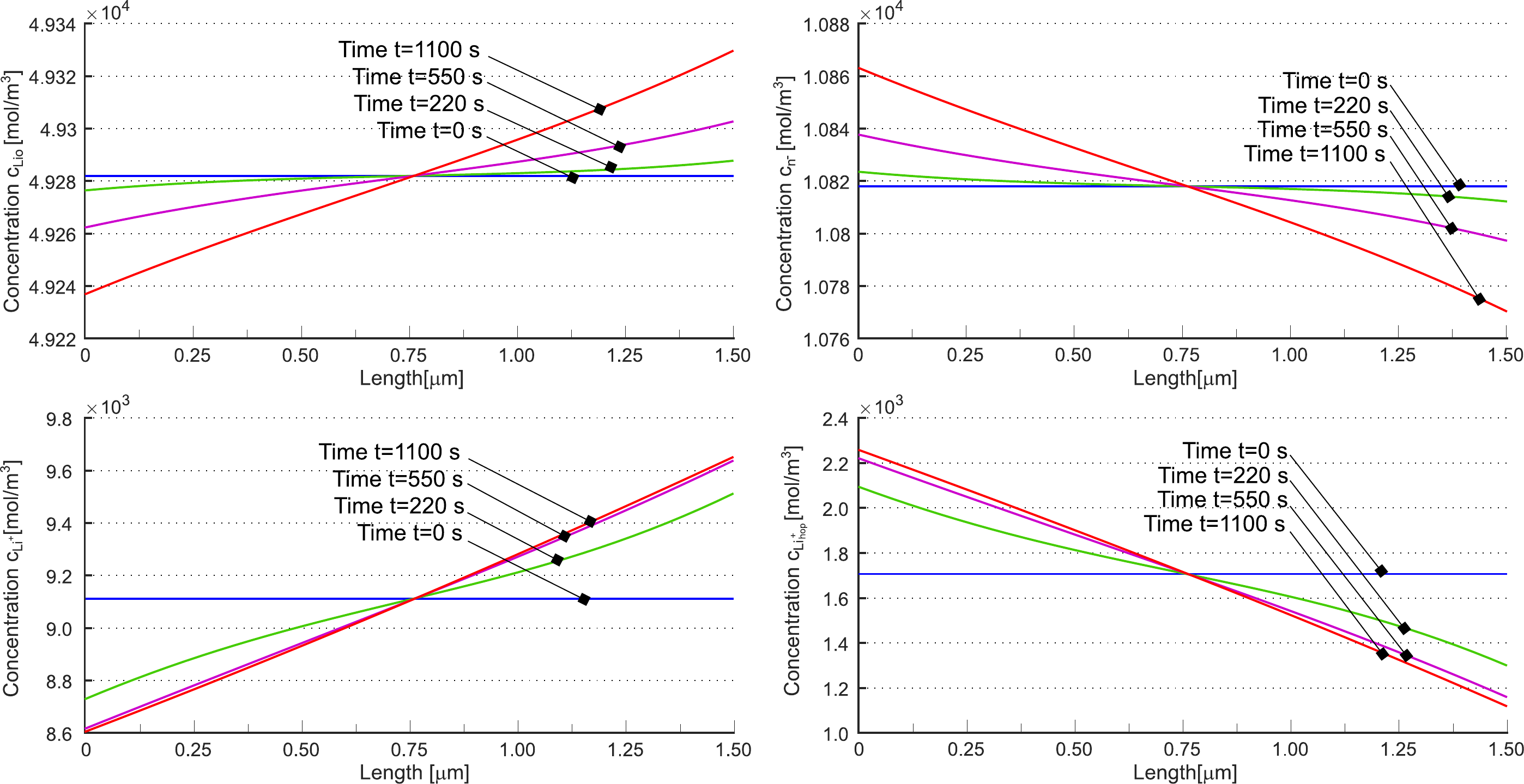}
        \caption{Concentration profiles}
        \label{fig:app:FluxBatt_a_2000}
    \end{subfigure}\hfill%
        \begin{subfigure}[!htb]{.33\textwidth}
        \centering
        \includegraphics[width=4.5cm]{Flussi_3_2_eq2000.pdf}
        \caption{Total flux}
        \label{fig:app:FluxBatt_b_2000}
    \end{subfigure}\hfill%
    \caption{\em a) Lithium concentration profiles of the different species $c_{\rm LiO}$, $c_{\rm n^-}$, $c_{\rm Li^+}$ and $c_{\rm Li^+_{hop}}$ inside the solid electrolyte at different time steps for a $C\!\!-\!rate\!=\!3.2$ and $k_{eq_1}=2000$. b) Lithium fluxes inside the electrolyte for $C\!\!-\!rate\!=\!3.2$ and $k_{eq_1}=2000$.}
    \label{fig:app:Fluxes_eq2000}
\end{figure}

\end{document}

%% file: definitions.tex
\newcommand{\beq}{\begin{equation}}
\newcommand{\eeq}{\end{equation}}

\newcommand{\beqar}{\begin{eqnarray}}
\newcommand{\eeqar}{\end{eqnarray}}
\newcommand{\bit}{\begin{itemize}}
\newcommand{\eit}{\end{itemize}}
\newcommand{\benum}{\begin{enumerate}}
\newcommand{\eenum}{\end{enumerate}}
\newcommand{\barr}{\begin{array}}
\newcommand{\earr}{\end{array}}

\def\XXint#1#2#3{{\setbox0=\hbox{$#1{#2#3}{\int}$}
   \vcenter{\hbox{$#2#3$}}\kern-.5\wd0}}

\def\b0{\boldsymbol{0}}

\newcommand{\bsigma}{\mbox{\boldmath $\sigma$}}

\newcommand{\bvarepsilon}{\mbox{\boldmath $\varepsilon$}}

\def\f0{\ensuremath{\mathbb{O}}}

